\documentclass[structabstract,longauth]{aa}

\usepackage{graphicx}
\usepackage{lscape}
\usepackage{amsmath}
\usepackage{txfonts}
\usepackage{url}
\usepackage{tabularx}
\usepackage{longtable}
\usepackage{multirow}
\usepackage{subcaption} 
\usepackage{upquote} 

\usepackage{array}    
\usepackage{booktabs} 

\usepackage{hyperref}

\makeatletter
\renewcommand*{\@fnsymbol}[1]{\ensuremath{\ifcase#1\or *\or \dagger\or \ddagger\or
    \mathsection\or \mathparagraph\or \|\or **\or \dagger\dagger
    \or \ddagger\ddagger \else\@ctrerr\fi}}
\makeatother

\begin{document} 

\title{\textit{Gaia} Data Release 2}
\subtitle{Summary of the variability processing and analysis results}

\titlerunning{\textit{Gaia} DR2 variability results}
\author{
B.~Holl\inst{\ref{inst1},\ref{inst4}}\fnmsep\thanks{Corresponding author: B. Holl
(\href{mailto:berry.holl@unige.ch}{\tt berry.holl@unige.ch})} \and  
M.~Audard\inst{\ref{inst1},\ref{inst4}} \and 
K.~Nienartowicz\inst{\ref{inst3}} \and  
G.~Jevardat~de~Fombelle\inst{\ref{inst3}} \and  
O.~Marchal\inst{\ref{inst4},\ref{inst25}} \and  
N.~Mowlavi\inst{\ref{inst1},\ref{inst4}} \and  
G.~Clementini\inst{\ref{inst22}} \and  
J.~De~Ridder\inst{\ref{inst6}} \and  
D.W.~Evans\inst{\ref{inst2}} \and  
L.P.~Guy\inst{\ref{inst4},\ref{inst28}} \and  
A.C.~Lanzafame\inst{\ref{inst7},\ref{inst8}} \and  
T.~Lebzelter\inst{\ref{inst21}} \and  
L.~Rimoldini\inst{\ref{inst4}} \and  
M.~Roelens\inst{\ref{inst1},\ref{inst4}} \and  
S.~Zucker\inst{\ref{inst19}} \and  
E.~Distefano\inst{\ref{inst8}} \and 
A.~Garofalo\inst{\ref{inst43},\ref{inst22}} \and 
I.~Lecoeur-Ta\"ibi\inst{\ref{inst4}} \and 
M.~Lopez\inst{\ref{inst10}} \and  
R.~Molinaro\inst{\ref{inst12}} \and  
T.~Muraveva\inst{\ref{inst22}} \and 
A.~Panahi\inst{\ref{int19b}} \and 
S.~Regibo\inst{\ref{inst6}} \and 
V.~Ripepi\inst{\ref{inst12}} \and 
L.M.~Sarro\inst{\ref{inst9}} \and 
C.~Aerts\inst{\ref{inst6},\ref{inst101}} \and 
R.I.~Anderson\inst{\ref{inst30}} \and  
J.~Charnas\inst{\ref{inst4}} \and  
F.~Barblan\inst{\ref{inst1}} \and 
S.~Blanco-Cuaresma\inst{\ref{inst1},\ref{inst44},\ref{inst45}} \and   
G.~Busso\inst{\ref{inst2}} \and  
J.~Cuypers\inst{\ref{inst5}}\thanks{Deceased on 28 February 2017} \and  
F.~De Angeli\inst{\ref{inst2}} \and  
F.~Glass\inst{\ref{inst1}} \and  
M.~Grenon\inst{\ref{inst1}} \and  
\'{A}.L.~Juh\'{a}sz\inst{\ref{inst11},\ref{inst27}} \and 
A.~Kochoska\inst{\ref{inst16},\ref{inst50}} \and 
P.~Koubsky\inst{\ref{inst15}} \and  
A.F.~Lanza\inst{\ref{inst8}} \and 
S.~Leccia\inst{\ref{inst12}} \and  
D.~Lorenz\inst{\ref{inst21}} \and 
M.~Marconi\inst{\ref{inst12}} \and 
G.~Marschalk\'{o}\inst{\ref{inst11},\ref{inst100}} \and  
T.~Mazeh\inst{\ref{int19b}} \and 
S.~Messina\inst{\ref{inst8}} \and 
F.~Mignard\inst{\ref{inst14}} \and 
A.~Moitinho\inst{\ref{inst18}} \and 	
L.~Moln\'{a}r\inst{\ref{inst11}} \and 
S.~Morgenthaler\inst{\ref{inst20}} \and 
I.~Musella\inst{\ref{inst12}} \and 
C.~Ordenovic\inst{\ref{inst14}} \and 
D.~Ord\'o\~nez\inst{\ref{inst4}} \and  
I.~Pagano\inst{\ref{inst8}} \and 
L.~Palaversa\inst{\ref{inst2},\ref{inst1}} \and  
M.~Pawlak\inst{\ref{inst40},\ref{inst41}} \and 
E.~Plachy\inst{\ref{inst11}} \and  
A.~Pr\v{s}a\inst{\ref{inst16}} \and  
M.~Riello\inst{\ref{inst2}} \and  
M.~S\"{u}veges\inst{\ref{inst29}} \and  
L.~Szabados\inst{\ref{inst11}} \and 
E.~Szegedi-Elek\inst{\ref{inst11}} \and 
V.~Votruba\inst{\ref{inst15}} \and
L.~Eyer\inst{\ref{inst1},\ref{inst4}}
%
} 
\authorrunning{Holl et al.}
\institute{Department of Astronomy, University of Geneva, Ch. des Maillettes 51, CH-1290 Versoix, Switzerland\label{inst1}
%
\and 
Department of Astronomy, University of Geneva, Ch. d'Ecogia 16, CH-1290 Versoix, Switzerland\label{inst4}
\and 
SixSq, Rue du Bois-du-Lan 8, CH-1217 Geneva, Switzerland\label{inst3}
\and
G\'EPI, Observatoire de Paris, Universit\'{e} PSL, CNRS, Place Jules Janssen 5, F-92195 Meudon, France\label{inst25} 
\and
INAF Osservatorio di Astrofisica e Scienza dello Spazio di Bologna, Via Gobetti 93/3, I - 40129 Bologna, Italy\label{inst22}
%
\and
Instituut voor Sterrenkunde, KU Leuven,  Celestijnenlaan 200D, 3001 Leuven, Belgium\label{inst6}
%
\and
Institute of Astronomy, University of Cambridge, Madingley Road, Cambridge CB3 0HA, UK\label{inst2} 
\and
Large Synoptic Survey Telescope, 950 N. Cherry Avenue, Tucson, AZ 85719, USA\label{inst28} 
\and
Universit\`a di Catania, Dipartimento di Fisica e Astronomia, Sezione Astrofisica, Via S. Sofia 78, I-95123 Catania, Italy\label{inst7} 
\and
INAF-Osservatorio Astrofisico di Catania, Via S. Sofia 78, I-95123 Catania, Italy\label{inst8}
\and
University of Vienna, Department of Astrophysics, Tuerkenschanz\-strasse 17, A1180 Vienna, Austria\label{inst21}
\and
Department of Geosciences, Tel Aviv University, Tel Aviv 6997801, Israel\label{inst19}
\and 
Dipartimento di Fisica e Astronomia, Universit\`{a} di Bologna, Via Piero Gobetti 93/2, 40129 Bologna, Italy \label{inst43} 
\and
Departamento de Astrof\'isica, Centro de Astrobiolog\'ia (INTA-CSIC), PO Box 78, E-28691 Villanueva de la Ca\~nada, Spain\label{inst10}
\and
School of Physics and Astronomy, Tel Aviv University, Tel Aviv 6997801, Israel\label{int19b}
\and
INAF-Osservatorio Astronomico di Capodimonte, Via Moiariello 16, 80131, Napoli, Italy\label{inst12}
\and
Dpto. Inteligencia Artificial, UNED, c/ Juan del Rosal 16, 28040 Madrid, Spain\label{inst9} 
\and 
Department of Astrophysics/IMAPP, Radboud University, P.O.Box 9010, 6500 GL Nijmegen, The Netherlands\label{inst101}
\and 
European Southern Observatory, Karl-Schwarzschild-Str. 2, D-85748 Garching b. M\"{u}nchen, Germany\label{inst30} 
\and 
Laboratoire d'astrophysique de Bordeaux, Univ. Bordeaux, CNRS, B18N, all{\'e}e Geoffroy Saint-Hilaire, 33615 Pessac, France \label{inst44}
\and 
Harvard-Smithsonian Center for Astrophysics, 60 Garden Street, Cambridge, MA 02138, USA\label{inst45}
\and
Royal Observatory of Belgium, Ringlaan 3, B-1180 Brussels, Belgium\label{inst5}
\and
Konkoly Observatory, Research Centre for Astronomy and Earth Sciences, Hungarian Academy of Sciences, H-1121 Budapest, Konkoly Thege Mikl\'{o}s \'{u}t 15-17, Hungary\label{inst11}
\and 
Department of Astronomy, E\"otv\"os Lor\'and University, P\'azm\'any P\'eter s\'et\'any 1/a, H-1117, Budapest, Hungary\label{inst27} 
\and
Villanova University, Department of Astrophysics and Planetary Science, 800 Lancaster Ave, Villanova PA 19085, USA \label{inst16}
\and
Faculty of Mathematics and Physics, University of Ljubljana, Jadranska ulica 19,
1000 Ljubljana, Slovenia \label{inst50} 
\and
Academy of Sciences of the Czech Republic, Fricova 298, 25165 Ondrejov, Czech Republic\label{inst15}
\and
Baja Observatory of University of Szeged, Szegedi \'{u}t III/70, H-6500 Baja, Hungary\label{inst100}
\and
Universit\'e C{\^o}te d'Azur, Observatoire de la C{\^o}te d'Azur, CNRS, Laboratoire Lagrange, Bd de l'Observatoire, CS 34229, 06304 Nice cedex 4, France \label{inst14}
\and
CENTRA FCUL, Campo Grande, Edif. C8, 1749-016 Lisboa, Portugal\label{inst18}
\and
EPFL SB MATHAA STAP, MA B1 473 (Bâtiment MA), Station 8, CH-1015 Lausanne, Switzerland\label{inst20}
\and
Warsaw University Observatory, Al. Ujazdowskie 4, 00-478 Warszawa, Poland \label{inst40} 
\and
Institute of Theoretical Physics, Faculty of Mathematics and Physics, Charles University in Prague, Czech Republic \label{inst41} 
%
%
\and
Max Planck Institute for Astronomy, K\"{o}nigstuhl 17, 69117 Heidelberg, Germany\label{inst29} 
%
%
%
%
}


\date{Received ; accepted}

\abstract
   { 
   The \textit{Gaia} Data Release 2 (DR2) contains more than half a million sources that are identified as variable stars. 
   }
   {
   We summarise the processing and results of the identification of variable source candidates of RR Lyrae stars, Cepheids, long-period variables (LPVs), rotation modulation (BY Dra-type) stars, $\delta$ Scuti and SX Phoenicis stars, and short-timescale variables. 
 In this release we aim to provide useful but not necessarily complete samples of candidates.
   }
%
   { 
    The processed \textit{Gaia} data consist of the $G$, $G_{\rm BP}$, and $G_{\rm RP}$  photometry during the first 22~months of operations as well as 
positions and parallaxes. Various methods from classical statistics, data mining, and time-series analysis were applied and tailored to the specific properties of \textit{Gaia} data, as were various visualisation tools to interpret the data. 
   }
%
   {
 The DR2 variability release contains
   228\,904 RR Lyrae stars, 11\,438~Cepheids, 151\,761~LPVs, 147\,535~stars with rotation modulation, 8\,882~$\delta$ Scuti and SX Phoenicis stars, and 3\,018~short-timescale variables. These results are distributed over a classification and various Specific Object Studies tables in the \textit{Gaia} archive, along with the three-band time series and associated statistics for the underlying 550\,737 unique sources. We estimate that about half of them are newly identified variables.
   The variability type completeness varies strongly as a function of sky position as a result of the non-uniform sky coverage and intermediate calibration level of these data. 
   The probabilistic and automated nature of this work implies certain completeness and contamination rates that are quantified 
   so that users can anticipate their effects. This means that even well-known variable sources can be missed or misidentified in the published data.
   }
   {
The DR2 variability release only represents a small subset of the processed  
data. Future releases will include more variable sources and data products; however, DR2 shows the (already) very high quality of the data and great promise for  variability studies.
   }

\maketitle

%
\section{Introduction
\label{sec:introduction}}
The Coordination Unit 7 (CU7) of the \textit{Gaia} Data Processing and Analysis Consortium (DPAC) is tasked to process the calibrated data of variable objects detected by \textit{Gaia}.
In \textit{Gaia} Data Release~1 \cite[DR1,][]{GDR1}, we published the light curves and properties of a sample of Cepheids and RR Lyrae stars detected in the Large Magellanic Cloud (LMC), see \cite{DPACP-13}.
We refer to \cite{EyerDR1} for a detailed description of the CU7 framework and processing pipeline.
In this second data release \cite[DR2,][]{DPACP-36}, we extend the
variability types for which we publish light curves to the following list of classes: Cepheids, RR Lyrae
stars, long-period variables (LPV), short-timescale variables, stars with
rotation modulation, and $\delta$ Scuti and SX Phoenicis stars. For some
of them, additional specific properties are inferred and published.
This constitutes roughly 100 times as many variable sources as were published in DR1 \citep{EyerDR1, DPACP-13}.
Moreover, it covers the whole sky. 
The variability analysis for this release focused primarily on large-amplitude variable stars and rotation modulated stars.
Other classes of variable stars are aimed to be introduced in later releases. We note that eclipsing binaries
were identified as well, but will be treated separately and only delivered
in future releases. We emphasize that specific selection criteria were
applied to each variability type to limit contamination. Completeness
within each variability type was not aimed at for DR2.

The various variability type-specific methods that were used for the published DR2 results are discussed in full length in the following dedicated papers:
\begin{itemize}
\item \textit{DR2: Variable stars in the Gaia colour-magnitude diagram}: \cite{LE-035}
\item \textit{DR2: All-sky classification of high-amplitude pulsating stars}: \cite{Rimoldini18Geq2} \
\item \textit{DR2: Specific characterisation and validation of all-sky Cepheids and RR
Lyrae stars}: \cite{Clementini18CepRrl}
\item \textit{DR2: Rotation modulation in late-type dwarfs}: \cite{Lanzafame18RotMod}
\item \textit{DR2: Short-timescale variability processing and analysis}:  \cite{Roelens18Sts}
\item \textit{DR2: The first Gaia catalogue of long-period variable candidates}: \cite{Mowlavi18Lpv}
\item \textit{DR2: Validation of the classification of RR Lyrae and Cepheid variables with the \textit{Kepler} and \textit{K2} missions: 
\cite{Molnar18CepRrl}}
\end{itemize}
This overview paper focuses on presenting 
the general properties of the exported sample of variable sources in the following way:
Section~\ref{sec:data} introduces the time-series photometric data and their relevant properties. Section~\ref{sec:overviewVarProcessing} presents a summarised overview of the data processing and analyses. Section~\ref{sec:results} gives an overview of the results, including our time-series filters, published time-series, classification, and Specific Object Studies (SOS) tables containing the results. Section~\ref{sec:conclusions} presents the final conclusions. Additional information regarding the \textit{Gaia} archive queries used in this paper is provided in Appendix~\ref{sec:archiveQueries}. 

The data are publicly 
available in the online \textit{Gaia} archive at \href{http://gea.esac.esa.int/archive/}{\tt http://gea.esac.esa.int/archive/} containing the `tables' and `fields' referred to in the rest of this article, as well as all of the `DR2 documentation' 
 and catalogue `data~model' 
 listed at \href{http://gea.esac.esa.int/archive/documentation/}{\tt http://gea.esac.esa.int/archive/documentation/}.


\section{Data
\label{sec:data}}

The results in DR2 are based on the first 22 months of \textit{Gaia} data, taken between 25~July 2014 and 23~May 2016.
For the variability analysis we made use of the photometric time series in the $G$, $G_{\rm BP}$, and $G_{\rm RP}$ bands, using field-of-view (FoV) averaged transit photometry.
An exception to this was the analysis of short-timescale variables,
for which some per-CCD $G$-band photometry was also analysed, but not published.
The photometry is described in detail in \cite{DWE-052} and \cite{MR-023}.
We emphasise that owing to the tight DR2 data-processing schedule, we only had access to a {preliminary} version of the DR2 astrometric solution \citep{DPACP-51}, on which
our results are based. No radial velocity data or astrophysical parameters were available during our processing.

\subsection{Relative completeness\label{sec:skycompleteness}}
\begin{figure*}[bh!]
\centering
\vspace{-0.2cm}
\includegraphics[angle=0,width=0.8\textwidth]{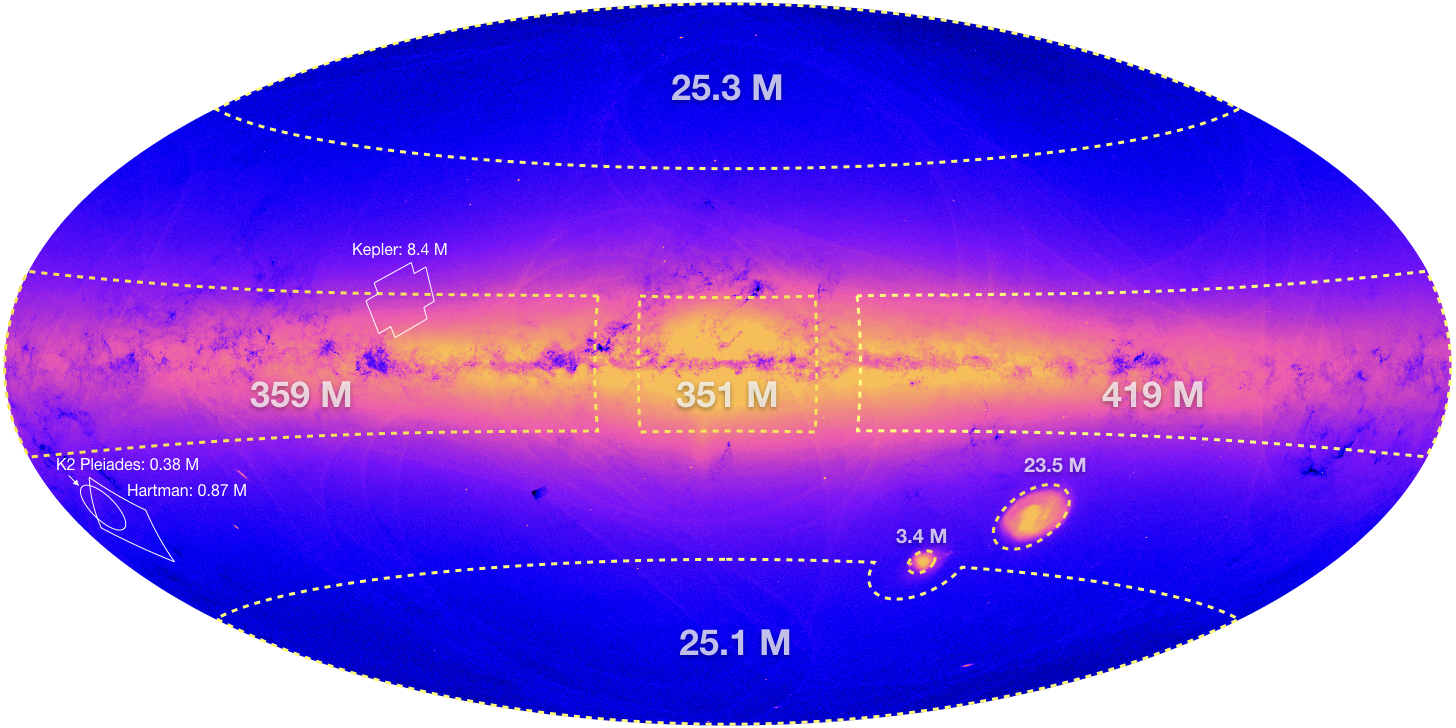} \\
\flushleft 
\vspace{-7.8cm}
\hspace{1.0cm}{\fontfamily{phv}\selectfont All sources $\geq$2 FoV transits}   \\
\hspace{1.0cm}{\fontfamily{phv}\selectfont 100\% (1607~M)} \vspace{6.8cm}\\
\centering
\includegraphics[angle=0,width=0.8\textwidth]{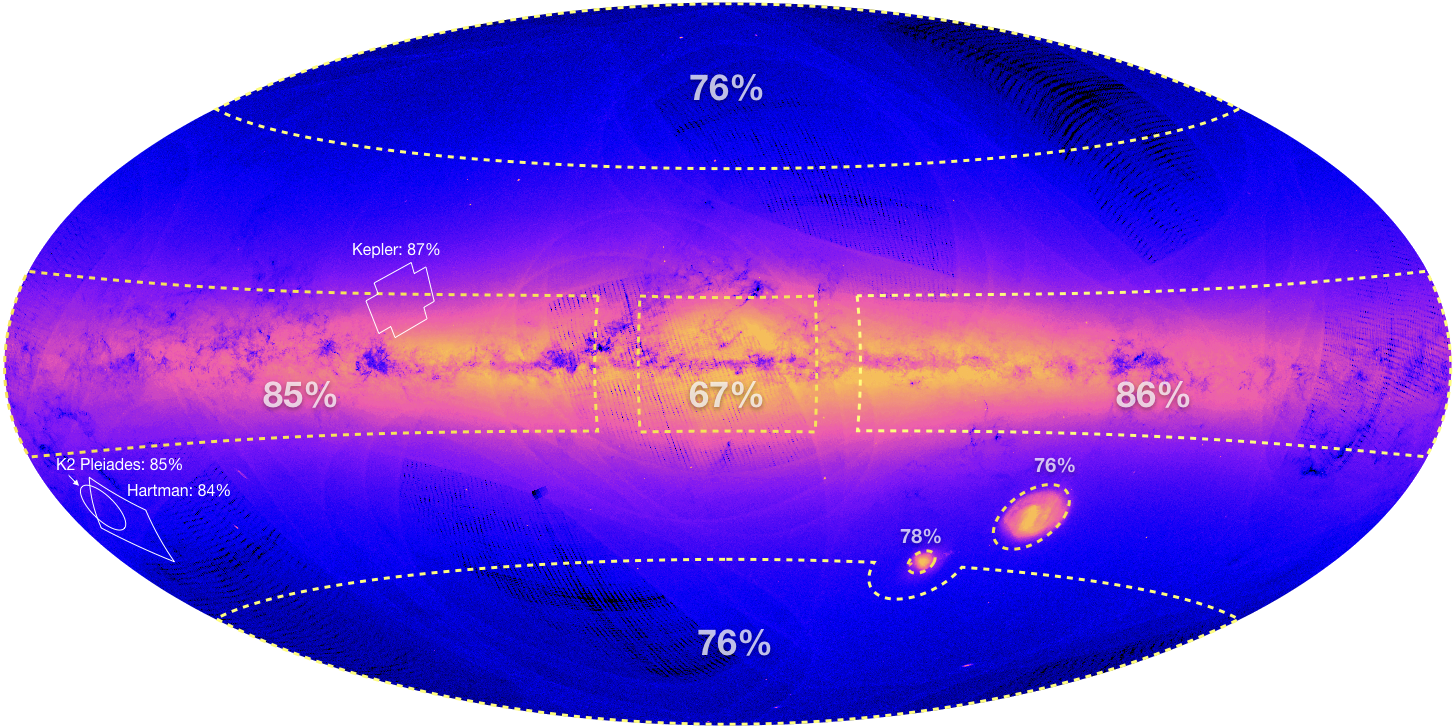} \\
\flushleft 
\vspace{-7.8cm}
\hspace{1.0cm}{\fontfamily{phv}\selectfont  All sources $\geq$12 FoV transits }  \\
\hspace{1.0cm}{\fontfamily{phv}\selectfont  80\% (1283~M) }\vspace{6.8cm}\\
\centering
\includegraphics[angle=0,width=0.8\textwidth]{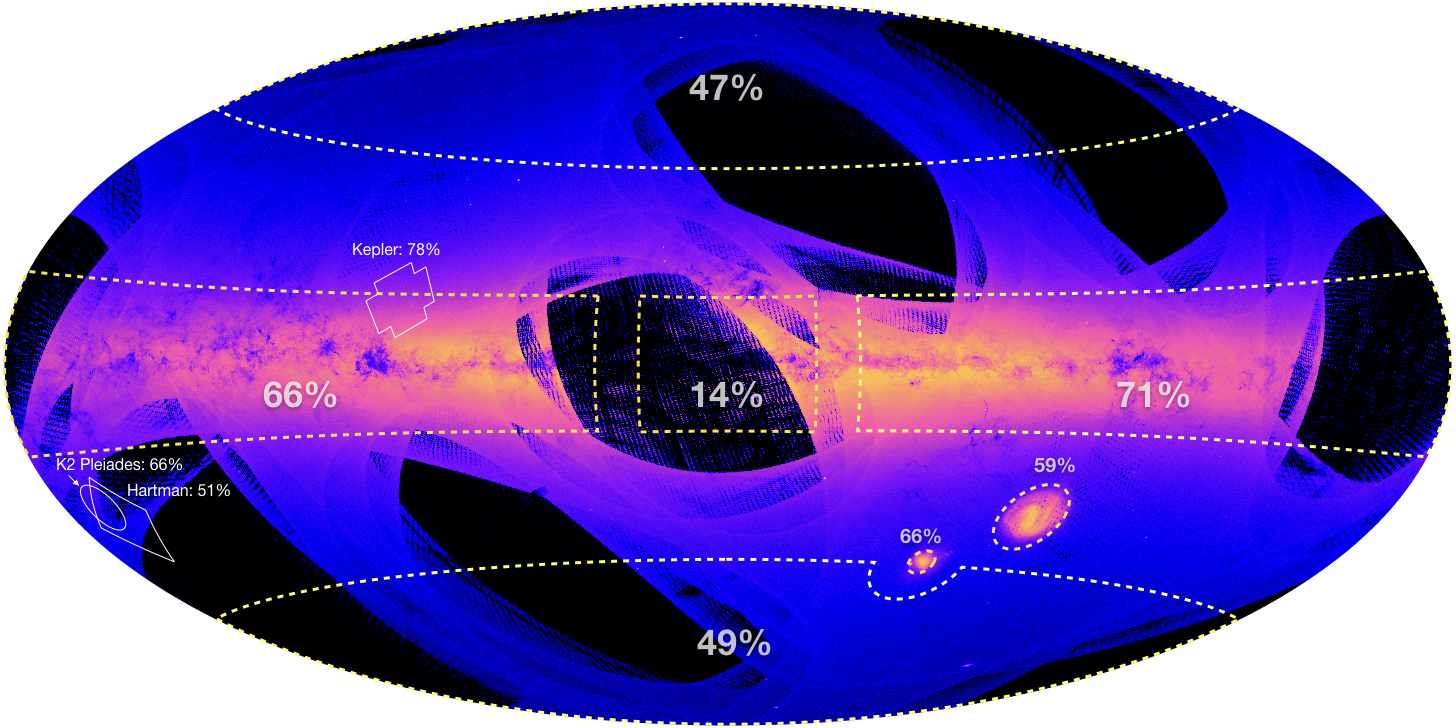}
\flushleft 
\vspace{-7.7cm}
\hspace{1.0cm}{\fontfamily{phv}\selectfont  All sources $\geq$20 FoV transits}   \\
\hspace{1.0cm}{\fontfamily{phv}\selectfont  51\% (826~M) }\vspace{6.8cm}\\
\centering
\includegraphics[angle=0,width=0.8\textwidth]{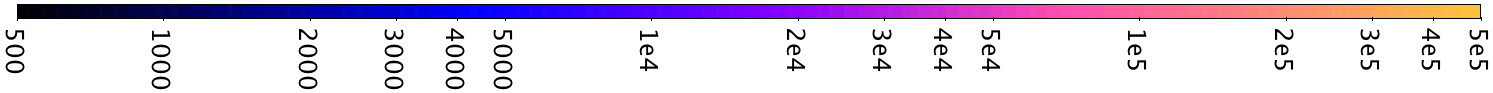} \\
\vspace{-1.35cm}\footnotesize{\fontfamily{phv}\selectfont  \ source density}\hspace{10.55cm}\footnotesize{\fontfamily{phv}\selectfont  [count deg$^{-2}$] }\\
\vspace{0.7cm}
\caption{
Relative source counts with respect to sources having $\geq$2 $G$-band FoV transits, illustrating the drop in relative completeness for source-sets with $\geq$12 and $\geq$20 $G$-band FoV transits, as applied in our processing (see Fig.~\ref{fig:variabilityProcessingOverview}). Regional values are listed in Table~\ref{tab:skyCoverageCounts} and discussed in Sect.~\ref{sec:skycompleteness}. Values for external catalogs are \textit{Gaia} source counts in the approximate sky footprint, not source cross-match counts. Note that actual time series are published only for the 550K variable sources in DR2 (see Fig.~\ref{fig:globalNumbers}).
}
\label{fig:skyDensitiesGeqCuts}
\end{figure*}

As a consequence of the \textit{Gaia} scanning law, the sky-coverage is rather non-uniform for the 22~months of DR2 data, resulting in sky-coverage gaps when selecting a certain minimum number of FoV transits. In our processing we used cuts of $\geq2$, $\geq12$, and $\geq$~$20$ $G$-band FoV transits, which we discuss in Sect.~\ref{sec:overviewVarProcessing}. Figure~\ref{fig:skyDensitiesGeqCuts} shows the effect of these cuts on the sample of all available time series at the time of processing. Assuming that all sources with $\geq$~$2$ FoV transits are bona fide sources, these figures show that a cut of $\geq$~$12$ FoV transits results in a relative  completeness of 80\%, and a cut of $\geq20$ FoV transits results in a relative completeness of only 51\%. It can generally be
expected that these {relative} completenesses are an upper limit to the {absolute} (sky) completeness, which is assessed in detail in the follow-up variability papers mentioned in the introduction. \cite{DPACP-39} conclude that the catalogue is mostly complete, in the absolute sense, roughly between magnitudes 7 and 17, and incomplete beyond. For a first-order relative completeness assessment 
in certain general Galactic directions, we introduce the following regions and compute the \textit{Gaia} source count in each of them:
\begin{itemize}
\item \textbf{Galactic plane} $30\degr<l<330\degr$ \ \& \ $|b|<15$\degr 
\item \textbf{Galactic centre}  $20\degr>l>340\degr$ \ \& $|b|<15$\degr 
\item \textbf{High Galactic latitudes}  $|b|>45\degr$ , excluding a 10\degr\  radius around the SMC 
\item \textbf{LMC} region within a radius of 7.5\degr\  from Galactic coordinates $l=280.47$\degr, $b=-32.89$\degr
\item \textbf{SMC} region within a radius of 2.5\degr\  from Galactic coordinates $l=302.81$\degr, $b=-44.33$\degr
\end{itemize}
We repeat this for the (approximate) sky footprints of various external catalogues, list the results in Table~\ref{tab:skyCoverageCounts} and display the regions in Fig.~\ref{fig:skyDensitiesGeqCuts}. It shows that a (maximum) completeness of between 70--85\% for $\geq12$  $G$-band FoV transits and a relative completeness of
between 50--70\% for $\geq20$ $G$-band FoV transits can be expected. The bottom panel of Fig.~\ref{fig:skyDensitiesGeqCuts} shows
that various sky regions are not or only partially sampled for $\geq20$ FoV: the Galactic centre region, for example, has a relative completeness of only~14\%.

We remark that the numbers in Table~\ref{tab:skyCoverageCounts} and Fig.~\ref{fig:skyDensitiesGeqCuts} do not correspond exactly to the number of sources published in the \texttt{gaia\_source} table of DR2 because various source filters in the global \textit{Gaia} processing chain were applied before and after the variability processing. The derived percentages match very well, however, and they are relevant for the presented discussion.
\begin{table*}[h!]
\caption{
Relative source counts for sources with $\geq$2 $G$-band FoV transits, illustrating the decrease in relative completeness for source sets with $\geq$12 and $\geq$20 $G$-band FoV transits, as applied in our processing. The top part specifies the numbers for some generic directions in the Galaxy, as defined in Sect.~\ref{sec:skycompleteness} and illustrated in Fig.~\ref{fig:skyDensitiesGeqCuts}. The bottom part specifies the numbers for several external catalogues, listing the \textit{Gaia} source counts in their approximate sky footprints (not their source cross-match counts). In parenthesis, we provide the actual \textit{Gaia} source counts in millions. Estimates of {\textup{the absolute}} completeness of the published variable samples are presented in Tables~\ref{tab:skyCompletenessClas} and \ref{tab:skyCompletenessSos}.
}
\label{tab:skyCoverageCounts}     
\centering
\begin{tabular}{l r@{\,} r r@{\,} r r@{\,} r}
\hline\hline
Direction & \multicolumn{2}{c}{$\geq2$~FoV} & \multicolumn{2}{c}{$\geq12$~FoV}  & \multicolumn{2}{c}{$\geq20$~FoV}  \\     
\hline
ALL             & 100\% &(1607)                 & 80\% &(1283)                                         & 51\% &(826) \\

Galactic plane          & 100\% &(778)          &  85\% &(665)                                                 & 68\% &(532) \\

Galactic centre                 & 100\% &(351)          & 67\% &(236)                                                  & 14\% &(50.0) \\

High Galactic latitudes          & 100\% &(50.4)                &  76\% &(38.4)                                                   & 48\% &(24.4) \\

LMC region      & 100\% &(23.5)                 & 76\% &(17.9)                                                 & 59\% &(13.9) \\

SMC region      & 100\% &(3.4)          & 78\% &(2.6)                                                           & 66\% &(2.2) \\

\hline
Catalina DR2    & 100\% &(400)  & 80\% &(320)                                           & 53\% &(210) \\
Catalina DR1    & 100\% &(200)  & 80\% &(160)                                           & 50\% &(100) \\
OGLE-IV Bulge   & 100\% &(305)  & 66\% &(200)                                           & 10\% &(29) \\
OGLE-IV MCs     & 100\% &(33)           & 77\% &(25)                                    & 62\% &(20) \\
Kepler  & 100\% &(8.4)          & 87\% &(7.4)                                           & 78\% &(6.6) \\
\cite{2016AJ....152..113R} (K2 Pleiades)        & 100\% &(0.38)         & 85\% &(0.32)                                    & 66\% &(0.25) \\
\cite{2010MNRAS.408..475H} (Pleiades)   & 100\% &(0.87)         & 84\% &(0.73)                    & 51\% &(0.44) \\
\hline
\end{tabular}
\end{table*}

\subsection{Outliers and calibration errors} 
Various issues are visible in the current photometric data. Some of them are caused by instrumental effects, while others are sky related. Improved calibration strategies and careful flagging of these events are being designed and will be implemented in the chain of processing systems that convert the raw data into calibrated epoch photometry for the next \textit{Gaia} Data Release. Some of these issues are also described in \cite{DWE-052}:
\begin{itemize}
\item hot pixel columns are not yet treated,
\item poor background estimates exist for some observations,
\item observations close to bright sources can have biased background estimates or might even be spurious detections due to diffraction spikes,
\item spatially close sources can have scan-angle direction dependent outliers,
\item isolated sources can still be affected through overlap of the two FoVs.
\end{itemize}
It is expected that a certain fraction of observation affected by these issues has not been flagged and hence users should be aware of such unflagged outliers in the published data. We discuss variability flagging in more detail in Sect.~\ref{sec:obsFiltering}. 

In addition to the flagging of {\textup{observations}}, the \textit{Gaia} data processing consortium has applied various filters that remove {\textup{sources}} that are affected by specific calibration issues; see \cite{DPACP-36} for more details.

\section{DR2 variability data processing and analysis 
\label{sec:overviewVarProcessing}}
The general variability data processing has been described in detail in our DR1 paper \citep{EyerDR1}, to which we refer here.
We here only summarise the main aspects that are specific to the DR2 release. More exhaustive information is provided in the DR2 documentation.

\begin{figure*}[bh!]
\centering
\includegraphics[angle=0,width=0.8\textwidth]{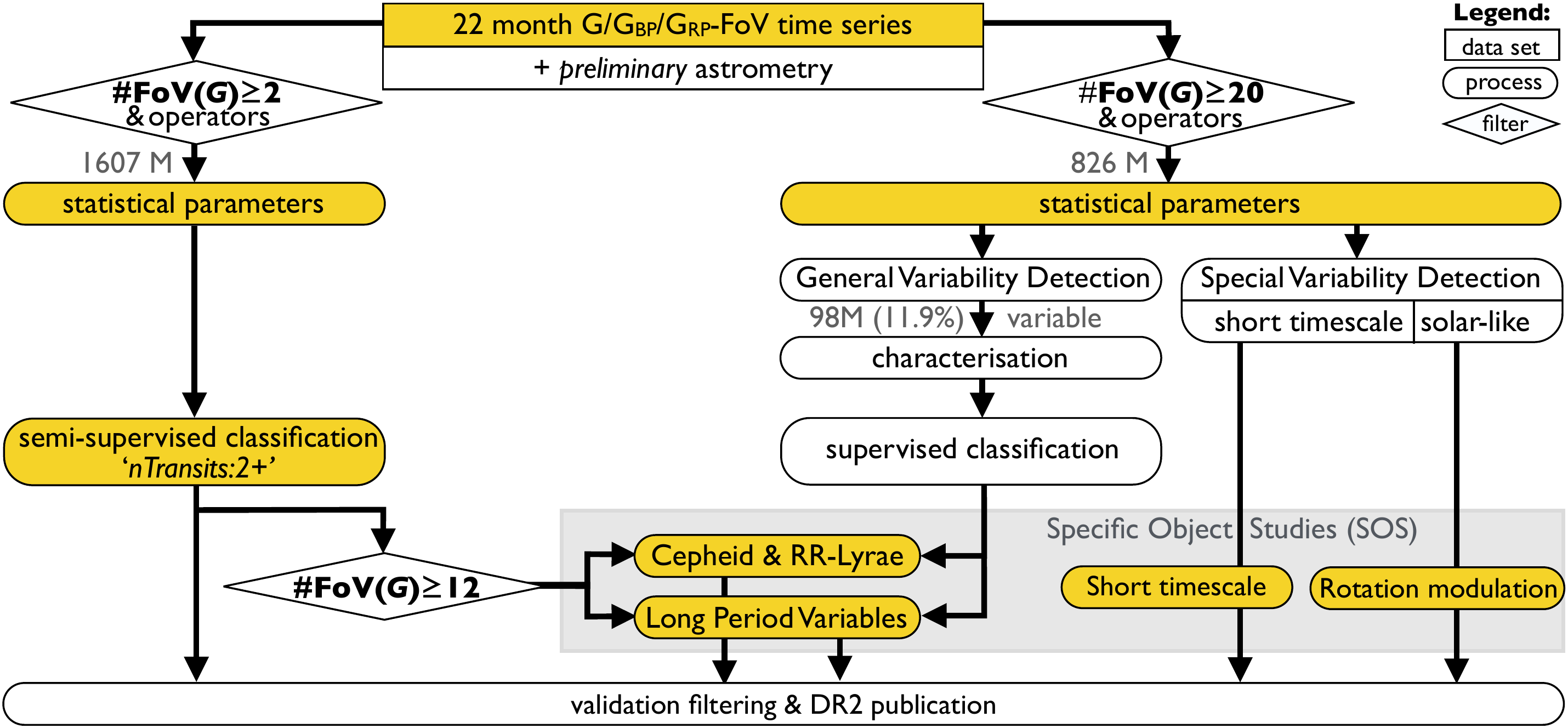}
\caption{DR2 variability processing overview. Data were published from the highlighted yellow boxes for the validation filtered sources. There were two main tracks: one starting from $\geq 2$ $G$-band FoV transits (left), and the other from  $\geq 20$ $G$-band FoV transits (right). The former resulted in the published \textit{nTransits:2+} classifier in the \texttt{vari\_classifier\_result} table, and the latter resulted in the published  SOS tables of  \texttt{vari\_short\_timescale} and \texttt{vari\_rotation\_modulation}. The published SOS tables of  \texttt{vari\_rrlyrae}, \texttt{vari\_cepheid} and \texttt{vari\_long\_period\_variable} result from a mixed feed of the published \textit{nTransits:2+} classifier (for sources with at least $\geq 12$ $G$-band FoV transits) and from the unpublished classifier of the $\geq 20$ track. 
}
\label{fig:variabilityProcessingOverview}
\end{figure*}

The  variability processing is summarised in Fig.~\ref{fig:variabilityProcessingOverview}, and the number of sources involved at various levels of the processing is indicated in the diagram.
The processing followed two main paths: one starting from 1607 million sources with $\geq 2$ $G$-band FoV transits (hereafter geq2), and the other from 826 million sources with $\geq 20$ $G$-band FoV transits (hereafter geq20).
The $\geq~X$ means a $G$-band time series with $X$ non-null and positive FoV flux observations, but {\textup{before}} the variability filters were applied that are described in Sects.~\ref{sec:obsFiltering} and \ref{sec:timeSeriesStatisticsAndFlag}.

The first path  resulted in the published \textit{nTransits:2+} classification results, as described in Sect.~\ref{sec:nTransits2class}, as well as in a split-off path with $\geq 12$ $G$-band FoV transits (hereafter geq12) that served as one of the inputs for the
SOS modules: SOS Cep and RRL \citep{DPACP-13, Clementini18CepRrl} and SOS LPV \citep{Mowlavi18Lpv}.

The second path leads to the majority of the pipelines and SOS results, as described in Sect.~\ref{sec:sos}. It 
had two different variability detection methods. The first is a `general' variability detection, defined by a 
classifier trained on known cross-matched constant objects\footnote{The least variable sources found in \url{ftp://ftp.astrouw.edu.pl/ ogle/ogle4/GSEP/maps/}} from OGLE4 \citep{2012AcA....62..219S}, Hipparcos \citep{HIPPARCOS_PERIODIC_ESA_1997}, the Sloan Digital
Sky Survey (SDSS) \citep{2007AJ....134..973I}, and classified constants, as well as many additional catalogues for variables \citep{Rimoldini18Geq2}. The second consists of multiple dedicated variability detection methods: one tuned to detect short-timescale variables and another to rotation modulation effects, which were subsequently followed by their associated SOS modules: SOS short timescale \citep{Roelens18Sts} and rotation modulation \citep{Lanzafame18RotMod}, analysing these sources in more detail. The general variability detection was followed by a   
generic multi-harmonic Fourier modelling based on the periodogram peak-frequency of an unweighted least-squares period search. Various classification {attributes} were derived from this model and the time-series statistical parameters, on which the classifier for the geq20 path was trained. 
The classified RR~Lyrae stars, Cepheids, and LPVs were subsequently fed into the corresponding
modules SOS Cep\&RRL and SOS LPV, being their second input source. All modules of the variability pipeline include validation rules to ensure that the output values are within acceptable ranges,
which allowed us to identify issues early on in the processing.

\begin{figure*}
\centering
\includegraphics[angle=0,width=\textwidth]{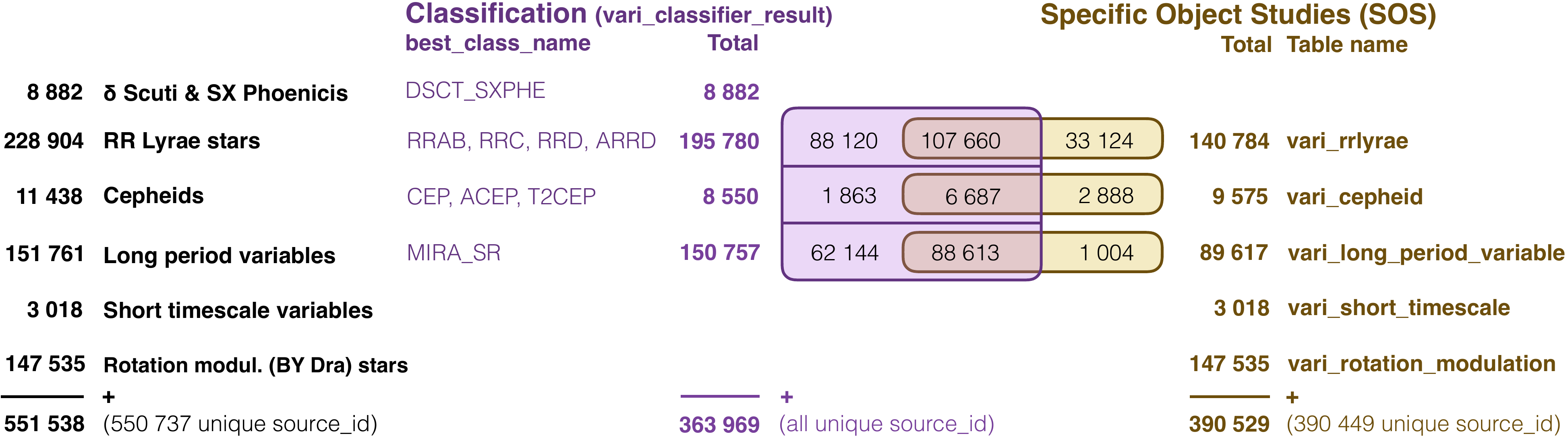}
\caption{
Accounting of the 550\,737 unique variable sources published in DR2. All have an entry in the \textit{Gaia} archive \texttt{gaia\_source} 
 and \texttt{vari\_time\_series\_statistics} tables. Their distributions in the additional variability tables is shown. The classification output types are mutually exclusive so that their sum matches the total number of entries in the published \texttt{vari\_classifier\_result} table. Eighty sources appear in more than one SOS table (not illustrated), which is detailed in Sect.~\ref{sec:sosOverlap}.
The overlap between similar-type classification and SOS is detailed in a Venn-like fashion showing the counts in the three possible subsets. A few hundred sources have different assigned types in classification and SOS tables (not illustrated), as explained in Sect.~\ref{sec:geq2SosDifferent}, hence the difference on the left between the total type count and the unique \texttt{source\_id} count. No SOS module treated $\delta$~Scuti \& SX Phoenicis stars.}
\label{fig:globalNumbers}
\end{figure*}

An accounting of the number of sources published in the classification and SOS modules is given in Fig.~\ref{fig:globalNumbers}.
It shows that of the 550\,737 unique variable sources, 363\,969~(66\%) appear in the classification results (Sect.~\ref{sec:nTransits2class}) and 390\,529~(71\%) in SOS (Sect.~\ref{sec:sos}). For several variability types, 203\,681 sources (37\%) appear in both, 
while the rest of the sources appear only in either classification 
(160\,288,~i.e.~29\%) 
 or SOS results
 (186\,768,~i.e.~34\%). 
Note that the number appearing in both classification and SOS tables (203\,681) is 721 more than the sum of overlapping types shown in Fig.~\ref{fig:globalNumbers} because these are cases of different assigned types in classification and SOS tables, which are omitted from the figure.

 Depending on the type of object,
there might only exist either a classification or SOS result.
SOS results have generally lower contamination than the classification results and contain various type-specific astrophysical parameters that can be used to obtain more detailed information (such as the period), although this information might be available for a less sky-uniform sample due to the minimum number of $G$-band FoV observations of 12 or 20, depending on the SOS type. 
The classification contains only a label and  `score' and has generally higher contamination rates, but it contains larger samples that are (more) uniformly distributed over the sky due to the minimum number of $G$-band FoV observations, which are 5 for RR Lyrae, 6~for $\delta$ Scuti and SX Phoenicis, 9 for Cepheids, and 12 for LPVs. These differences between classification and SOS in contamination rate and sample size are part of the pipeline design in which classification results feed into most of the SOS modules, as shown in Fig.~\ref{fig:variabilityProcessingOverview} and discussed in more detail in the DR1 paper \cite{EyerDR1}.

\begin{figure*}[h]
\begin{tabular}{@{}lll@{}}
\setlength{\tabcolsep}{0pt} 
\renewcommand{\arraystretch}{0} 
  \vspace{0cm}\\

 \includegraphics[width=0.48\textwidth]{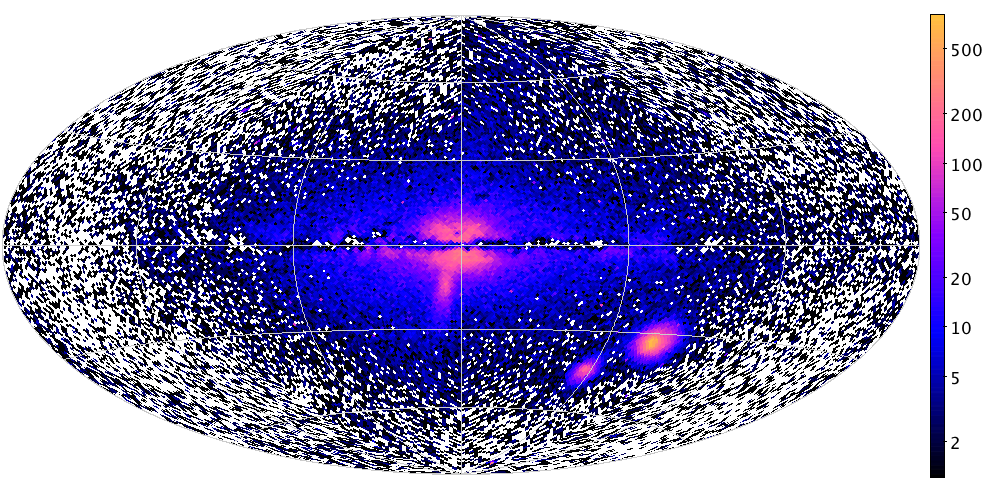}    & \includegraphics[width=0.48\textwidth]{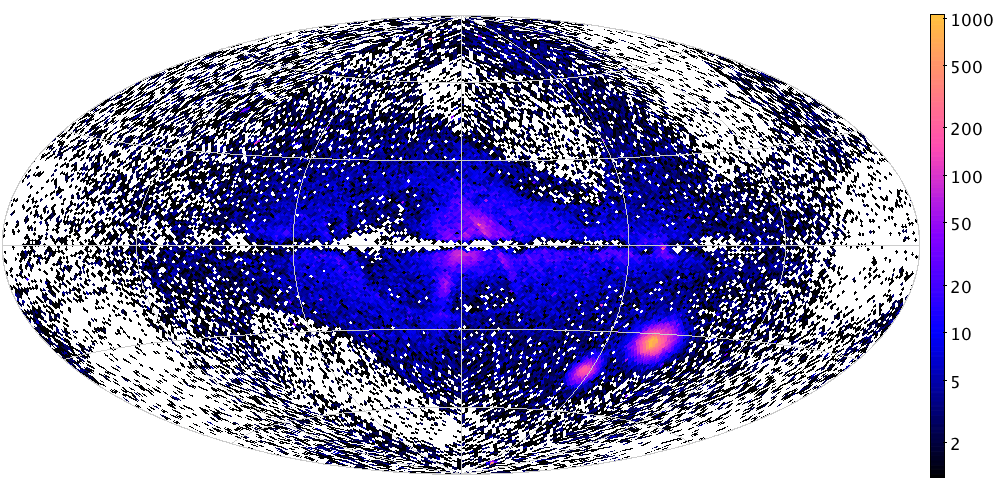}  \vspace{-4.7cm}
\\
\hspace{0.cm}{\fontfamily{phv}\selectfont \tiny Classif.: RR Lyrae stars \scriptsize{ (RRAB, RRC, RRD, ARRD)}}& \hspace{0.cm}{\fontfamily{phv}\selectfont \tiny SOS: RR Lyrae stars }\vspace{4.4cm} 
\\[6pt]

 \includegraphics[width=0.48\textwidth]{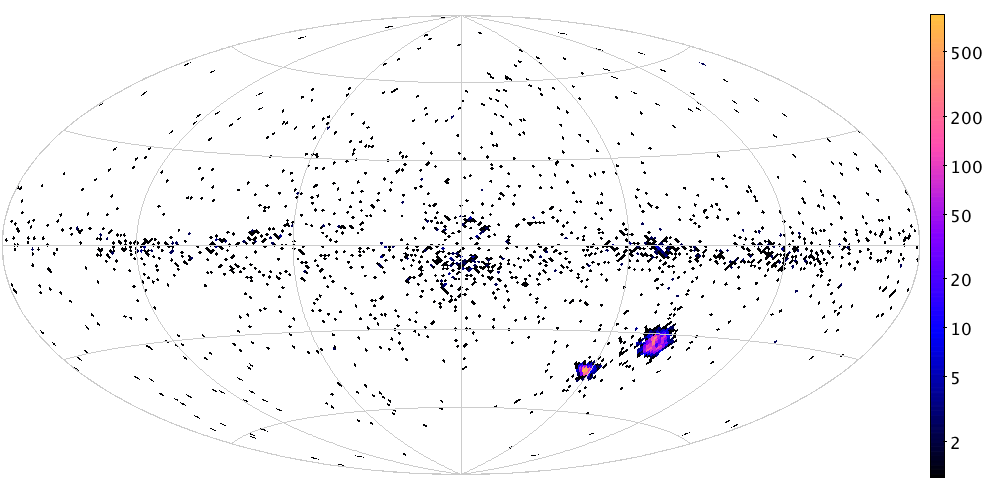}    & \includegraphics[width=0.48\textwidth]{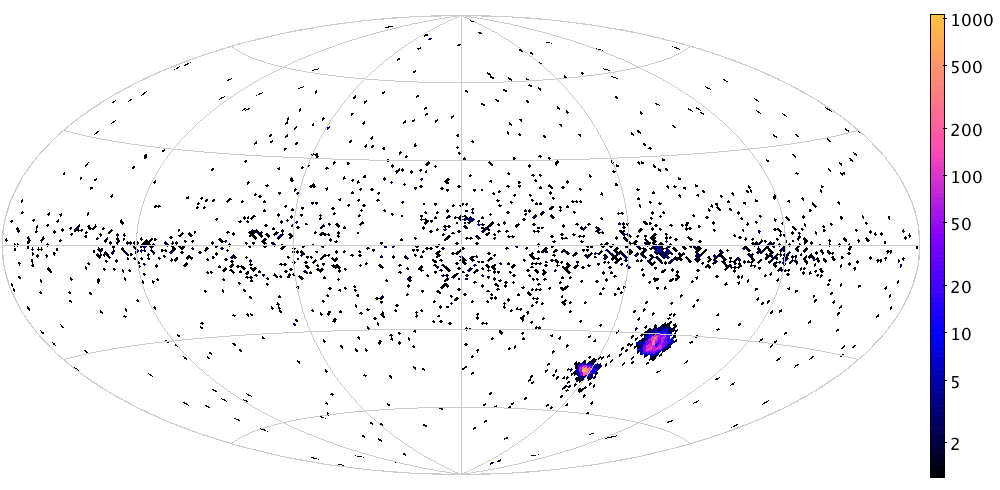}  \vspace{-4.7cm}
\\
\hspace{0.cm}{\fontfamily{phv}\selectfont \tiny Classif.: Cepheids  \scriptsize{ (CEP, ACEP, T2CEP)} }& \hspace{0.cm}{\fontfamily{phv}\selectfont \tiny SOS: Cepheids  }\vspace{4.4cm} 
\\[6pt]

 \includegraphics[width=0.48\textwidth]{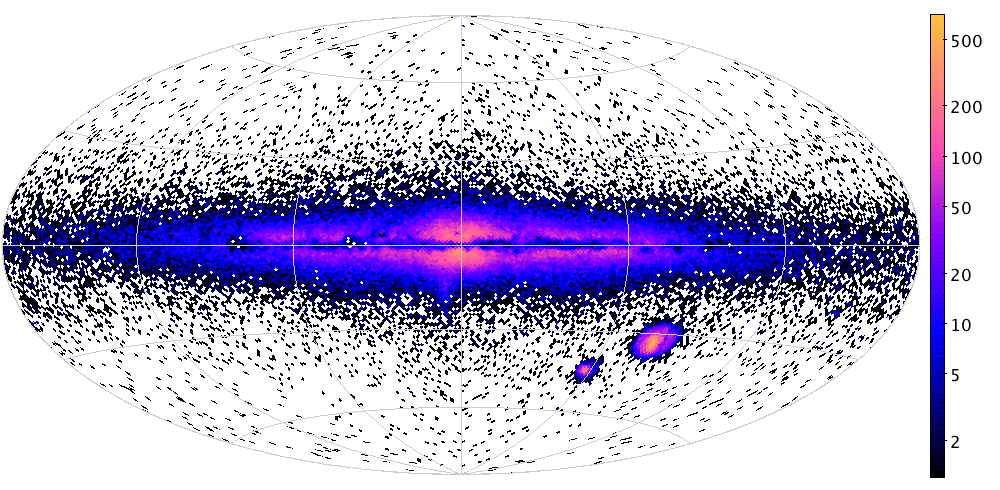}    & \includegraphics[width=0.48\textwidth]{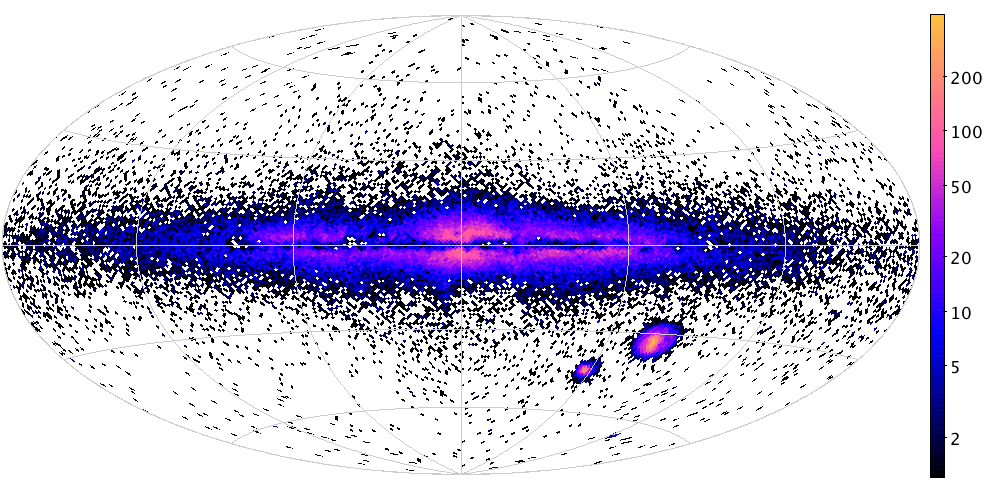}  \vspace{-4.7cm}
\\
\hspace{0.cm}{\fontfamily{phv}\selectfont \tiny Classif.: Long-period variables  \scriptsize{ (MIRA\_SR)} }& \hspace{0.cm}{\fontfamily{phv}\selectfont \tiny SOS: Long-period variables  }\vspace{4.4cm} 
\\[6pt]

  \includegraphics[width=0.48\textwidth]{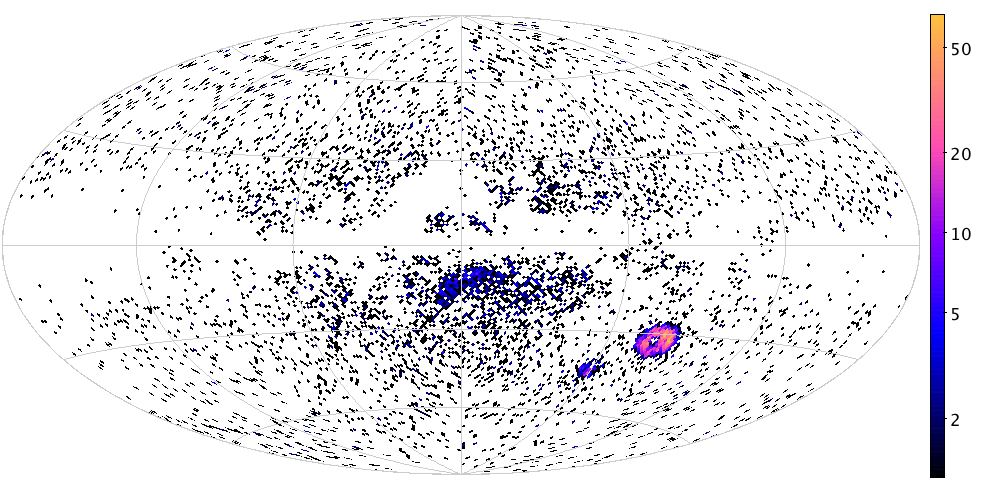}    & \includegraphics[width=0.48\textwidth]{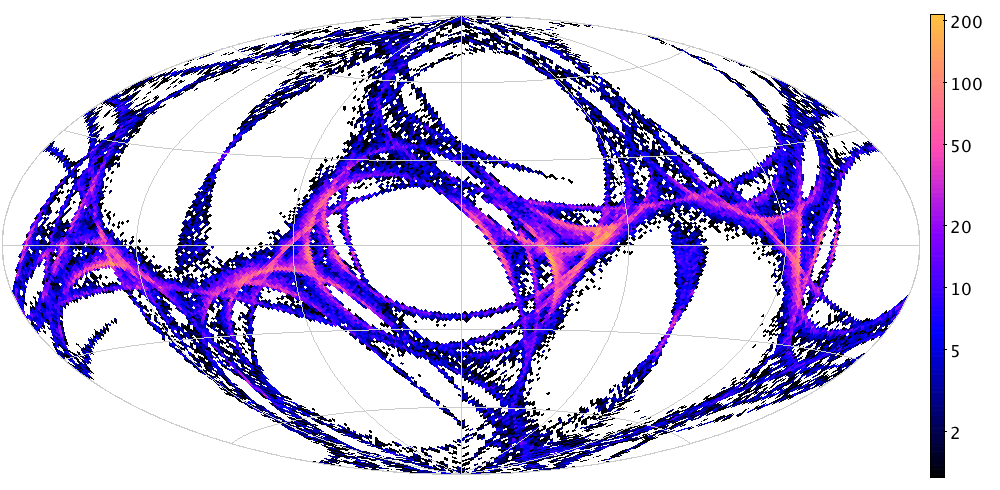}  \vspace{-4.7cm}
\\
\hspace{0.0cm}{\fontfamily{phv}\selectfont \tiny Classif.: $\delta$ Scuti and SX Phoenicis stars \scriptsize{ (DSCT\_SXPHE)} }& \hspace{0.0cm}{\fontfamily{phv}\selectfont \tiny SOS:  Solar-like stars with rotation modulation (BY Dra stars) } \vspace{4.4cm} 
\\[6pt]

\includegraphics[width=0.48\textwidth]{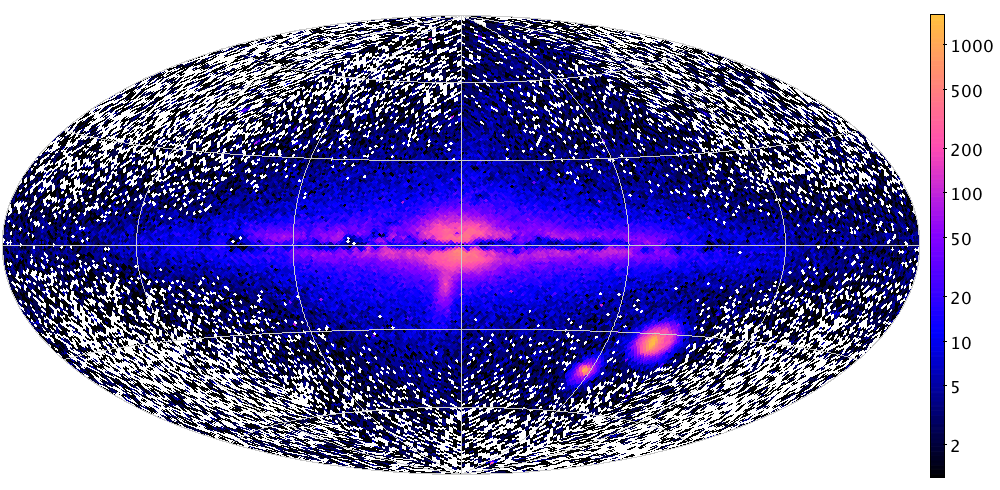}   
 & \includegraphics[width=0.48\textwidth]{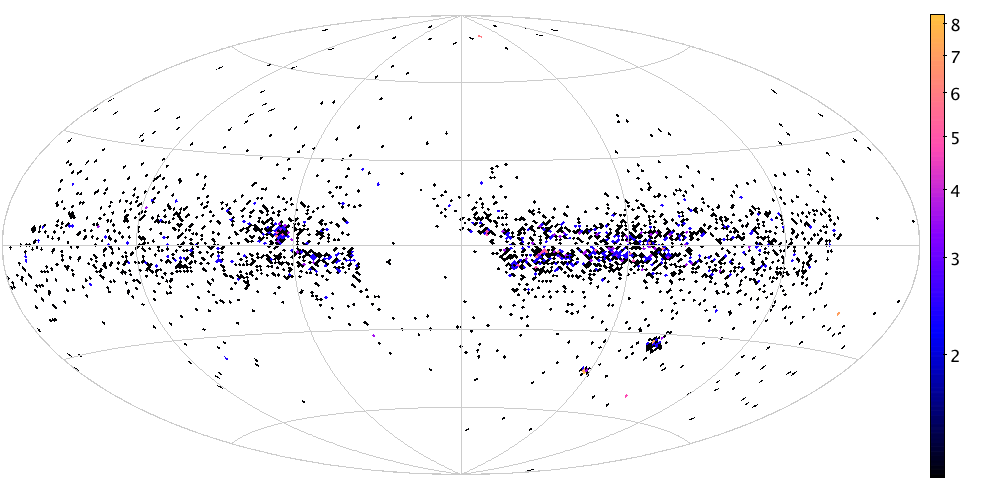} 
  \vspace{-4.7cm}
\\
\hspace{0.cm} {\fontfamily{phv}\selectfont \tiny Classif.: ALL }
& \hspace{0.cm}{\fontfamily{phv}\selectfont \tiny SOS: Short-timescale variables} 
\vspace{4.cm} 
\\[6pt]

\end{tabular}
\caption{Sky source densities [count deg$^{-2}$] in Galactic coordinates of the published sources in the classification table (left column, see Sect.~\ref{sec:nTransits2class}) and SOS tables (right column, see Sect.~\ref{sec:sos}). In the classification plots the \texttt{best\_class\_name} entries were grouped by main type, as listed in parentheses. Galactic longitude increases to the left side. 
}
\label{fig:skyDensities}
\end{figure*}

\begin{figure*}[h]
\begin{tabular}{@{}lll@{}}
\setlength{\tabcolsep}{0pt} 
\renewcommand{\arraystretch}{0} 
  \vspace{0cm}\\

 \includegraphics[width=0.48\textwidth]{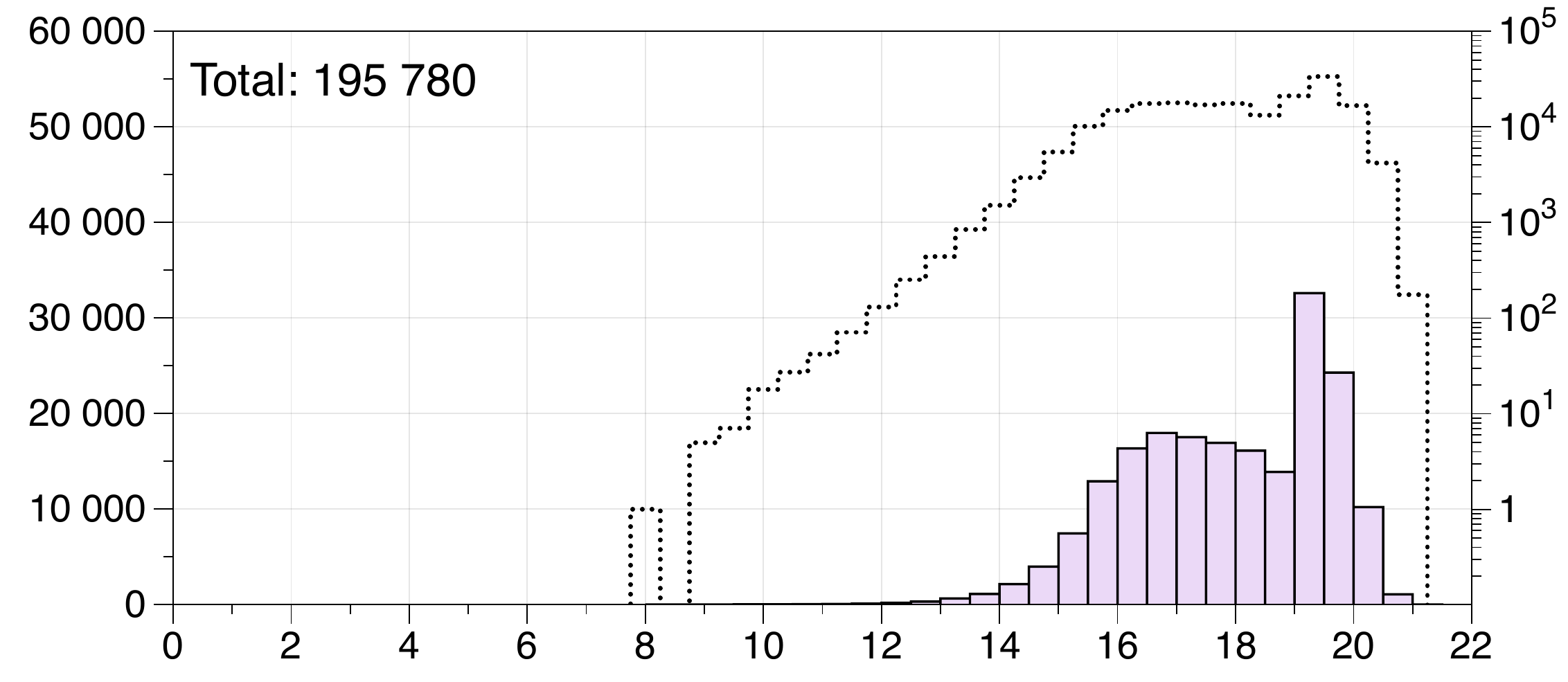}    & \includegraphics[width=0.48\textwidth]{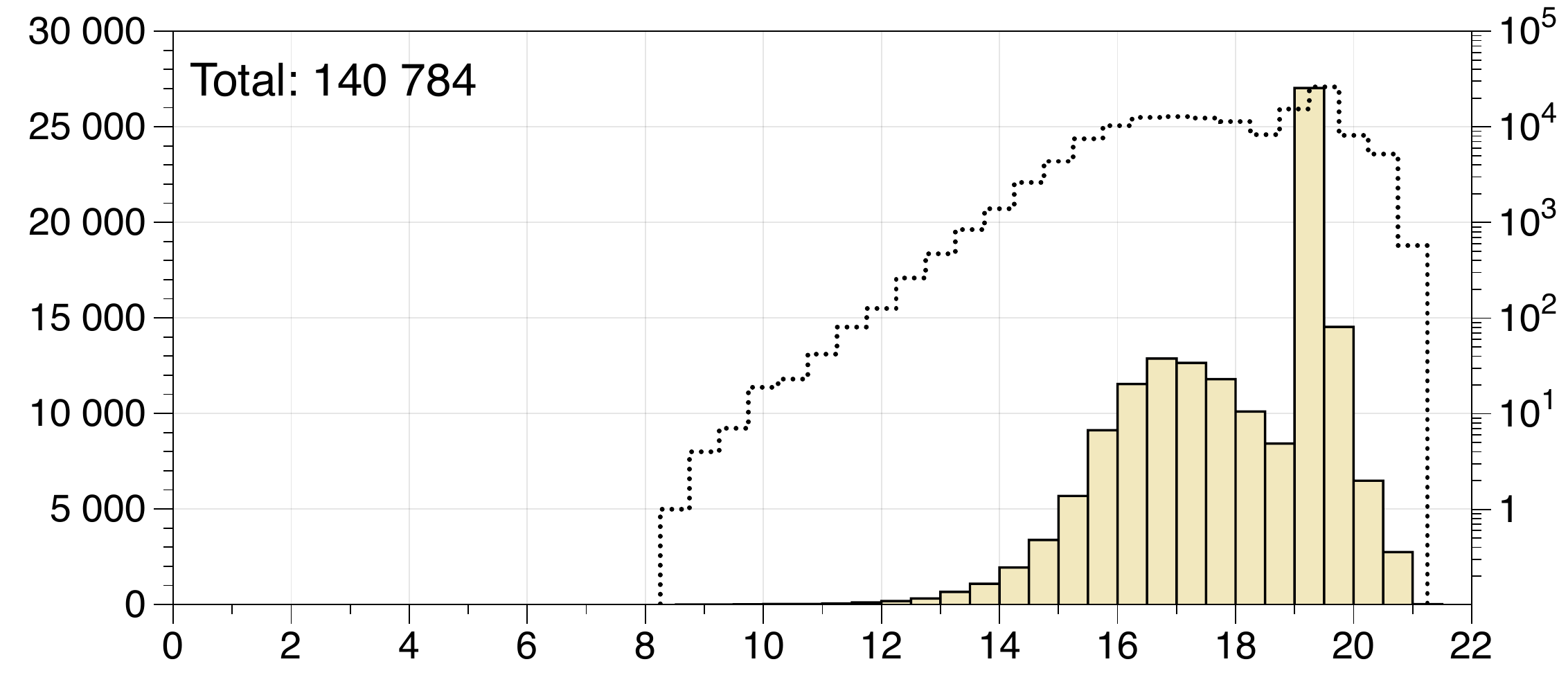}  \vspace{-4.2cm}
\\
\hspace{0.96cm}{\fontfamily{phv}\selectfont \tiny Classif.: RR Lyrae stars \scriptsize{ (RRAB, RRC, RRD, ARRD)} } & \hspace{0.96cm}{\fontfamily{phv}\selectfont \tiny SOS: RR Lyrae stars}\vspace{4.2cm} 
\\[6pt]

 \includegraphics[width=0.48\textwidth]{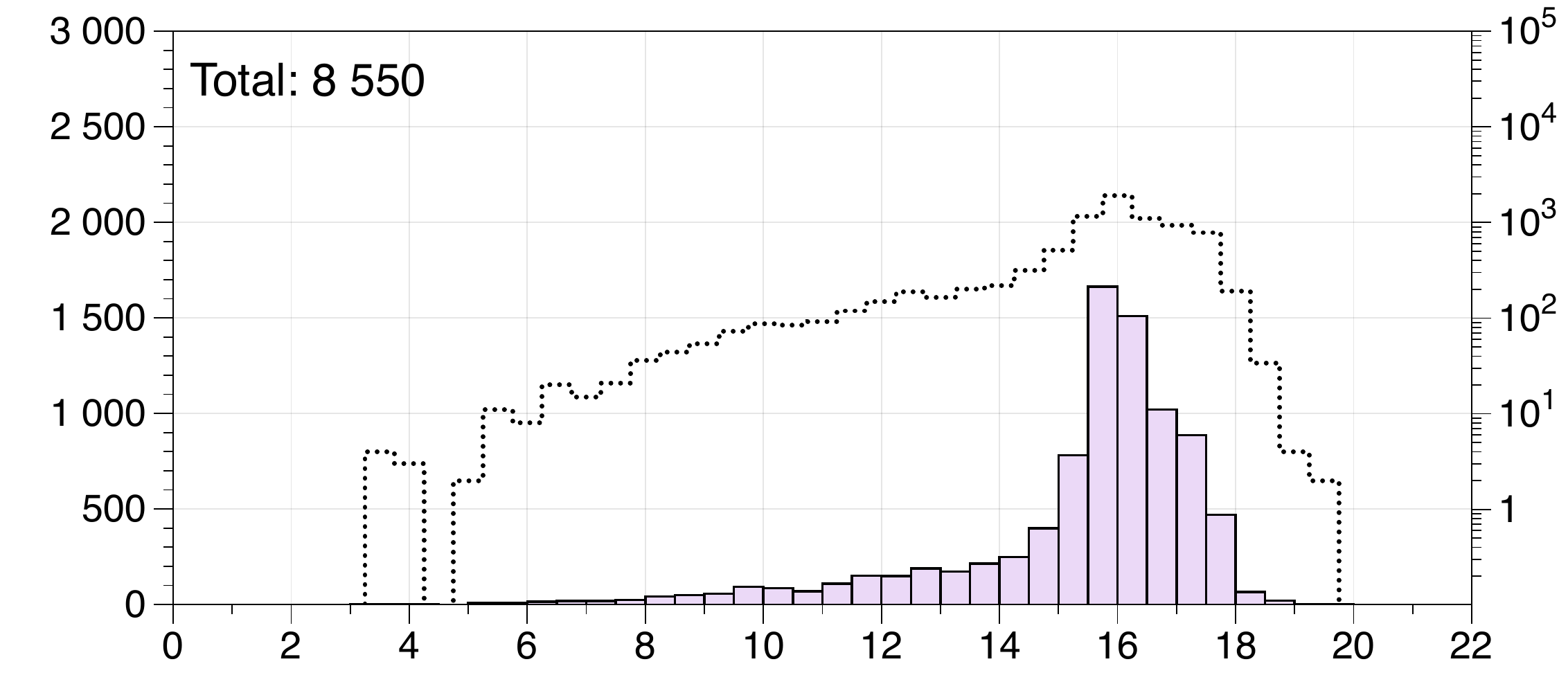}    & \includegraphics[width=0.48\textwidth]{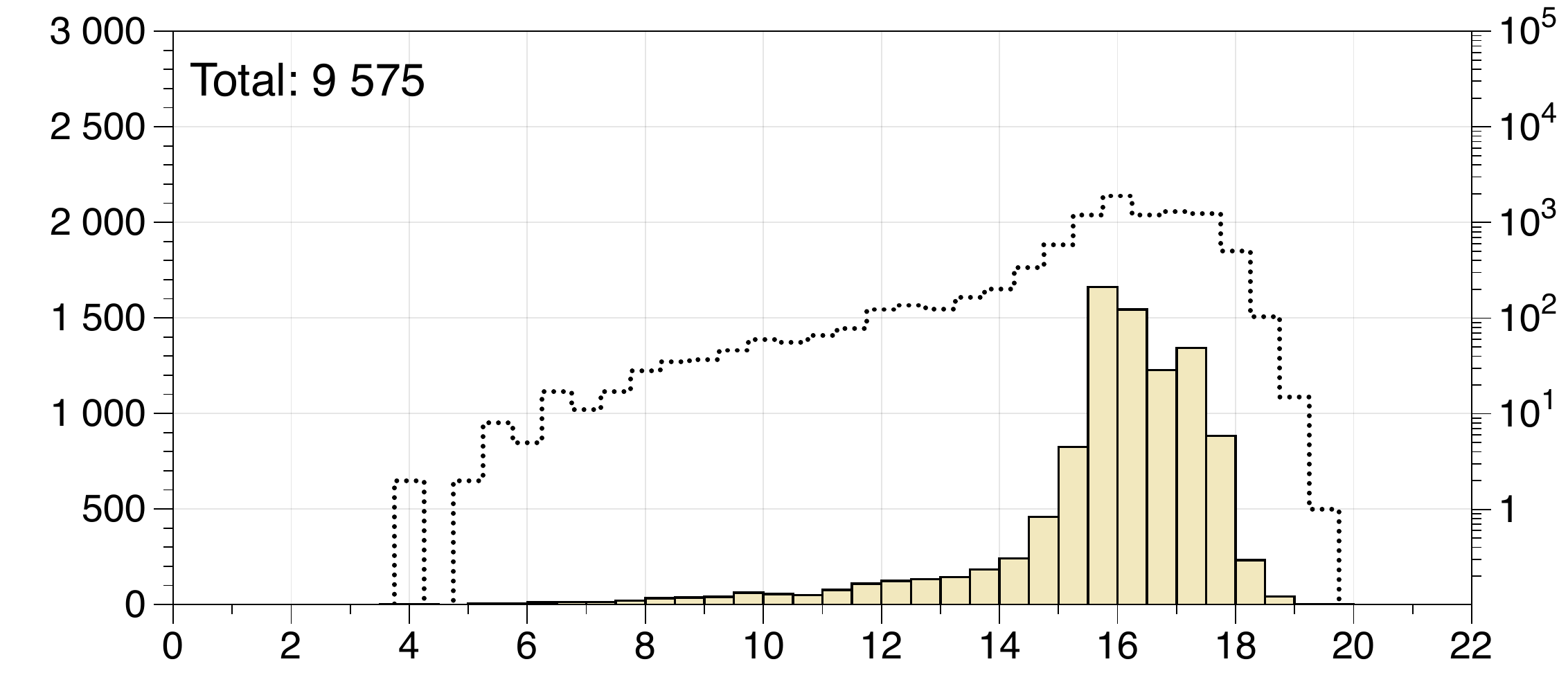}  \vspace{-4.2cm}
\\
\hspace{0.96cm}{\fontfamily{phv}\selectfont \tiny Classif.: Cepheids  \scriptsize{ (CEP, ACEP, T2CEP)} }& \hspace{0.96cm}{\fontfamily{phv}\selectfont \tiny SOS: Cepheids  \vspace{4.2cm} }\\[6pt]

 \includegraphics[width=0.48\textwidth]{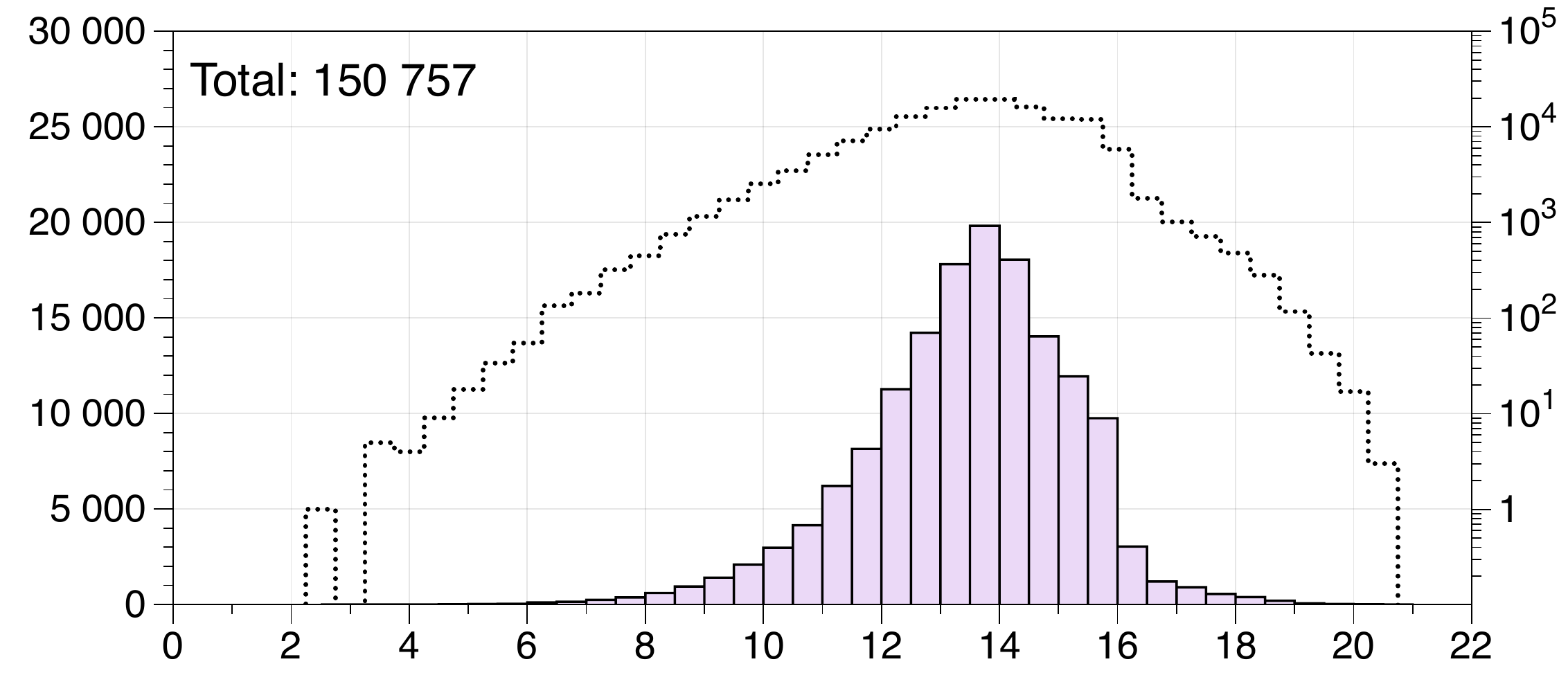}    & \includegraphics[width=0.48\textwidth]{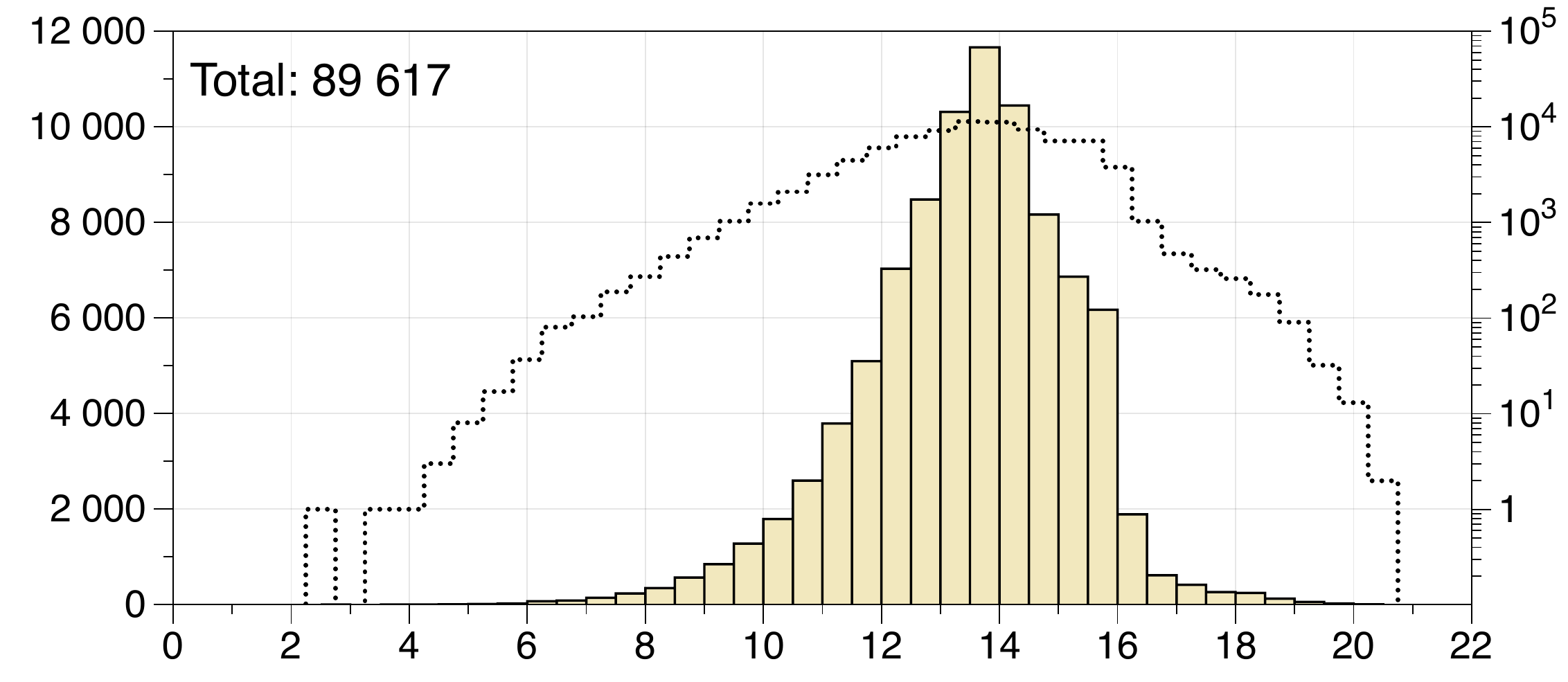}  \vspace{-4.2cm}
\\
\hspace{0.96cm}{\fontfamily{phv}\selectfont \tiny Classif.: Long-period variables  \scriptsize{ (MIRA\_SR)} }& \hspace{0.96cm}{\fontfamily{phv}\selectfont \tiny SOS: Long-period variables } \vspace{4.2cm} 
\\[6pt]

  \includegraphics[width=0.48\textwidth]{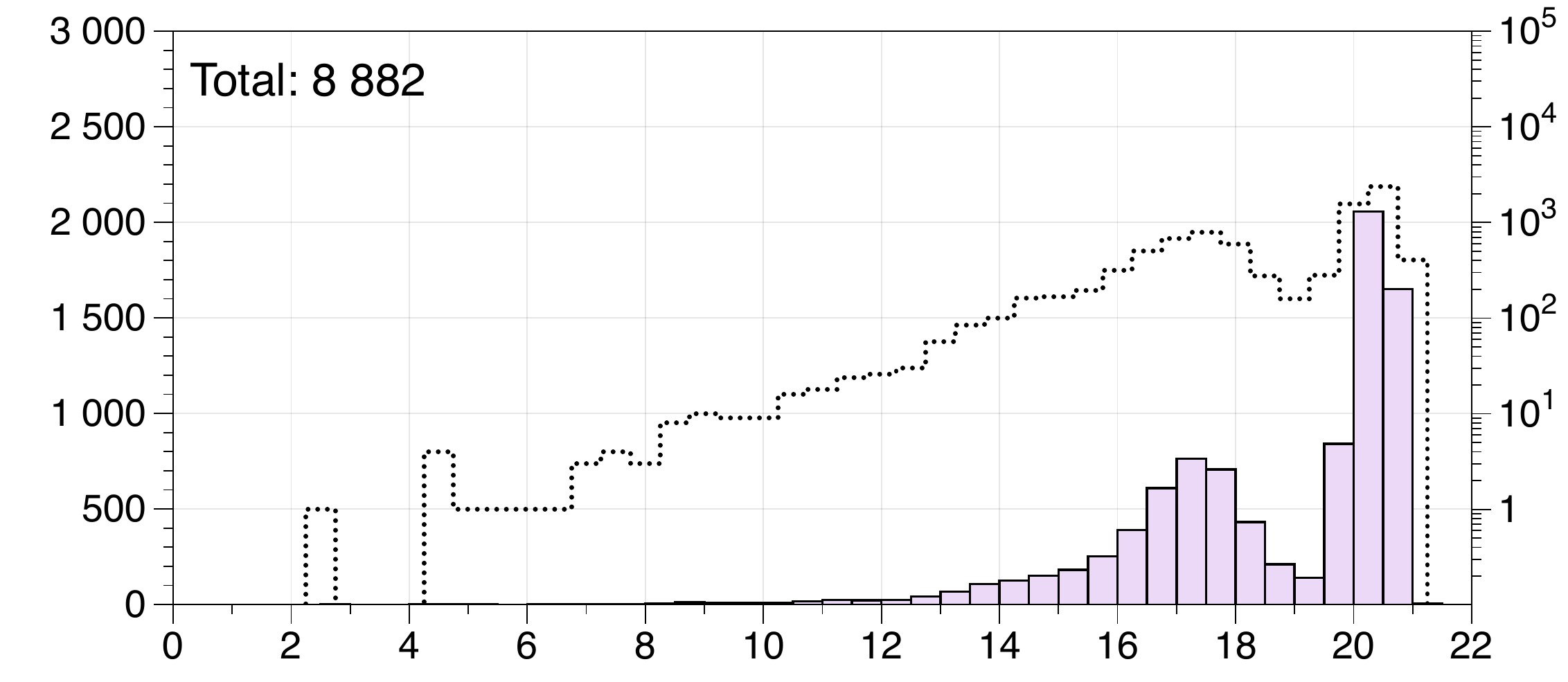}    & \includegraphics[width=0.48\textwidth]{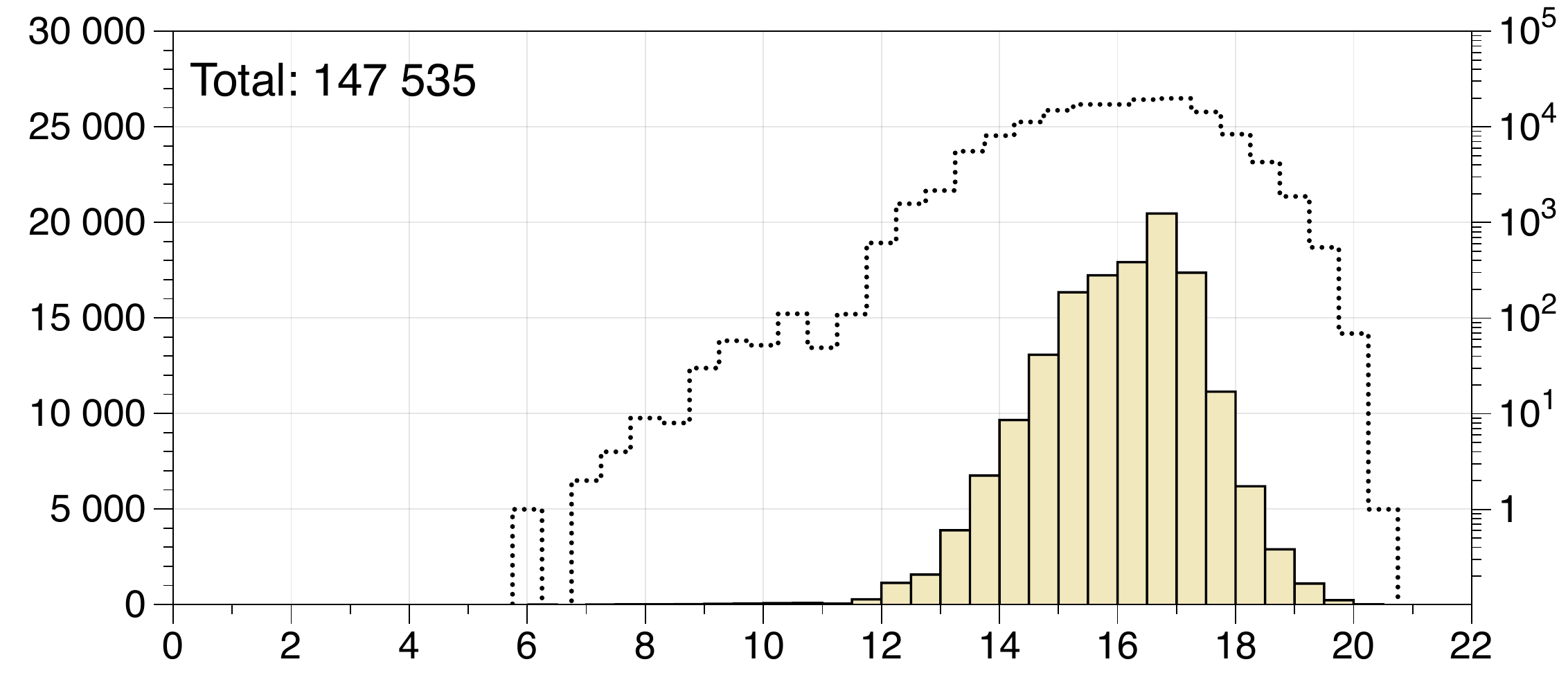}  \vspace{-4.2cm}
\\
\hspace{0.96cm}{\fontfamily{phv}\selectfont \tiny Classif.: $\delta$ Scuti and SX Phoenicis stars\scriptsize{ (DSCT\_SXPHE)} }& \hspace{0.96cm}{\fontfamily{phv}\selectfont \tiny SOS:  Solar-like stars with rotation modulation (BY Dra stars) }\vspace{4.2cm} 
\\[6pt]

 \includegraphics[width=0.48\textwidth]{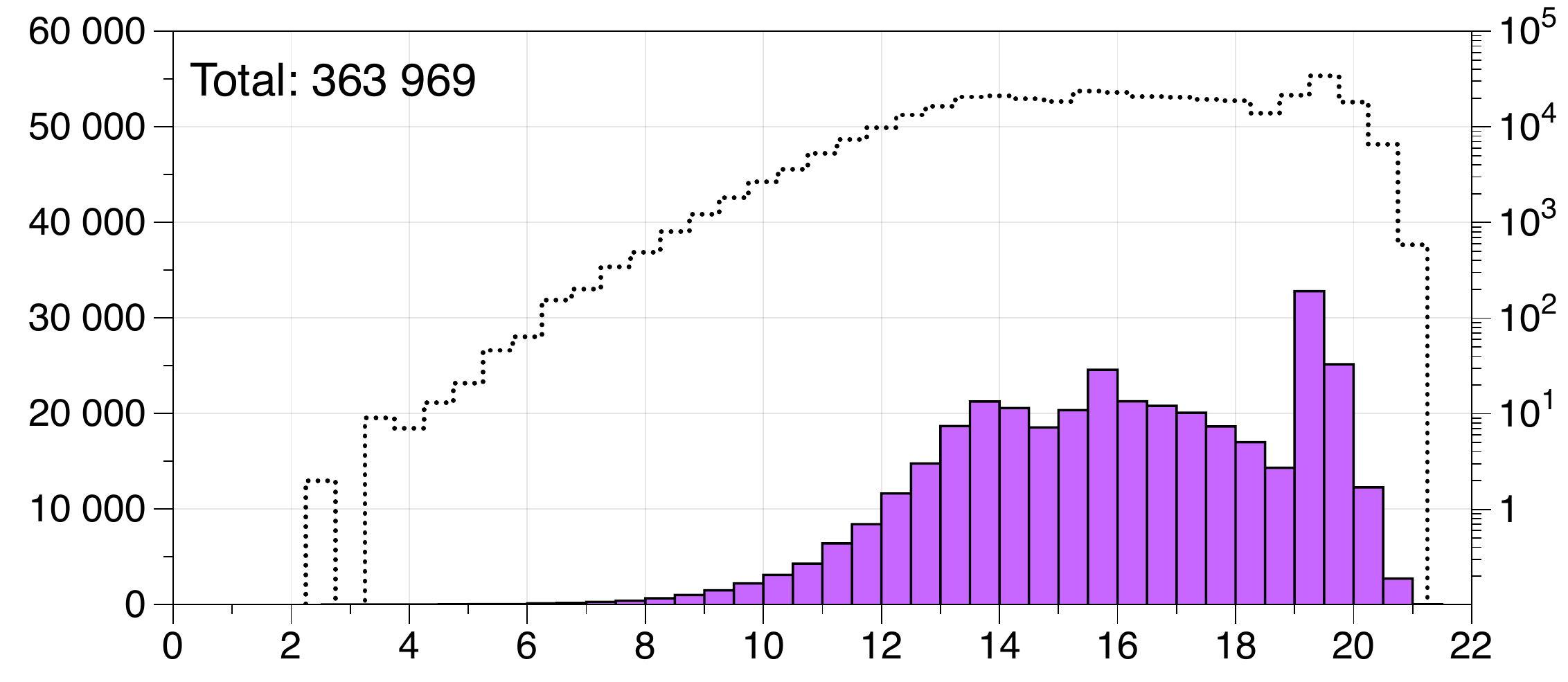}    & \includegraphics[width=0.48\textwidth]{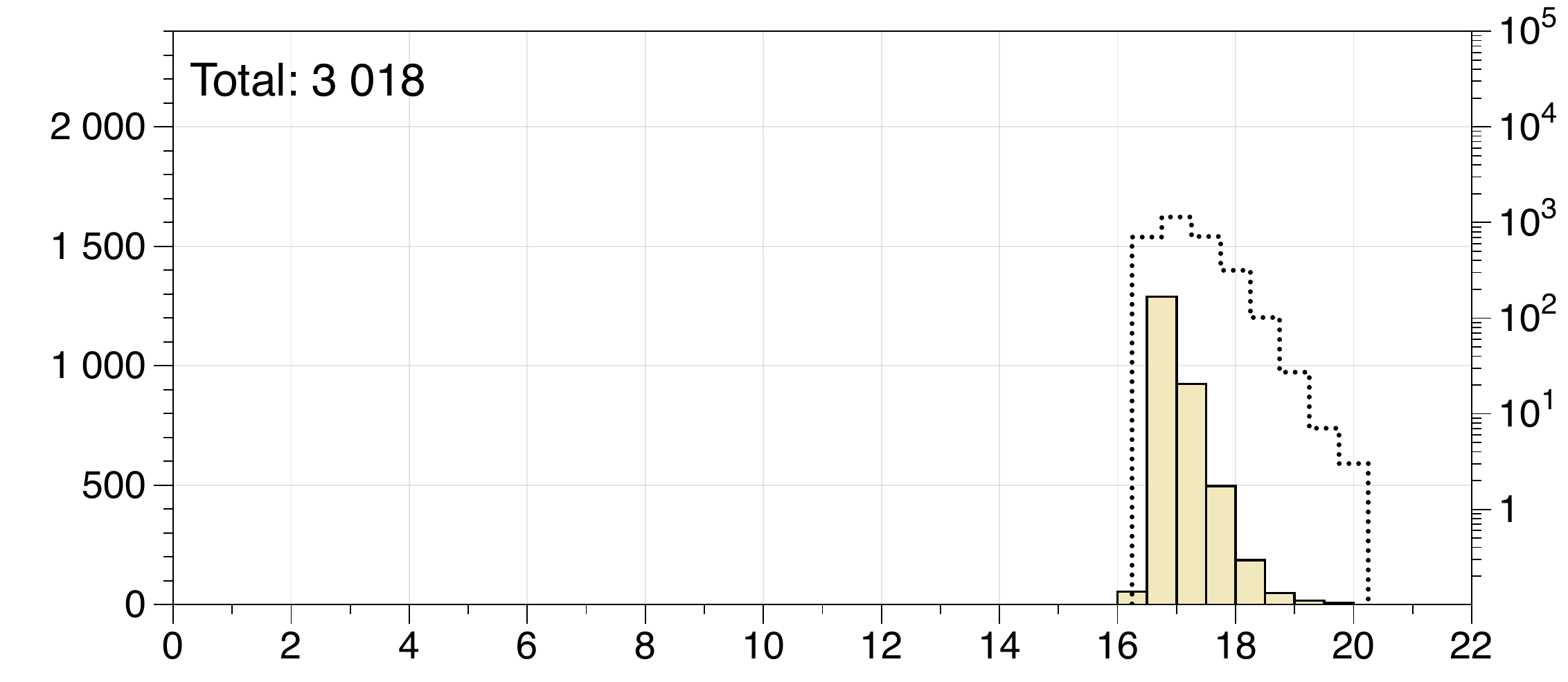} 
  \vspace{-4.2cm}
\\
\hspace{0.96cm}{\fontfamily{phv}\selectfont \tiny Classif.: ALL}    & \hspace{0.96cm}{\fontfamily{phv}\selectfont \tiny SOS: Short-timescale variables }
\vspace{3.6cm} 
\\[6pt]
\hspace{3.5cm}{\fontfamily{phv}\selectfont \tiny median $G$ [mag]}  & \hspace{3.5cm}{\fontfamily{phv}\selectfont \tiny median $G$ [mag}]\\[1pt]

\end{tabular}
\caption{Histogram counts of the median $G$-band magnitude distribution of the published classification classes (left, see Sect.~\ref{sec:nTransits2class}) and SOS tables (right, see Sect.~\ref{sec:sos}). In the classification plots the \texttt{best\_class\_name} entries were grouped by main type, as listed in parentheses. The filled bars correspond to the linear scale on the left side of the plots (which varies in scale), while the dotted line corresponds to the logarithmic scale on the right side of the plots (always in the same scale). The bin size is fixed to 0.5 mag.}
\label{fig:histMags}
\end{figure*}

\begin{figure*}[h]
\begin{tabular}{@{}lll@{}}
\setlength{\tabcolsep}{0pt} 
\renewcommand{\arraystretch}{0} 
  \vspace{0cm}\\
  
 \includegraphics[width=\textwidth]{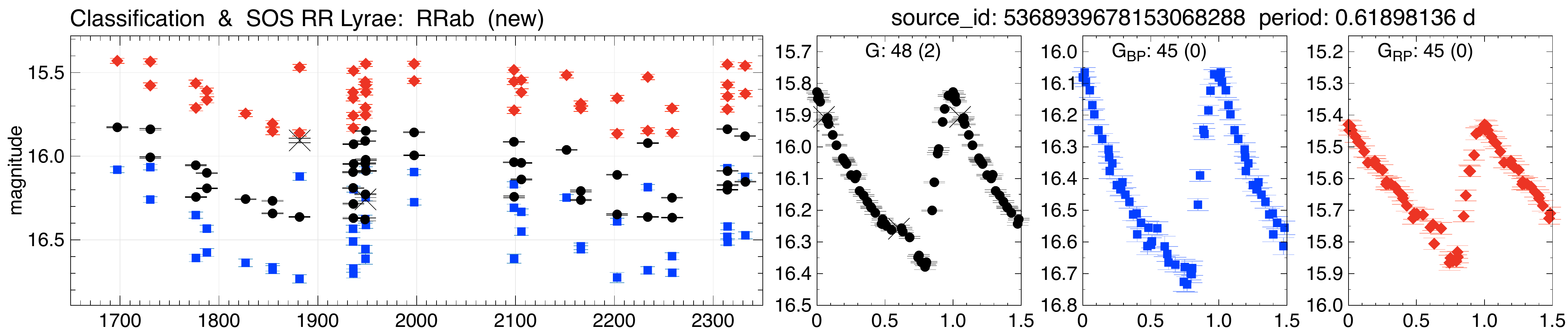}    
  \vspace{-3.9cm}
\\
\hspace{0.8cm}
\vspace{3.5cm} \\[6pt]

 \includegraphics[width=\textwidth]{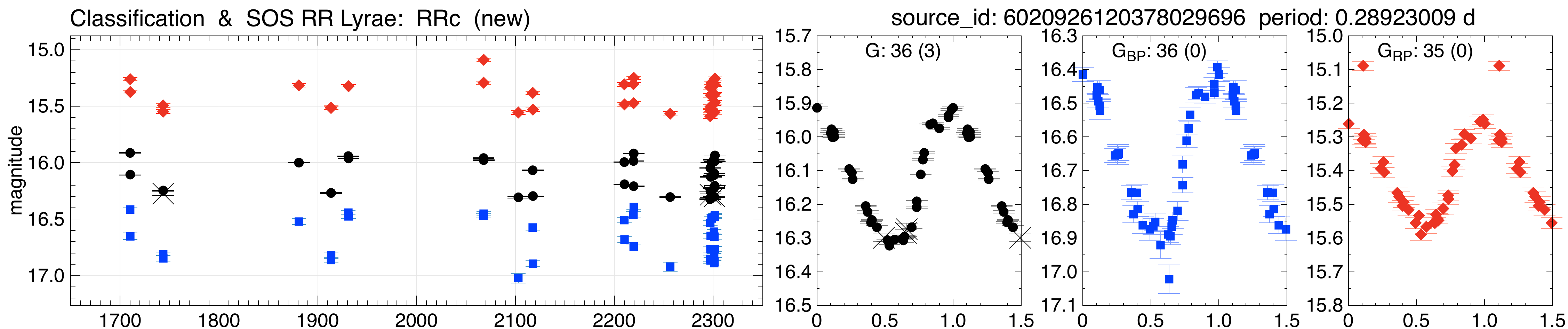}    
  \vspace{-3.9cm}
\\
\hspace{0.8cm}
\vspace{3.5cm} \\[6pt]
 \includegraphics[width=\textwidth]{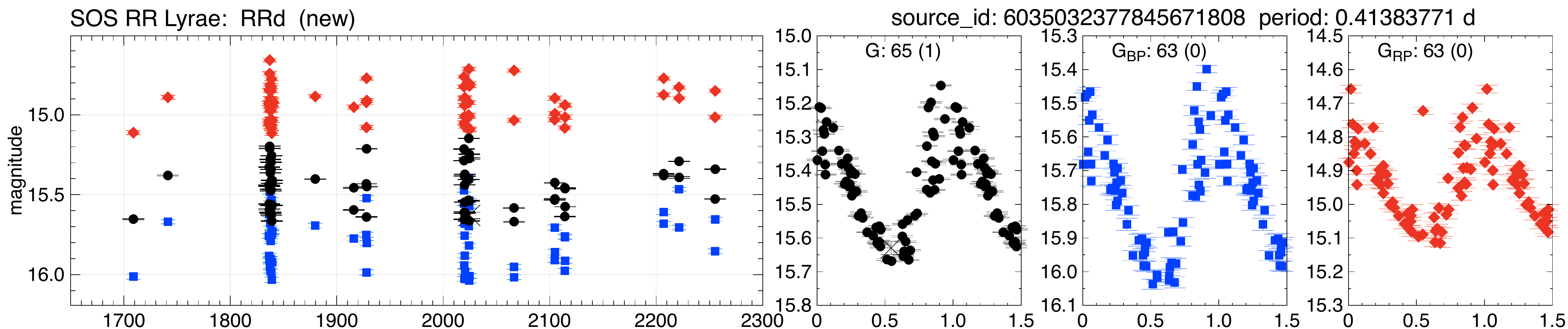}    
  \vspace{-3.9cm}
\\
\hspace{0.8cm}
\vspace{3.5cm} \\[6pt]

 \includegraphics[width=\textwidth]{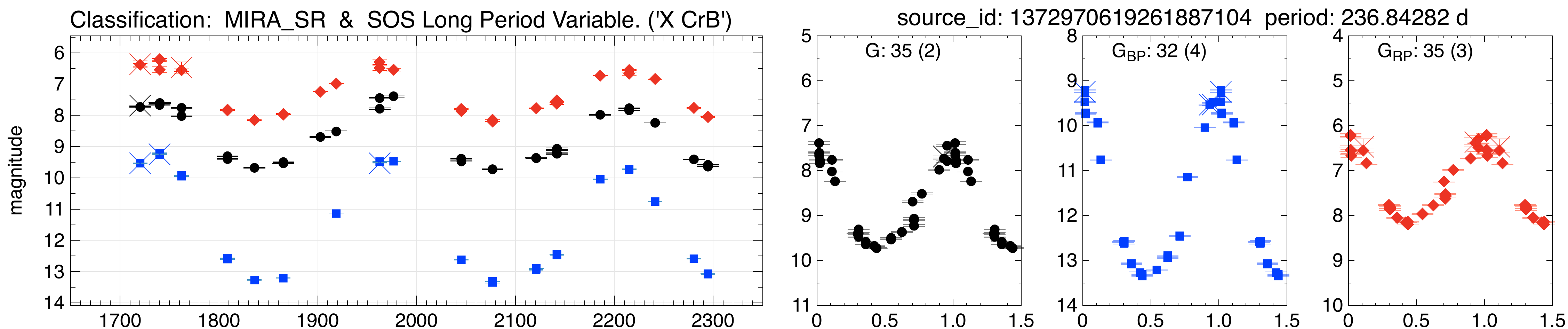}   
 \vspace{-3.9cm}
\\
\hspace{0.8cm}
\vspace{3.3cm} \\[6pt]

 \includegraphics[width=\textwidth]{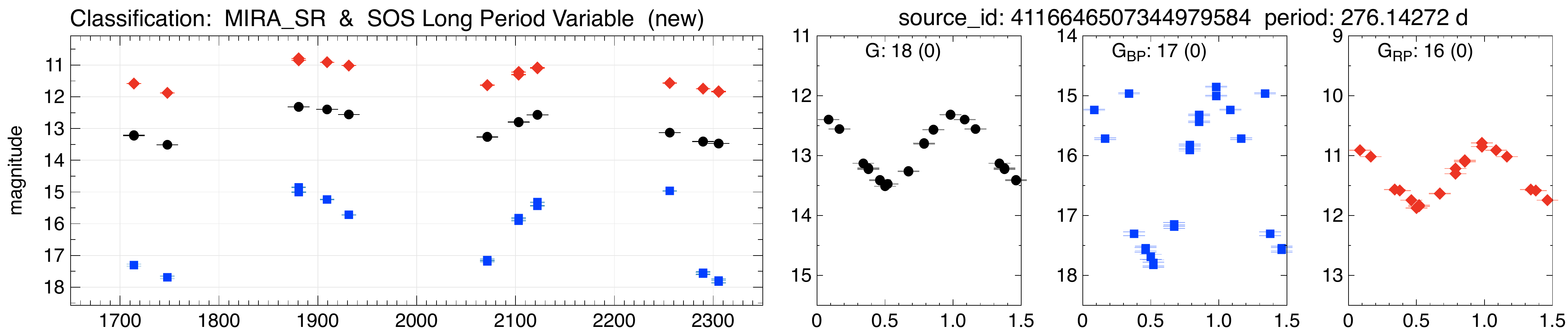}   
\vspace{-3.9cm}
\\
\hspace{0.8cm}
\vspace{3.3cm} \\[6pt]

\hspace{2.9cm}{\fontfamily{phv}\selectfont{\tiny BJD - 2455197.5 (TCB)} \hspace{4.5cm}{\fontfamily{phv}\selectfont \tiny phase} \hspace{2.25cm}{\fontfamily{phv}\selectfont \tiny phase} \hspace{2.25cm}{\fontfamily{phv}\selectfont \tiny phase}}\\[1pt]

\end{tabular}
\caption{Example light-curves of RR Lyrae stars and LPV in the data.
For each band we show the valid FoV transits, as well as variability-rejected FoV transits (number in parentheses, data plotted with crosses). In most cases, two FoV-transits occur within the 6~h spin period, causing indistinguishable overlapping points in the left diagrams. Some outliers are not flagged in the exported data (see e.g. the $G_{\rm RP}$ light curve of the RRc and RRd example); these transits were removed in the more strict outlier rejection used in the SOS Cep and RRL module, as mentioned in Sect.~\ref{sec:obsFiltering}.}
\label{fig:lightcurves1}
\end{figure*}

\begin{figure*}[h]
\begin{tabular}{@{}lll@{}}
\setlength{\tabcolsep}{0pt} 
\renewcommand{\arraystretch}{0} 
  \vspace{0cm}\\

 \includegraphics[width=\textwidth]{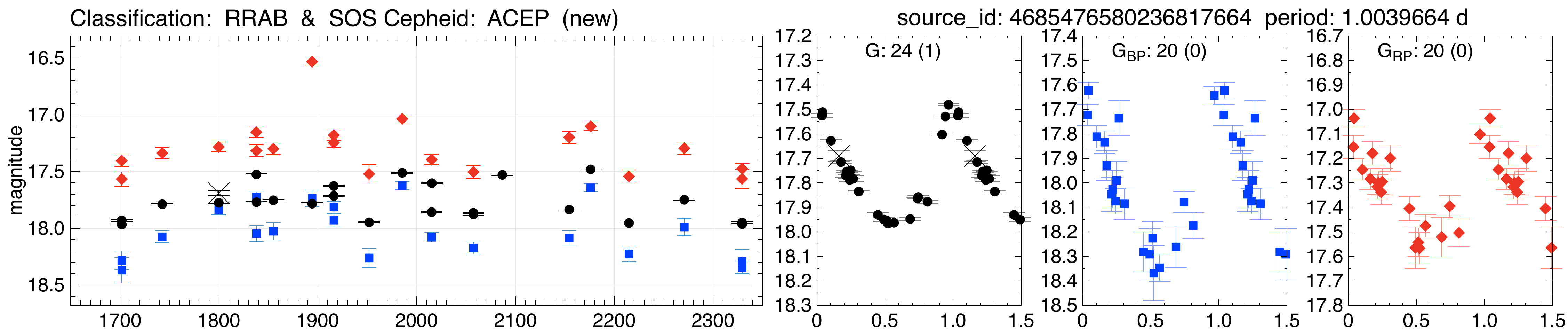}    
   \vspace{-3.9cm}
\\
\hspace{0.8cm}
\vspace{3.5cm} \\[6pt]

 \includegraphics[width=\textwidth]{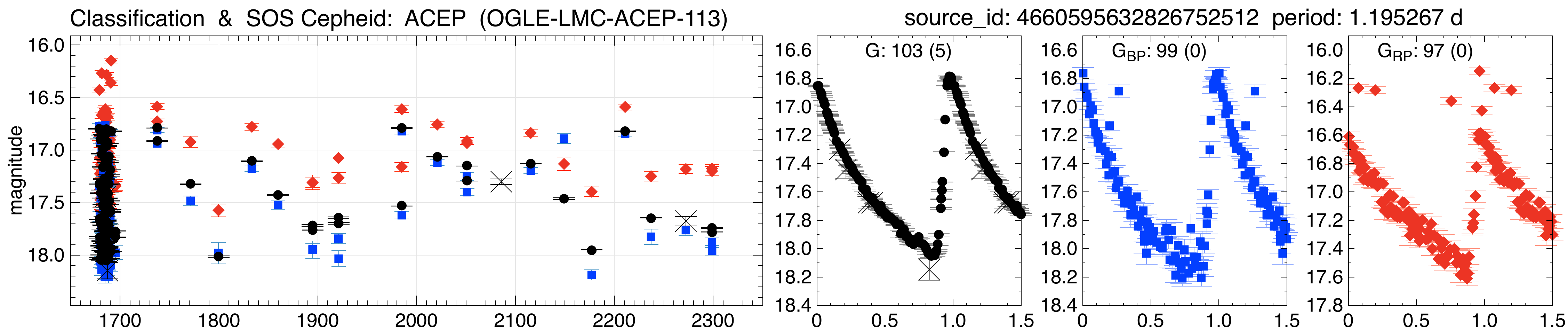}    
   \vspace{-3.9cm}
\\
\hspace{0.8cm}
\vspace{3.5cm} \\[6pt]

 \includegraphics[width=\textwidth]{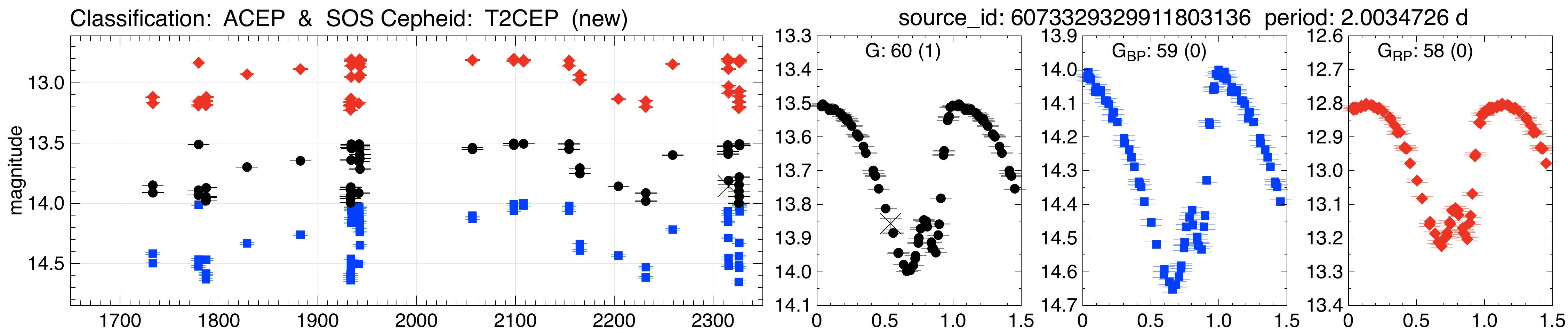}    
  \vspace{-4.cm}
\\
\hspace{0.9cm}
\vspace{3.6cm} \\[6pt]

 \includegraphics[width=\textwidth]{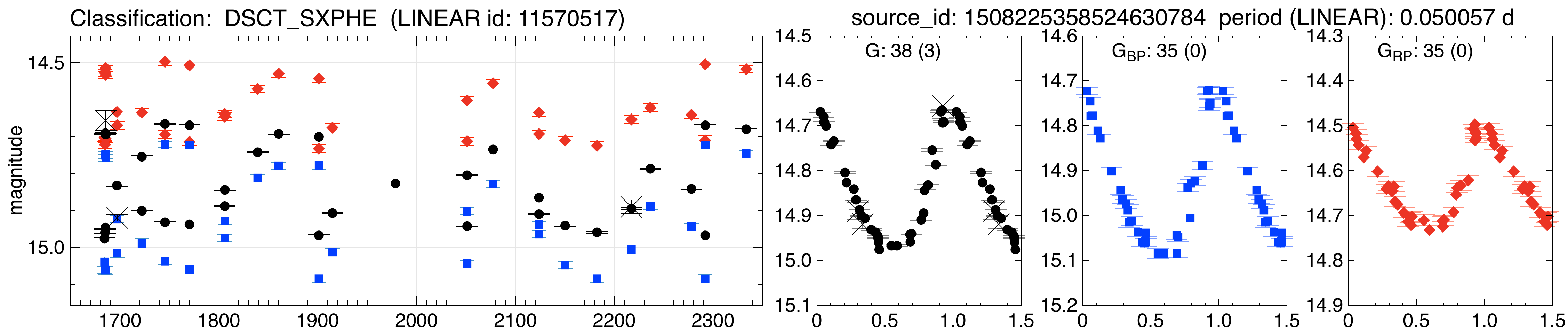}    
   \vspace{-3.9cm}
\\
\hspace{0.8cm}
\vspace{3.5cm} \\[6pt]

 \includegraphics[width=\textwidth]{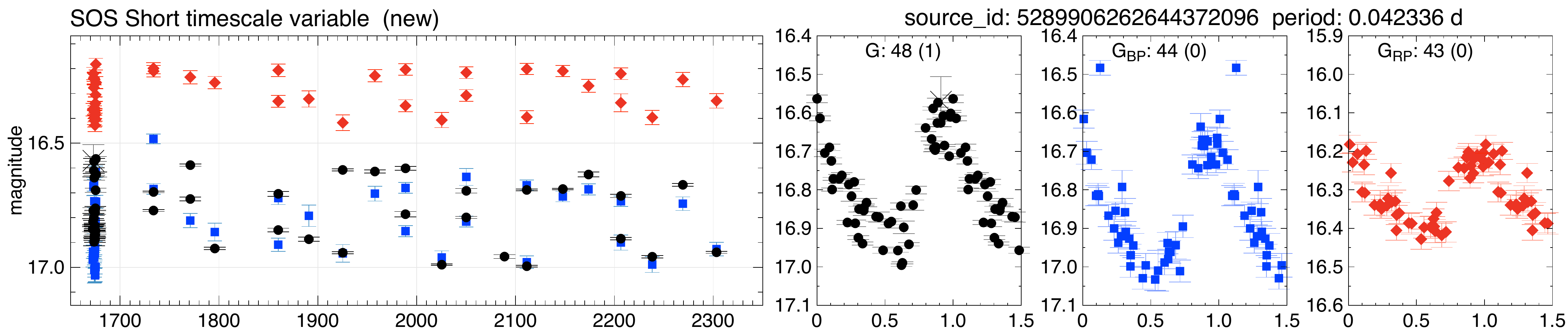}    
  \vspace{-3.9cm}
\\
\hspace{0.8cm}
\vspace{3.3cm} \\[6pt]

\hspace{2.9cm}{\fontfamily{phv}\selectfont{\tiny BJD - 2455197.5 (TCB)} \hspace{4.5cm}{\fontfamily{phv}\selectfont \tiny phase} \hspace{2.25cm}{\fontfamily{phv}\selectfont \tiny phase} \hspace{2.25cm}{\fontfamily{phv}\selectfont \tiny phase}}\\[1pt]

\end{tabular}
\caption{Example light-curves of the Cepheid, $\delta$ Scuti, and short-timescale variables.
For each band we show the valid FoV transits, as well as variability-rejected FoV transits (number in parentheses, data plotted with crosses). In most cases, two FoV-transits occur within the 6~h spin period, causing indistinguishable overlapping points in the left diagrams. Some outliers are not flagged in the exported data (see e.g. the $G_{\rm BP}$ and $G_{\rm RP}$ light curve of the \texttt{ACEP} example); these transits were removed in the more strict outlier rejection used in the SOS Cep and RRL module, as mentioned in Sect.~\ref{sec:obsFiltering}.
No period is published for the classification results, therefore the period for the \texttt{DSCT\_SXPHE} example is taken from the LINEAR survey \citep{2013AJ....146..101P}.
\label{fig:lightcurves2}
}
\end{figure*} 

\begin{figure*}[h]
\begin{tabular}{@{}lll@{}}
\setlength{\tabcolsep}{0pt} 
\renewcommand{\arraystretch}{0} 
  \vspace{0cm}\\

 \includegraphics[width=\textwidth]{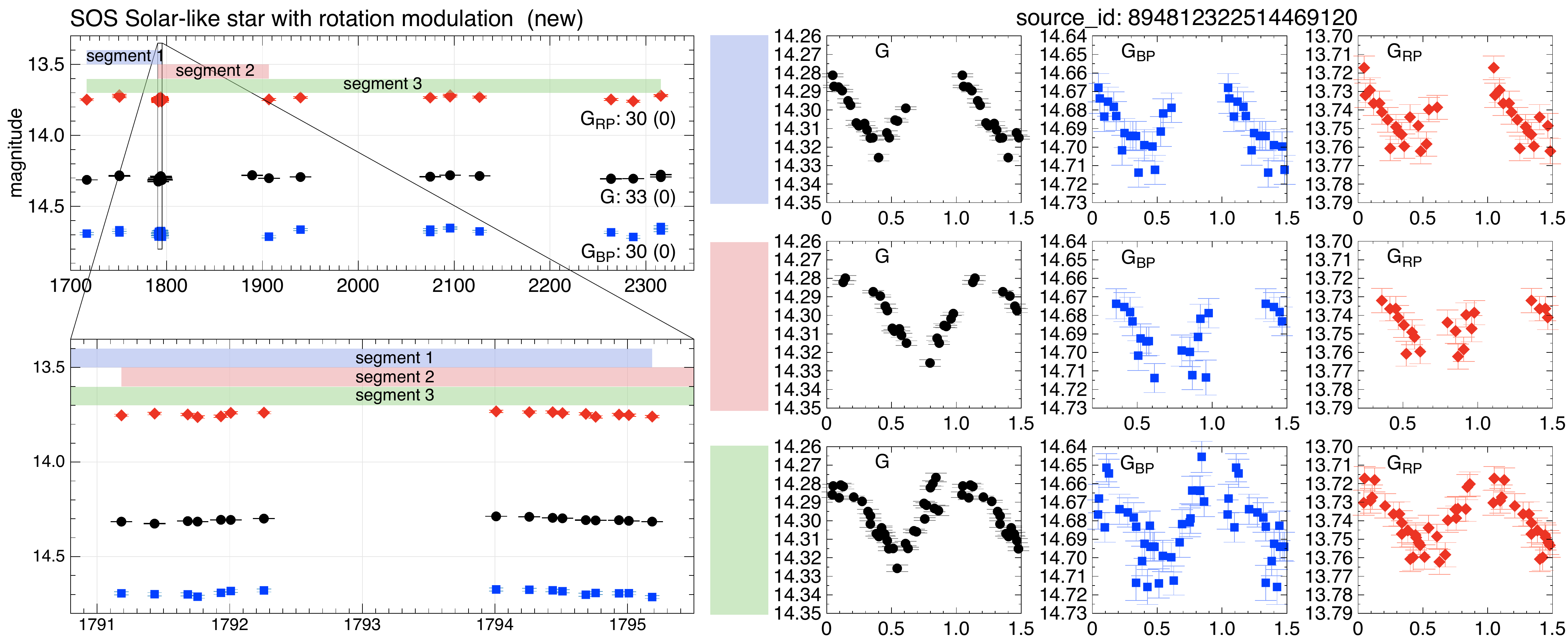}    
   \vspace{-7.5cm}
\\
\vspace{-0.23cm} \\[6pt]
\hspace{8.3cm} \rotatebox{90}{\tiny{\hbox{\ \ \ {\fontfamily{phv}\selectfont segment 1}}}} \rotatebox{90}{\tiny{{\fontfamily{phv}\selectfont period=3.926 d}}} \vspace{0.4cm} \\[1pt]
\hspace{8.3cm} \rotatebox{90}{\tiny{\hbox{\ \ \ {\fontfamily{phv}\selectfont segment 2}}}} \rotatebox{90}{\tiny{{\fontfamily{phv}\selectfont period=4.587 d}}} \vspace{0.35cm} \\[1pt]
\hspace{8.3cm} \rotatebox{90}{\tiny{\hbox{\ \ \ {\fontfamily{phv}\selectfont segment 3}}}} \rotatebox{90}{\tiny{{\fontfamily{phv}\selectfont period=3.874 d}}} \vspace{0.32cm} \\[1pt]
\hspace{2.9cm}{\fontfamily{phv}\selectfont{\tiny BJD - 2455197.5 (TCB)} \hspace{4.5cm}{\fontfamily{phv}\selectfont \tiny phase} \hspace{2.25cm}{\fontfamily{phv}\selectfont \tiny phase} \hspace{2.25cm}{\fontfamily{phv}\selectfont \tiny phase}}\\[1pt]

\end{tabular}
\caption{Example light-curves of a solar-like star with rotation modulation. Period search is performed over multiple segments, as shown in the right panels; see Sect.~\ref{sec:sosRotMod} for more details. The bottom left panel expands a high-cadence time
range of the time series. The top left panel shows for each band the valid FoV transits and variability-rejected FoV transits (number in parentheses, here 0). 
\label{fig:lightcurveRot}
}
\end{figure*}


\section{Results\label{sec:results}}

\subsection{Observation filtering and operators\label{sec:obsFiltering}} 
The variability processing makes use of photometric data provided in units of fluxes (e/s), and then converts them into magnitudes using the magnitude zero-points defined in  \cite{DWE-052}. In our pipeline, this transformation is done by our \texttt{GaiaFluxToMagOperator}  {\textup{operator}}. Several additional operators were used to remove FoV transits by flagging and filtering\textup{} them when they did not meet the required criteria, that is, when they were of insufficient quality.
 The chain of consecutive {\textup{operators}} is described briefly here, for more details see the DR2 documentation:

\begin{itemize}

\item \textbf{RemoveNaNNegativeAndZeroValuesOperator} flagged and removed transits with NaN, negative, or zero flux values.
\item \textbf{RemoveDuplicateObservationsOperator} flagged and removed pairs of transits with too close observation times (essentially one transit from the pair was incorrectly assigned to the source). This operator was responsible for the majority of the transit removals for $G$ magnitudes brighter than $\sim$8, see \cite{Mowlavi18Lpv} for more details.  

\item \textbf{GaiaFluxToMagOperator} converted the Gaia fluxes into magnitude using the Gaia zero-point magnitudes.
\item \textbf{ExtremeValueCleaning} flagged and removed transits with unrealistically faint magnitudes. The cut values were chosen to be $G\geq 25$, $G_{\rm BP}\geq 24$, and $G_{\rm RP} \geq 22$. 
\item \textbf{ExtremeErrorCleaningMagnitudeDependent} flagged and removed transits with extreme magnitude errors. For $G$, we used both a lower (0.01\%) and upper (99.7\%) threshold based on the observed distribution of transit magnitude errors (as a function of magnitude), while for $G_{\rm BP}$ and $G_{\rm RP}$, we used only an upper threshold (99.9\%).
\item \textbf{RemoveOutliersFaintAndBrightOperator} flagged and removed the time-series outliers of the source. Different criteria were used to account for the distribution of magnitudes or of the magnitude errors in the time series. 
\end{itemize}

The time series following the above last operator was then the input for the different variability modules, 
and all transits that were flagged and filtered in the operator chain can be identified in the published time series
by the \texttt{rejected\_by\_variability} flag (see Sect.~\ref{sec:timeSeriesStatisticsAndFlag}). The resulting number of {\textup{selected}} (i.e. not-rejected) transits can be found in the  \texttt{num\_selected\_g\_fov/bp/rp} fields of the \texttt{vari\_time\_series\_statistics} table.
 In Figs.~\ref{fig:lightcurves1}, \ref{fig:lightcurves2}, and \ref{fig:lightcurveRot} the rejected transits are plotted as crosses. However, some SOS modules (e.g. SOS Cep and RRL, SOS rotation modulation) applied individual stricter conditions that
were tuned for their analysis.   We note that the SOS short timescale module also used another operator, \textbf{RemoveOutlierPerTransitOperator}, to remove CCD outliers per transit.

\begin{figure*}[h]
\begin{tabular}{@{}cc@{}}
\setlength{\tabcolsep}{0pt} 
\renewcommand{\arraystretch}{0} 
  \vspace{0cm}\\
\multicolumn{2}{c}{\hspace{-1.cm} {\fontfamily{phv}\selectfont \tiny $G$: mean \# FoV transits} } \\
\multicolumn{2}{c}{ \includegraphics[width=0.75\textwidth]{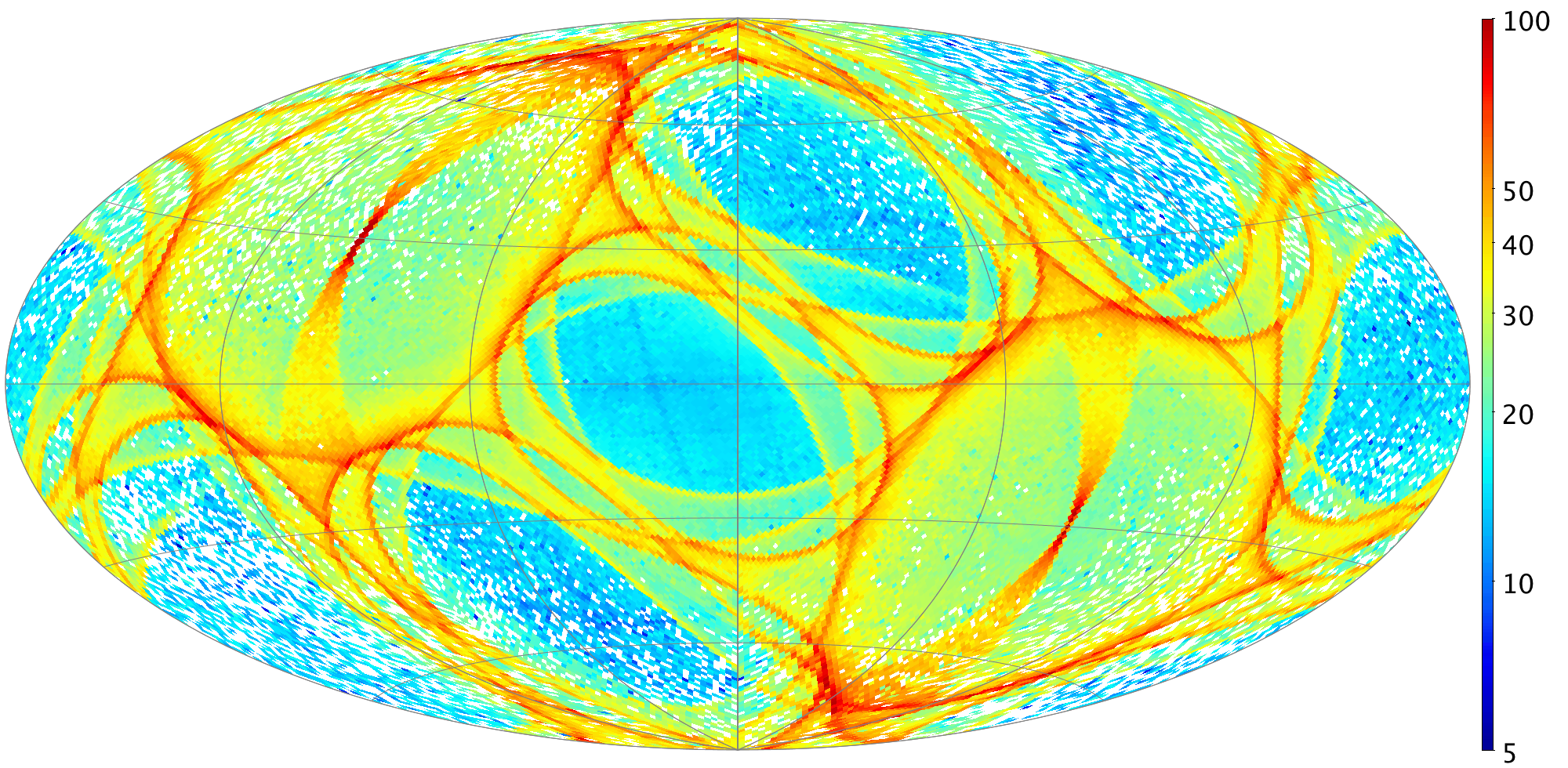}  } \\[6pt]
\hspace{-0.6cm}{\fontfamily{phv}\selectfont \tiny $G_{\rm BP}$: mean \# transits} 
& \hspace{-0.6cm}{\fontfamily{phv}\selectfont \tiny $G_{\rm RP}$: mean \# transits} \\
 \includegraphics[width=0.48\textwidth]{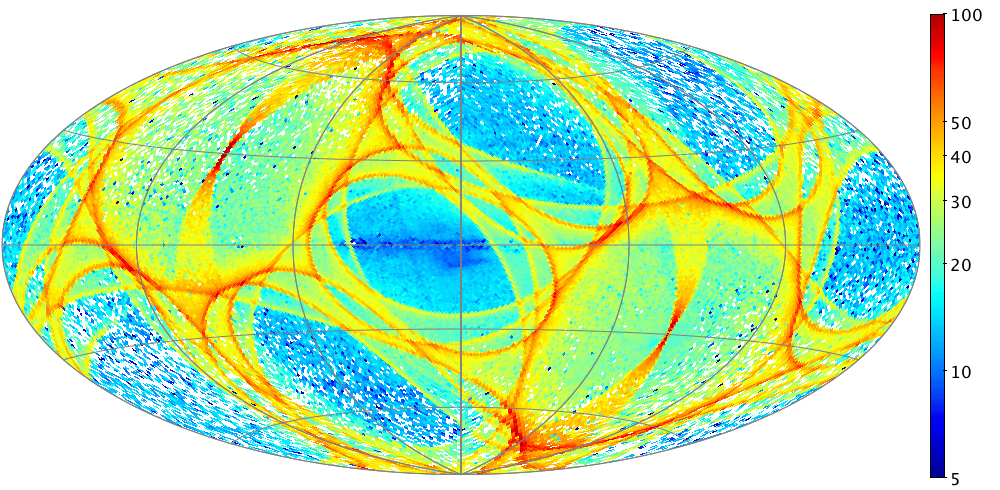}   &  \includegraphics[width=0.48\textwidth]{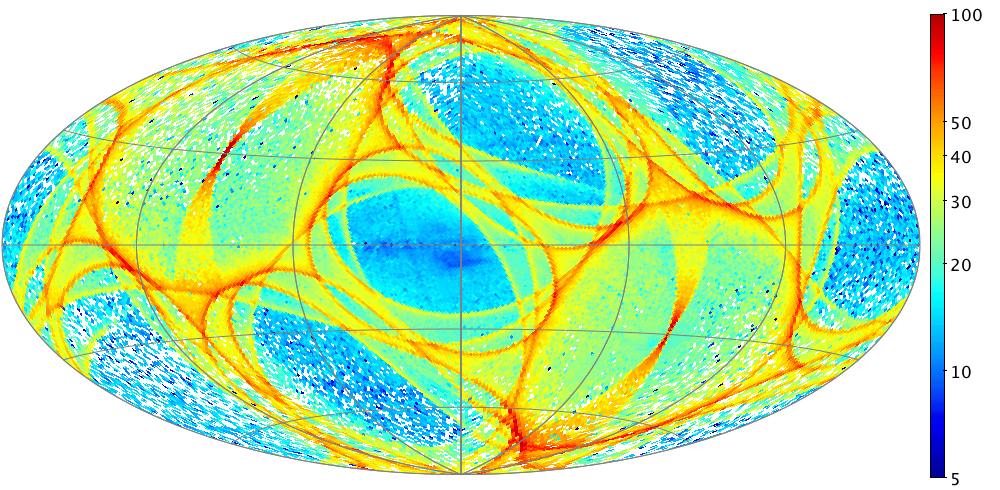}    \\[6pt]
\end{tabular}
\caption{Selected (i.e. not-rejected) number of transits in the time series of the $G$, $G_{\rm BP}$, and $G_{\rm RP}$ photometric bands in Galactic coordinates for all 550\,737 published sources. These numbers have been aggregated in sky bins of about 0.84~deg$^{2}$ (level 6 HEALPix) and their mean value is shown. The colour scale is clipped at 100, although the number of transits reaches up to 242, as shown in Fig.~\ref{fig:numTransitsHist}. The choice of the rainbow colour scale has been motivated by the convenient colour divisions between 5, 10, 20, 30, 40, 50, and 100 transits. Galactic longitude increases to the left side. 
}
\label{fig:numTransitsOnSky}
\end{figure*}

\begin{figure*}[h]
\begin{tabular}{@{}lll@{}}
\setlength{\tabcolsep}{0pt} 
\renewcommand{\arraystretch}{0} 
  \vspace{0cm}\\

 \includegraphics[width=0.99\textwidth]{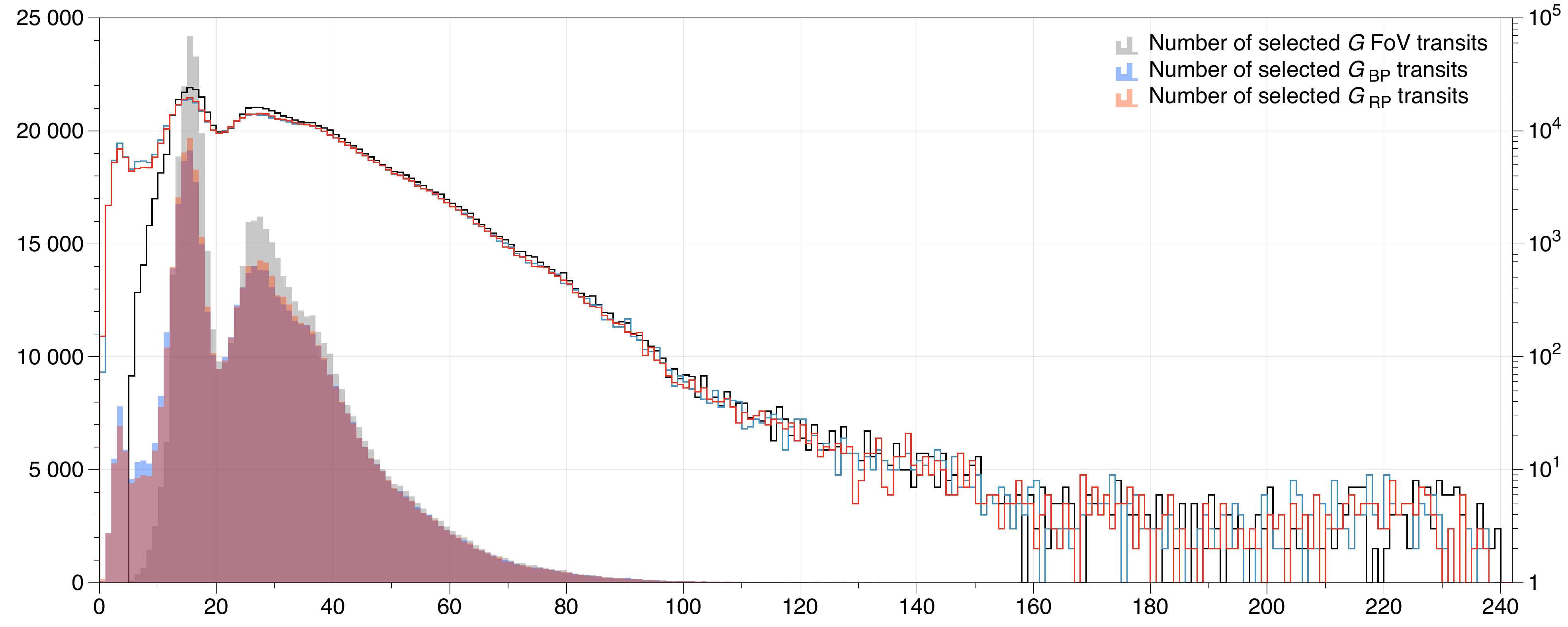}  \vspace{-0.4cm} \\[6pt]
\end{tabular}
\caption{Histogram of selected (i.e. not-rejected) number of transits in time series of the $G$, $G_{\rm BP}$, and $G_{\rm RP}$ photometric bands for the  550\,737 published variable sources. The filled bars correspond to the linear scale on the left side, while the solid lines corresponds to the logarithmic scale on the right side of the plot. The bin size is 1 FoV transit.
}
\label{fig:numTransitsHist}
\end{figure*}

The sky distribution of the  selected number of transits in the $G$, $G_{\rm BP}$, and $G_{\rm RP}$ time series are shown in Fig.~\ref{fig:numTransitsOnSky} for all 550\,737 published sources. The distribution of the mean values per sky-pixel is dominated by the scanning law, which is also visible in the large masked regions of the $\geq20$~FoV panel of Fig.~\ref{fig:skyDensitiesGeqCuts}. 
Figure~\ref{fig:numTransitsHist} shows a histogram of the time-series lengths. The minimum number of transits in $G$ is 5.

\subsection{Time series, statistics, and variability flag}
\label{sec:timeSeriesStatisticsAndFlag}
The full list of variable stars published in DR2 can be easily identified 
in the \texttt{gaia\_source} table by the field \texttt{phot\_variable\_flag,} which is set to \texttt{VARIABLE}, while for all other sources
it is set to \texttt{NOT\_AVAILABLE}. The time series of all
variable stars are published in $G$, $G_{\rm BP}$, and $G_{\rm RP}$. In contrast to DR1, the time series are no longer provided in a separate archive table, but instead are accessible in a
Virtual Observatory Table\footnote{See \href{http://www.ivoa.net/documents/VOTable/}{\tt http://www.ivoa.net/documents/VOTable/}} (VOTable) linked\footnote{For tutorials on using this datalink, see the Gaia archive help webpage \url{http://gea.esac.esa.int/archive-help}. Time-series bulk download is possible from \url{http://cdn.gea.esac.esa.int/Gaia/}.} via the field \texttt{epoch\_photometry\_url} in the \texttt{gaia\_source} table. For various reasons, a fraction of the transit observations was not photometrically processed in DR2, resulting in {\textup{null}}
values. The variability processing flags per FoV transit (see previous section) are provided in the column \texttt{rejected\_by\_variability} of the VOTable epoch photometry. Additional flags from the photometric processing are available for transits in the column \texttt{rejected\_by\_photometry}.  Unchanged from DR1 is the expression of observation time in units of barycentric JD (in TCB) in days $- 2\,455\,197.5$, see DR2 documentation or \citet{EyerDR1} for more details.

All time series are provided in both flux and magnitude, where the flux is provided with the associated 
standard error. The magnitude-transformed error value is omitted given its non-symmetric nature\footnote{Practical approximation: $mag_{error} \approxeq (2.5 / ln(10)) 
\cdot flux_{error} / flux_{value}$}. The Vega-system flux to magnitude zero-points are defined in \cite{DWE-052}.

Several basic statistical parameters of the cleaned\footnote{Selecting FoV transits with \texttt{rejected\_by\_variability=false}}  time series are listed in the \texttt{vari\_time\_series\_statistics} table. The variable stars will furthermore  have an entry in at least one of the \texttt{vari\_*} tables, see Fig.~\ref{fig:globalNumbers} for their numbers. An overview of time-series statistics per \textit{Gaia} band and types is provided in Tables~\ref{tbl:tsrStatGeq2} and \ref{tbl:tsrStatSOS}. 
The mean magnitudes provided in \texttt{gaia\_source} (\texttt{phot\_g/bp/rp\_mean\_mag}) differ from the mean magnitudes in \texttt{vari\_time\_series\_statistics} (\texttt{mean\_mag\_g\_fov/bp/rp}) because they are calculated
differently (see \citealt{DWE-052}) and because
the filtering of the light curves was different; furthermore, the \texttt{gaia\_source} mean photometry can be absent, but present in the \texttt{vari\_time\_series\_statistics} table (together
with the related time series in \texttt{epoch\_photometry\_url}) because various filters were applied by the photometry pipeline.
The sky and median magnitude distributions of these sources are shown in more detail in Figs.~\ref{fig:skyDensities} and \ref{fig:histMags}.

\subsection{Classification}
\label{sec:nTransits2class}
In DR2 we introduce the output of the semi-supervised classifier targeted to identify high-amplitude variable stars over the whole sky, resulting from the geq2 path described in Sect.~\ref{sec:overviewVarProcessing}.
This is the aim of the \textit{nTransits:2+} classification.
Owing to the limited 22 months of data and rejected observations, the number of FoV transits per source can be very small, prohibiting the use of Fourier modelling parameters for a large portion of the sources. Previous studies such as \citet{2000AJ....120..963I} and \cite{2016ApJ...817...73H} have shown that it is still possible to identify good RR Lyrae star candidates with just a few observations. Our classifier was trained 
to identify high-amplitude variables of the type of RR~Lyrae stars (anomalous~RRd, see \citealt{2016MNRAS.463.1332S} ; RRd, RRab, RRc), LPV (Mira-type and semi-regulars), Cepheids (anomalous Cepheids, $\delta$ Cepheids, type-II Cepheids), $\delta$ Scuti and SX Phoenicis  stars, amongst the many other \mbox{(non-)}variable source types that exist. These classes can be found in the \textit{Gaia} archive \texttt{vari\_classifier\_result} table with \texttt{best\_class\_name} =
\texttt{ARRD, RRD, RRAB, RRC, MIRA\_SR, ACEP, CEP, T2CEP,and DSCT\_SXPHE}, respectively. Each entry has an associated \texttt{best\_class\_score} between 0 and 1.
Additional (non-published) classes were included in the prediction types, see \cite{Rimoldini18Geq2} for all details. Descriptions of the classifier and the published classes can be found in the \textit{Gaia} archive \texttt{vari\_classifier\_definition} and \texttt{vari\_classifier\_class\_definition} table,  respectively.

The sky and magnitude distribution of the published classes, grouped by main type, is shown in the left panels of Figs.~\ref{fig:skyDensities} and \ref{fig:histMags}, and some time-series examples are shown in Figs.~\ref{fig:lightcurves1} and \ref{fig:lightcurves2}.

The published source samples were cleaned to reduce contamination levels caused by data and processing artefacts, as well as by
genuine variability of other types.
The results and details of this procedure are fully detailed in \cite{Rimoldini18Geq2}.

The output of this classifier\, for sources containing at least 12 measurements in their $G$-band time series was used as one of the two inputs to the SOS of RR Lyrae stars, Cepheids, and LPV, as shown in Fig.~\ref{fig:variabilityProcessingOverview}, hence accounting for the overlap between classification and SOS results in Fig.~\ref{fig:globalNumbers}. Some SOS modules may reclassify the type in the SOS tables, see Sect. \ref{sec:geq2SosDifferent}
An overview of the inferred completeness and contamination with respect to several cross-match catalogues of the different types is provided in Table~\ref{tab:skyCompletenessClas}.

\begin{table*}
\caption{Classification (absolute) completeness and contamination estimates with respect to available cross-matched reference catalogues. For more details on the validation of the classification results, we refer to \cite{Rimoldini18Geq2}.}
\label{tab:skyCompletenessClas}      
\centering
\begin{tabular}{l l l r r}
\hline\hline
Group & Class (\texttt{best\_class\_name}) & Catalogue (and region) & Completeness & Contamination \\     
\hline
Cepheids & \texttt{CEP, ACEP, T2CEP} & OGLE-IV (LMC) & 72\% &>5\% \\ 
Cepheids & \texttt{CEP, ACEP, T2CEP} & OGLE-IV (SMC) & 68\% & >5\% \\ 
$\delta$ Scuti and SX Phoenicis & \texttt{DSCT\_SXPHE} & Catalina \citep{2014ApJS..213....9D} &  99\% & $>$13\% \\
RR Lyrae stars & \texttt{RRAB, RRC, RRD, ARRD} & OGLE-IV (LMC) & 60\% & >9\% \\ 
RR Lyrae stars & \texttt{RRAB, RRC, RRD, ARRD} & OGLE-IV (SMC) & 64\% & 15\% \\
RR Lyrae stars & \texttt{RRAB, RRC, RRD, ARRD} & OGLE-IV (Bulge) & 49\% & 44\% \\
RR Lyrae stars & \texttt{RRAB, RRC, RRD, ARRD} & Catalina \citep{2014ApJS..213....9D} &  63\% & $>$ 5\% \\
LPV & \texttt{MIRA\_SR} & OGLE-III\tablefootmark{a} (LMC+SMC) & 40--50\% & $<5$\% \\
LPV & \texttt{MIRA\_SR} & \textit{Gaia} DR2 with\tablefootmark{b} $\varpi/\sigma_{\varpi}>10$ & -- & $\sim 7$\% \\
\hline
\end{tabular}
\tablefoot{
        \tablefoottext{a}{Estimates are restricted to sources with peak-to-peak amplitude >0.2 mag.}
        \tablefoottext{b}{$\varpi$ and $\sigma_\varpi$ refer to the DR2 parallax and associated error.}
}
\end{table*}

\subsection{Specific Object Studies}
\label{sec:sos}
In the geq20 path described in Sect.~\ref{sec:overviewVarProcessing}, potential variable type candidates that are identified in the supervised classification and special variability detection (SVD) parts of the pipeline were analysed in SOS modules, which often derived a model and/or additional astrophysical properties. The SOS modules can also decide whether an object is not of the expected type and as a result not produce any output for it. We recall that a sub-sample of the sources processed in the geq2 path was processed by some SOS modules (see Fig.~\ref{fig:variabilityProcessingOverview}). We present short descriptions of the SOS modules used for DR2
below . 

The sky and magnitude distribution of each published type is shown in the right panels of Figs.~\ref{fig:skyDensities} and \ref{fig:histMags}, and example light curves are presented in 
Figs.~\ref{fig:lightcurves1}, \ref{fig:lightcurves2}, and \ref{fig:lightcurveRot}.
Because SOS modules can be fed by (partially) independent pipeline-paths, it is not excluded that a source is listed in multiple SOS tables, as discussed in Sect.~\ref{sec:sosOverlap}, or has a different classification and SOS type, as discussed in Sect.~\ref{sec:geq2SosDifferent}.
Despite extensive validation efforts, the nature of our processing is an automated and statistical characterisation of the sources, hence identified sources that are not yet known in the literature should be considered to be {\textup{candidates}}, unless they are specifically validated in one of the SOS articles referenced below.

An overview of the inferred (absolute) sky completeness and contamination with respect to existing catalogues of the different types is provided in Table~\ref{tab:skyCompletenessSos}; see sections below and the references therein for more details.
Many of the listed reference catalogues have a (very) limited sky coverage, and hence the contamination might be locally biased. To assess contamination in the whole SOS published samples, we performed the following additional visual inspection for each SOS output separately: we took a {\textup{random}} sample of 500 sources from the published results, as well as a {\textup{sky-uniform}\footnote{Approximate sky-uniform sampling is achieved by first grouping the sources in 12\,288~bins \citep[level 5 HEALPix, see][]{2005ApJ...622..759G} of about 3.6~deg$^{2}$ and then randomly drawing non-empty pixels and extracting a source at random from it.}} sample of 500~sources (non-overlapping with the random sample). The former sample follows the general sky distribution as shown in Fig.~\ref{fig:skyDensities}, while the latter achieves an almost sky-uniform sampling and hence also draws samples from very low density regions. The results are summarised below for each SOS result:
\begin{itemize}
\item \textbf{Cepheids}: 5\% of the random sample was visually rejected, of which 3\% might be due to binaries or ellipsoidals. Most of this sample is located in the Magellanic Clouds.

Of the sky-uniform sample, 15\%  was visually rejected, of which 11\% might be due to binaries or ellipsoidals.

\item \textbf{RR Lyrae stars}: 9\% of the random sample was visually rejected, of which 4\% are faint ($20.0 < G < 20.7$~mag).

Of the sky-uniform sample, 12\%  was visually rejected, of which 5\% are faint ($19.9 < G < 20.7$~mag).

\item \textbf{Long-period variables}: 
0.1\% (1) was visually rejected; it looks like a young stellar object, confirmed by its position in the Hertzsprung-Russel (HR)
diagram.
\item \textbf{Short-timescale variability}: 
0.3\% was visually rejected, and about 24\% seem to exhibit short-timescale periodic variability, but this can currently not be confirmed with a high confidence.
\item \textbf{Rotation modulation}: 
0.3\% was visually rejected.
\end{itemize}
These identified contamination rates are compatible\footnote{For Cepheids and RR Lyrae stars, the contamination rates quoted in Table~\ref{tab:skyCompletenessSos} were taken from these analyses because contaminants from the OGLE-IV catalogues were already removed from the published results; see \cite{Clementini18CepRrl} for more details.} with the cross-match sample contamination rates found in Table~\ref{tab:skyCompletenessSos}.

\begin{table*}
\caption{SOS (absolute) completeness and contamination estimates with respect to cross-matched reference catalogues. Processing paths are based on the number of selected FoV transits, as shown in Fig.~\ref{fig:variabilityProcessingOverview} and discussed in Sect.~\ref{sec:overviewVarProcessing}. See Sect.~\ref{sec:sos} for additional contamination rate estimates for random and sky-uniform samples. For more details on the validation of the SOS type results, see Sect.~\ref{sec:sos} and the references therein.}
\label{tab:skyCompletenessSos}      
\centering
\begin{tabular}{l l l r r}
\hline\hline
Processing path & SOS Type & Catalogue (and region) & Completeness & Contamination \\     
\hline
geq12 \& geq20 & Cepheids & OGLE-IV (LMC, limited region) & 74\% &   $\sim$ 5\%\tablefootmark{a}\\
geq12 \& geq20 & Cepheids & OGLE-IV (Bulge)  & 3\% & \tablefootmark{b} \\
geq12 \& geq20 & RR Lyrae stars & OGLE-IV (LMC, limited region) & 64\% &  $\sim$ 9\%\tablefootmark{a} \\
geq12 \& geq20 & RR Lyrae stars & OGLE-IV (Bulge)  & 15\% & \tablefootmark{b} \\

geq12 \& geq20 & LPV & OGLE-III\tablefootmark{c} (SMC+LMC) &$\sim$ 30\% & < few\%\ \ \ \\
geq12 \& geq20 & LPV & OGLE-III\tablefootmark{c} (SMC+LMC) &$\sim$ 30\% & < few\%\ \ \ \\

geq20 & Short timescale & OGLE-II, III, IV (MCs) & 0.05\% & 10--20\%\ \ \ \\
geq20 & Rotational modulation &The Pleiades \citep{2010MNRAS.408..475H} & 0.7 -- 5.0\% & 
< 50\%\tablefootmark{d} \\\hline
\end{tabular}
\tablefoot{
        \tablefoottext{a}{Estimated from non-cross-match random samples inspection discussed in Sect.~\ref{sec:sos}, extending partially beyond OGLE-IV footprint.}
        \tablefoottext{b}{No reliable contamination rate for the Bulge has been derived, but in the {\textup{Galactic disk}} footprint \citep[see][]{2017EPJWC.15201002U}, contamination of the SOS Cepheids and RR Lyrae stars might be of the order of several tens of percent (A. Udalski, priv. comm. based on unpublished OGLE-IV data).}
        \tablefoottext{c}{Estimates are restricted to sources with a peak-to-peak amplitude >0.2 mag and periods longer than 60 days.}
        \tablefoottext{d}{The 50\% `contamination' quantifies the non-occurrence in the reference catalogue; the actual contamination of grazing binaries is estimated to be $<1\%$, see Sect.~\ref{sec:sosRotMod}.} 
}\end{table*}

\subsubsection{Cepheids and RR Lyrae stars}

The SOS Cepheids and RR Lyrae module \cite[SOS Cep and RRL,][]{DPACP-13,Clementini18CepRrl}, confirmed,  characterised, and sub-classified in type 140\,784
RR~Lyrae  stars and 9\,575 
Cepheids of the {\textup{candidates}} provided by the classifiers. The module can reclassify the type provided by the classifiers from Cepheid to RR Lyrae star, and vice versa. 
Amongst the RR Lyrae stars, the following number of sub-types were identified: 
98\,026 fundamental mode pulsators (RRab),
40\,380 first-overtone pulsators (RRc), and 
2378 double-mode pulsators (RRd). 
Amongst the Cepheids, there were
8890 $\delta$ Cepheids,
100 anomalous Cepheids, and 
585 type II Cepheids, of which 
223 were classified as BL Her,
253 as W Vir, and 
109 as RV Tau. 

A rough estimate of completeness and contamination was obtained by comparing our samples with OGLE catalogues of RR~Lyrae stars and Cepheids in the Large Magellanic Cloud (LMC). We selected an area of the  LMC centred at $\alpha=83\degr$ and $\delta$=$-67.5\degr$ and extending from $67.5\degr$ to $97.5\degr$ in right ascension and from $-63.5\degr$ to $-73\degr$ in declination. Completeness of our  RR Lyrae sample in this area is 64\%, and contamination is less than 0.1\%. In the same area, the Cepheid sample is 74\% complete with 100\% purity.
A more in-depth description and analyses of the results, as well as the origin of the Bulge completeness rates quoted in Table~\ref{tab:skyCompletenessSos}, can be found in \citealt{Clementini18CepRrl}.
Of the 599 Cepheids and 2595 RR Lyrae  stars that were identified in DR1 (\citealt{DPACP-13, EyerDR1}),
533 Cepheids and 2517 RR Lyrae stars are also available in this DR2 release.

\subsubsection{Long-period variables}

Long-period variables are red giants that are characterised by long periods and variability amplitudes of up to a few magnitudes.
They are identified by classification as \texttt{MIRA\_SR}.
On the basis of data available for DR2, CU7 selected  151\,761 LPV candidates that fulfil the selection criteria for LPVs, which included a minimum of 12 points in $G$, a $G_{\rm BP}-G_{\rm RP}$ colour  higher than 0.5~mag, a variability amplitude (quantified with the 5\%-95\% trimmed range in $G$) larger than 0.2 mag, and a correlation between $G$ and $G_{\rm BP}-G_{\rm RP}$ higher than 0.5.
Of these, 150\,757 are published in the classification table, and 89\,617 have SOS results.
The detailed distribution into these two tables is indicated in Fig.~\ref{fig:globalNumbers}.
The SOS LPV module processes 
LPV candidates when their period is  longer than 60~days.
The steps applied include period search, the determination of bolometric correction from $G_{\rm BP}$ and $G_{\rm RP}$ fluxes, the identification of red supergiants, and the computation of absolute magnitude.
It must be noted that all LPV specific attributes published in DR2 that depend on the parallax must be recomputed based on the published parallaxes, because these LPV-specific attributes were computed based on an older version of the parallaxes that was available at the time of variability processing within the consortium. We refer to \cite{Mowlavi18Lpv} for additional information.
The light curves of all LPV candidates, whether identified in the classification table or in the SOS table, are published in DR2.

The completeness and contamination numbers for the LPV candidates towards the Magellanic Clouds are computed relative to a sub-sample of the OGLE-III catalogue of LPVs \citep{2009AcA....59..239S, 2011AcA....61..217S}.
The sub-sample consists of all OGLE-III LPVs with $I$-band amplitude larger than 0.2~mag for the classification completeness and contamination estimates, and is further restricted to the OGLE-III LPVs with periods  longer than 60~d for the SOS completeness and contamination estimates.
Additionally, an all-sky contamination estimate is provided for \textit{Gaia} LPV candidates that have small relative parallax uncertainties based on their position in the HR diagram.
The main contaminants in this sample consist of young stellar objects.
Details of SOS LPV processing and estimates of completeness and contamination can be found in \cite{Mowlavi18Lpv}, where the DR2 data set of LPV candidates are presented and further analysed.

\subsubsection{Short-timescale variability}

Short-timescale variable sources are astrophysical objects showing any variability in their optical light-curve with a characteristic timescale of the variation shorter than $1\,$d \citep{Roelens2017}. A variety of astronomical objects are known to exhibit such fast photometric variations, including both periodic and transient variability, involving amplitudes from a few millimagnitudes to a few magnitudes, with different variability characteristics and phenomena at the origin of variation, from pulsations to flares to eclipsing systems. The diverse variability types targeted by the short-timescale module range from short-period binary stars to pulsating white dwarfs (e.g. ZZ Ceti stars or V777 Herculis stars) and cataclysmic variables such as AM CVn stars.
However, for the \textit{Gaia}  DR 2, the short-timescale variability processing is oriented towards periodic variability, and the exact variable type of the detected short-timescale candidates is not yet determined. Both the investigation of transient variability  \citep{wevers18} and the further classification of the candidates are foreseen for the future \textit{Gaia} data releases. 

As mentioned previously, the identification of the short-timescale variable candidates published in \textit{Gaia} DR2 involves particular variability detection methods that are specifically tuned to detect fast photometric variability, complementarily to the general variability detection approach (which may miss some of these events because they are
so fast). These methods are applied to a preselected subset of the \textit{Gaia} data set \citep{EyerDR1}: for \textit{Gaia} DR 2, the short-timescale variability analysis is restricted to relatively faint sources ($G \sim 16.5 - 20\,$mag) whose $G$ per-CCD time series indicates that they are likely to show rapid variations \citep{Roelens18Sts}.
Consequently, the candidates resulting from the short-timescale module are expected to overlap with other short-period SOS variables such as RR Lyrae and $\delta$~Scuti stars with periods shorter
than $1\,$d, even though not there is no complete overlap with these SOS variable candidates.

The SOS short-timescale  module 
  identified 3018 short-timescale, suspected periodic candidates. 
These bona fide, short-timescale, suspected periodic candidates
were identified based on the combination of the variogram analysis \citep{Eyer1999, Roelens2017}, least-squares frequency search \citep{2009A&A...496..577Z}, analysis of the environment of candidate sources over the sky, and a series of selection criteria from various statistics, such as the Abbe or inter-quartile range (IQR) values in the three bands ($G$, $G_{\rm BP}$ and $G_{\rm RP}$).

Completeness and contamination numbers for the short-timescale variable candidate sample, estimated by comparing OGLE-II, III, IV, and \textit{Gaia} data in the Magellanic Clouds, are listed in Table~\ref{tab:skyCompletenessSos}. Furthermore, about 50\% of the known OGLE-III and IV variables belonging to the short-timescale sample in these regions are longer period variables (i.e. with period $>$~1d). However, most of these contaminant sources have periods of a few days and quite high amplitudes, hence although they are no true short-period variables, their presence in the short-timescale published sample is justified.
A more detailed description and analysis of the results can be found in \cite{Roelens18Sts}.

\subsubsection{Rotational modulation}\label{sec:sosRotMod}
Stellar flux modulation induced by surface inhomogeneities and rotation is searched for in a region of the  HR diagram, broadly embracing stars of spectral type later than F on the main sequence. 
This type of variability, often indicated in the literature as BY~Draconis stars (BY~Dra) regardless of possible binarity, is an indication of active regions (dark spots and bright faculae on the stellar photosphere) that are produced by stellar magnetic activity.
The evolution of stellar magnetic fields, similar to solar fields, produces variability phenomena on a wide range of timescales \citep[see e.g.][and references therein]{2004A&A...425..707L,2012MNRAS.421.2774D, 2006AN....327...21L}; one of the most prominent phenomena is the rotational modulation itself, from which the stellar rotation period can be inferred.

The variability pipeline comprises two packages dedicated to the detection and characterisation of solar-like stars with rotational modulation.
First, solar-like variable candidates are selected, and if they
are confirmed, they are studied in more detail to determine stellar rotation periods and other properties. Fig.~\ref{fig:lightcurveRot} shows one of the published light-curves together with the segments in which a significant (similar) period was detected.

We detected some $7 \times 10^5$ periodic variables in the pre-selected HR diagram region, which excludes areas populated by pulsating variables. The remaining expected main contaminants are eclipsing binaries and spurious detections derived from an incomplete sampling. To filter out these cases as well as possible, we applied filters that exclude sources whose period-folded light curves have a
significantly uneven distribution in phase, with significant gaps, and are far from being sinusoidal \citep[see][for details]{Lanzafame18RotMod}. The final DR2 clean sample contains 147\,535 rotational modulation candidates and fills 38\% of the whole sky when divided into bins of $\approx$0.84~deg$^{2}$ (level 6 HEALPix), see for example Fig.\,\ref{fig:skyDensities}.

An estimate of the final completeness and contamination of the DR2 rotational modulation variables is hampered by the fact that the occurrence rate of the BY Dra phenomenon is largely unknown
so far. 
It is expected that all low-mass dwarfs are magnetically active to some extent, and the rotational modulation detectability essentially depends on the instrumental sensitivity, and also on the active region distribution and the phase of the magnetic cycle at the epoch of observations.
New low-mass dwarfs displaying rotational modulation are continuously detected with the ever-wider span and increasing sensitivity of modern surveys, none of which has the full sky coverage capabilities of \textit{Gaia}, however.
At this stage, it is possible to perform some meaningful comparison only with the \cite{2010MNRAS.408..475H} and \cite{2016AJ....152..113R} observations of the Pleiades, with which, nevertheless, the DR2 geq20 sky coverage still overlaps only marginally. 
Assuming that the \cite{2010MNRAS.408..475H} catalogue lists all the BY Dra in its FoV down to $G\approx14.5$, we estimate that the completeness of the  BY~Dra sample is 14\% down to $G\approx16.5$ in the overlapping field.
Assuming that this value is uniform over the whole sky, we estimate a completeness upper limit of 5\%.
At the other extreme, we may assume that all low-mass dwarfs are BY~Dra variables.
Then, comparing with all stars observed by \textit{Gaia} in the same sky region and magnitude range, we estimate a lower completeness limit of 0.7\%. 
 
We may roughly estimate an upper limit for contamination by assuming that the \textit{Gaia} BY~Dra detections that are not present in the \cite{2010MNRAS.408..475H} or in the \cite{2016AJ....152..113R} sample in the common region on the sky are variables of other types.
This upper limit is obviously largely overestimated given that no BY~Dra catalogue can be deemed complete to date, as also testified by the fact that there are sources in the \cite{2010MNRAS.408..475H} sample that are not contained in the \cite{2016AJ....152..113R} sample, and vice versa.

At the other extreme, we can make an educated guess regarding the contamination of close grazing binaries that would be incorrectly classified as BY~Dra because of the similarity of the light curves. If the orbital period is long, then the 
grazing eclipse occurs only in a small part of the  orbit, and it does not look like rotation modulation. 
We assume an upper limit period of 10~days for this contaminating effect. The fraction of stars with such short-period companions is only about 2--5\% \citep{2010ApJS..190....1R},  of which only 2--5\% are grazing binaries (at most), which means 0.04--0.25~\% of the stars, while we did not yet take into account that for most binaries, the secondaries are substantially smaller. Even though we cannot consider the rotation modulation sample a random subset of G and later-type stars (which is the assumption in above estimate), it seems unlikely that this type of contamination would exceed 1\%. The same conclusion is reached from our internal validation against known and newly identified (but not published) eclipsing binaries, based on which, we estimate an upper limit of 0.5\% grazing contamination.
From the estimated grazing binary contamination  and non-occurrence in the reference catalogue discussed above, we estimate a contamination level ranging from $<1\%$ to $\approx$50\%, as listed in Table~\ref{tab:skyCompletenessSos}.
More details on the SVD solar-like and SOS rotational modulation  packages and results can be found in \cite{Lanzafame18RotMod}.

\subsection{Sources with multiple SOS types \label{sec:sosOverlap}}
 There are 80 sources in DR2 with an entry in more than one SOS table.
These cases are all overlaps between short-timescale objects and RR Lyrae stars (72), Cepheids (5), and rotation modulation objects (3), all of which are justified by their overlapping type definition. No overlap is found with long-period variables.

    



\subsection{Sources with different classification and SOS type \label{sec:geq2SosDifferent}}
Differences are found between classification types (as obtained by supervised classification in the geq2 path) and the
types that are eventually derived in SOS results (either from the geq12 or geq20 paths). In particular, one source was classified as \texttt{DSCT\_SXPHE,}
but also appears in SOS rotational modulation. Similarly, one source was classified as \texttt{MIRA\_SR}, was not confirmed by SOS LPV, but appears in SOS  rotational modulation.
Of the sources classified as a Cepheid (any type), 6 appear in more than one SOS table. One is a confirmed SOS Cep and also appears in SOS short timescale,
while 3 appear as SOS rotational modulation and 2 as SOS short timescale. Of the sources classified as an RR Lyrae star (any type), 70 appear in the SOS short timescale table, one
was instead confirmed as a Cepheid by SOS (shown in the top panel of Fig.~\ref{fig:lightcurves2}), and many were confirmed as RR Lyrae star by SOS. Overall, a reassignment of the type from RRL into Cep by SOS
occurred in 618 cases, but only in 77 cases from Cepheid to RR Lyrae star by SOS. We have decided to retain the original classification type
despite the possible reassignment by the SOS Cep\&RRL module or in cases where they were confirmed as another type by SOS modules.

\section{Conclusions
\label{sec:conclusions}}
In this second \textit{Gaia} data release, a sample of 550\,737 variable sources with their three-band time series and analysis results has been released to the astronomical community, showcasing samples of high-amplitude variable candidates distributed over (a large portion of) the whole sky. It demonstrates the immense potential of \textit{Gaia} data to provide an unbiased all-sky photometric survey together with astrometric and spectroscopic data. Although the variability analyses in DR2 did only mildly rely on the astrometric data and on none of the spectroscopic data, this dependence will become much heavier in future data releases. 

Several caveats have been outlined in this paper:
\begin{itemize}
\item Most provided stellar samples are rather incomplete as
a result of the current state of data calibration (e.g. even well-known variables might be missing) or the limited scanning law coverage for the minimum number of selected observations,
\item known incomplete flagging of outliers in the time series,
\item existence of uncalibrated systematics or spurious calibration error signals in some of the time series,
\item the probabilistic and automated nature of this work implies certain completeness and contamination rates (Tables~\ref{tab:skyCompletenessClas} and \ref{tab:skyCompletenessSos}), hence even well-known literature sources could be misidentified.
\end{itemize}

Overall, we estimate\footnote{From occurrence in a great variety of cross-match catalogues (most of which are documented in the DR2 documentation), and taking into account that our results have a varying degree of contamination.} to have identified the following numbers of new variables: a few hundred Cepheids, a few tens of thousand RR Lyrae stars, about one hundred thousand stars with rotation modulation, several tens of thousand long-period variables, a few thousand~$\delta$ Scuti and SX Phoenicis stars, and a few thousand short-timescale variables.  In total, this is about half of the variable sources released in this DR2.
Compared to the DR1 example, we increased the sample size about 100 times, which is expected to result in a large number of novel studies, regardless of whether they are combined with existing data sets. 

The next variability release will be based on data that are better calibrated yet again and on a longer time baseline. It  will
also include BP/RP spectrophotometry and radial velocities from RVS, which will allow us to improve the quality and quantity of the \textit{Gaia} DR3 release.

\begin{acknowledgements}\\
We dedicate this paper to our dear friend and colleague Jan~Cuypers (1956-2017), whose passing has left a great void in our variability group. Thank you, Jan, for your friendship and lifelong contribution to science. \\ 

We would like to thank the referee, Andrzej Udalski, for various suggestions and providing so far unpublished OGLE-IV estimates that improved the contents of this paper.
We thank Xavier Luri, Antonella Vallenari, and Annie Robin for valuable feedback during the preparation of this paper. 
We also acknowledge Mark Taylor for creating and implementing new requested features in the astronomy-oriented data handling and visualization software TOPCAT
\citep{2005ASPC..347...29T}. This work has made use of data from the ESA space mission \textit{Gaia}, processed by the \textit{Gaia} Data Processing and Analysis Consortium (DPAC).\\ 

Funding for the DPAC has been provided by national institutions, some of which participate in the \textit{Gaia} Multilateral Agreement, 
which include, 
for Switzerland, the Swiss State Secretariat for Education, Research and Innovation through the ESA PRODEX program, the ``Mesures d'accompagnement'', the ``Activit\'{e}s Nationales Compl\'{e}mentaires'', the Swiss National Science Foundation, and the Early Postdoc. Mobility fellowship;
 Belgium, the BELgian federal Science Policy Office (BELSPO) through PRODEX grants;
for Italy, Istituto Nazionale di Astrofisica (INAF) and the Agenzia Spaziale Italiana (ASI) through grants I/037/08/0,  I/058/10/0,  2014-025-R.0, and 2014-025-R.1.2015 to INAF (PI M.G. Lattanzi);
for France, the Centre National d'Etudes Spatiales (CNES). This research has received additional funding from: the Israeli Centers for Research Excellence (I-CORE, grant No.~1829/12); 
the Hungarian Academy of Sciences through the Lend\"ulet Programme LP2014-17 and the J\'anos Bolyai Research Scholarship (L.~Moln\'ar and E.~Plachy); 
the Hungarian National Research, Development, and Innovation Office through grants OTKA/NKFIH K-113117, K-115709, PD-116175, and PD-121203; 
the \'UNKP-17-3 program of the Ministry of Human Capacities of Hungary (\'A.L.J.); 
the ESA Contract No.~4000121377/17/NL/CBi; and the Polish National Science Center through ETIUDA grant 2016/20/T/ST9/00170; Charles University in Prague through PRIMUS/SCI/17 award; and from the EuropeanResearch Council (ERC) under the European Union's Horizon 2020
research and innovation programme (grant agreement N$^\circ$670519: MAMSIE).

\end{acknowledgements}

\bibliographystyle{aa}
\raggedbottom
\bibliography{local}

%
\begin{sidewaystable*}
{\footnotesize\begin{minipage}[t][180mm]{\textwidth}
\caption{Statistical parameters for the \textit{nTransits:2+} classification types published in DR2. The 1\%, 50\% (median), and 99\% percentile
    are provided for each type.\label{tbl:tsrStatGeq2}}
\begin{tabular}{lr@{\,\,\,\,}r@{\,\,\,\,}rr@{\,\,\,\,}r@{\,\,\,\,}rr@{\,\,\,\,}r@{\,\,\,\,}rr@{\,\,\,\,}r@{\,\,\,\,}r}
\hline\hline
\relax\\[-1.7ex]
$G$ band  & \multicolumn{3}{c@{}}{Cepheids\tablefootmark{a}} & \multicolumn{3}{c@{}}{RR Lyrae stars\tablefootmark{b}} & \multicolumn{3}{c@{}}{Long-period var.\tablefootmark{c}}
        & \multicolumn{3}{c@{}}{$\delta$ Sct and SX Phe\tablefootmark{d}}\\
\cmidrule(l){2-4}
\cmidrule(l){5-7}
\cmidrule(l){8-10}
\cmidrule(l){11-13}
Filtered number of FoV transits & $13$ & $27$ & $120$ & $7$ & $19$ & $69$ & $12$ & $24$ & $65$ & $9$ & $21$ & $64$ \\ 
Time duration (d) & $513.15$ & $627.70$ & $667.73$ & $383.37$ & $588.36$ & $652.89$ & $426.47$ & $591.70$ & $638.78$ & $402.04$ & $572.46$ & $654.34$ \\ 
Mean observation time (d) & $1741.28$ & $2018.52$ & $2112.72$ & $1800.95$ & $2026.53$ & $2156.08$ & $1895.78$ & $2037.13$ & $2155.47$ & $1862.23$ & $2023.30$ & $2142.49$ \\ 
Min. magnitude & $7.48$ & $15.61$ & $17.60$ & $13.35$ & $17.75$ & $19.81$ & $8.01$ & $13.25$ & $17.07$ & $11.06$ & $19.50$ & $20.59$ \\ 
Max. magnitude & $8.03$ & $16.15$ & $18.21$ & $14.10$ & $18.42$ & $20.73$ & $8.84$ & $14.01$ & $17.79$ & $11.36$ & $19.80$ & $21.09$ \\ 
Mean magnitude & $7.80$ & $15.91$ & $17.97$ & $13.81$ & $18.15$ & $20.34$ & $8.47$ & $13.63$ & $17.35$ & $11.25$ & $19.69$ & $20.84$ \\ 
Median magnitude & $7.77$ & $15.93$ & $18.02$ & $13.85$ & $18.19$ & $20.43$ & $8.45$ & $13.61$ & $17.34$ & $11.25$ & $19.69$ & $20.86$ \\ 
Magnitude range & $0.19$ & $0.56$ & $1.14$ & $0.30$ & $0.74$ & $1.19$ & $0.21$ & $0.41$ & $3.03$ & $0.14$ & $0.34$ & $0.65$ \\ 
Standard deviation (mag) & $0.06$ & $0.18$ & $0.38$ & $0.10$ & $0.23$ & $0.39$ & $0.06$ & $0.12$ & $1.04$ & $0.04$ & $0.10$ & $0.20$ \\ 
Skewness of magnitudes & $-1.68$ & $-0.27$ & $1.09$ & $-2.14$ & $-0.53$ & $0.42$ & $-1.35$ & $0.16$ & $1.74$ & $-2.12$ & $-0.54$ & $0.34$ \\ 
Kurtosis of magnitudes & $-1.88$ & $-1.12$ & $2.81$ & $-2.04$ & $-0.90$ & $5.43$ & $-1.90$ & $-0.73$ & $3.98$ & $-1.72$ & $-0.51$ & $6.34$ \\ 
MAD (mag) & $0.05$ & $0.18$ & $0.47$ & $0.06$ & $0.20$ & $0.48$ & $0.02$ & $0.12$ & $1.36$ & $0.03$ & $0.09$ & $0.23$ \\ 
Abbe & $0.12$ & $0.65$ & $1.19$ & $0.49$ & $0.97$ & $1.47$ & $0.06$ & $0.31$ & $0.64$ & $0.53$ & $1.02$ & $1.55$ \\ 
IQR (mag) & $0.09$ & $0.28$ & $0.73$ & $0.11$ & $0.33$ & $0.74$ & $0.04$ & $0.18$ & $2.03$ & $0.05$ & $0.13$ & $0.36$  \\
\cmidrule(l){2-13}\\[-3mm]
$G_{\rm BP}$ band\tablefootmark{e}  \\
Filtered number of FoV transits & $11$ & $27$ & $120$ & $2$ & $16$ & $65$ & $9$ & $23$ & $63$ & $2$ & $18$ & $62$ \\ 
Time duration (d) & $495.78$ & $627.60$ & $667.05$ & $90.71$ & $574.38$ & $653.54$ & $409.84$ & $589.70$ & $638.71$ & $209.69$ & $563.26$ & $654.81$ \\ 
Mean observation time (d) & $1741.55$ & $2018.59$ & $2117.48$ & $1786.25$ & $2024.36$ & $2197.56$ & $1894.31$ & $2034.85$ & $2164.59$ & $1843.47$ & $2021.86$ & $2171.98$ \\ 
Min. magnitude & $7.93$ & $15.74$ & $17.53$ & $12.97$ & $17.39$ & $20.17$ & $10.11$ & $15.69$ & $19.04$ & $11.21$ & $17.64$ & $20.29$ \\ 
Max. magnitude & $8.67$ & $16.52$ & $18.65$ & $14.39$ & $18.89$ & $22.66$ & $11.21$ & $17.08$ & $22.49$ & $11.57$ & $20.16$ & $22.85$ \\ 
Mean magnitude & $8.33$ & $16.21$ & $18.11$ & $14.06$ & $18.41$ & $20.99$ & $10.78$ & $16.49$ & $20.53$ & $11.44$ & $19.55$ & $20.94$ \\ 
Median magnitude & $8.30$ & $16.24$ & $18.16$ & $14.12$ & $18.44$ & $20.88$ & $10.73$ & $16.47$ & $20.50$ & $11.45$ & $19.60$ & $20.79$ \\ 
Magnitude range & $0.27$ & $0.79$ & $2.85$ & $0.29$ & $1.16$ & $5.80$ & $0.34$ & $0.91$ & $5.95$ & $0.18$ & $1.34$ & $6.14$ \\ 
Standard deviation (mag) & $0.08$ & $0.23$ & $0.57$ & $0.12$ & $0.35$ & $1.64$ & $0.10$ & $0.25$ & $1.87$ & $0.06$ & $0.34$ & $1.67$ \\ 
Skewness of magnitudes & $-3.97$ & $-0.35$ & $1.26$ & $-3.73$ & $-0.54$ & $1.73$ & $-3.28$ & $0.02$ & $1.68$ & $-4.39$ & $-0.66$ & $1.86$ \\ 
Kurtosis of magnitudes & $-1.82$ & $-0.86$ & $19.92$ & $-2.01$ & $-0.02$ & $17.61$ & $-1.88$ & $-0.51$ & $14.54$ & $-1.81$ & $1.02$ & $22.96$ \\ 
MAD (mag) & $0.06$ & $0.22$ & $0.57$ & $0.05$ & $0.29$ & $1.11$ & $0.05$ & $0.23$ & $2.36$ & $0.04$ & $0.22$ & $1.14$ \\ 
Abbe & $0.14$ & $0.69$ & $1.24$ & $0.41$ & $0.99$ & $1.56$ & $0.09$ & $0.39$ & $1.11$ & $0.43$ & $1.00$ & $1.55$ \\ 
IQR (mag) & $0.10$ & $0.34$ & $0.87$ & $0.13$ & $0.46$ & $2.07$ & $0.08$ & $0.35$ & $3.51$ & $0.06$ & $0.33$ & $1.95$  \\
\cmidrule(l){2-13}\\[-3mm]
$G_{\rm RP}$ band\tablefootmark{e}  \\
Filtered number of FoV transits & $11$ & $27$ & $118$ & $2$ & $17$ & $66$ & $10$ & $23$ & $63$ & $3$ & $18$ & $62$ \\ 
Time duration (d) & $498.77$ & $627.60$ & $667.05$ & $94.90$ & $575.03$ & $654.29$ & $425.11$ & $589.95$ & $638.71$ & $214.30$ & $563.59$ & $654.80$ \\ 
Mean observation time (d) & $1741.55$ & $2017.73$ & $2117.87$ & $1785.99$ & $2023.48$ & $2194.71$ & $1894.72$ & $2034.49$ & $2161.65$ & $1841.08$ & $2020.79$ & $2167.98$ \\ 
Min. magnitude & $6.83$ & $14.89$ & $16.86$ & $11.43$ & $16.25$ & $18.66$ & $6.66$ & $11.74$ & $15.24$ & $10.27$ & $16.96$ & $19.90$ \\ 
Max. magnitude & $7.42$ & $15.58$ & $17.70$ & $13.60$ & $17.48$ & $20.44$ & $7.44$ & $12.49$ & $16.09$ & $11.09$ & $19.92$ & $21.91$ \\ 
Mean magnitude & $7.14$ & $15.37$ & $17.36$ & $13.36$ & $17.18$ & $19.34$ & $7.09$ & $12.13$ & $15.66$ & $10.95$ & $19.29$ & $20.54$ \\ 
Median magnitude & $7.12$ & $15.38$ & $17.42$ & $13.37$ & $17.21$ & $19.38$ & $7.07$ & $12.11$ & $15.65$ & $10.95$ & $19.36$ & $20.53$ \\ 
Magnitude range & $0.19$ & $0.58$ & $4.84$ & $0.09$ & $0.76$ & $7.34$ & $0.16$ & $0.39$ & $2.95$ & $0.14$ & $1.43$ & $7.68$ \\ 
Standard deviation (mag) & $0.06$ & $0.16$ & $0.90$ & $0.05$ & $0.22$ & $1.63$ & $0.05$ & $0.11$ & $0.97$ & $0.04$ & $0.38$ & $1.74$ \\ 
Skewness of magnitudes & $-5.88$ & $-0.40$ & $1.17$ & $-5.04$ & $-0.59$ & $1.77$ & $-3.74$ & $0.14$ & $1.78$ & $-5.24$ & $-0.78$ & $2.05$ \\ 
Kurtosis of magnitudes & $-1.83$ & $-0.54$ & $44.60$ & $-2.44$ & $0.54$ & $28.74$ & $-1.91$ & $-0.62$ & $16.83$ & $-1.76$ & $1.59$ & $30.45$ \\ 
MAD (mag) & $0.04$ & $0.15$ & $0.39$ & $0.02$ & $0.18$ & $0.59$ & $0.02$ & $0.10$ & $1.25$ & $0.03$ & $0.20$ & $1.06$ \\ 
Abbe & $0.14$ & $0.75$ & $1.25$ & $0.41$ & $1.00$ & $1.56$ & $0.08$ & $0.36$ & $1.06$ & $0.45$ & $1.01$ & $1.55$ \\ 
IQR (mag) & $0.07$ & $0.22$ & $0.61$ & $0.05$ & $0.27$ & $1.37$ & $0.03$ & $0.16$ & $1.86$ & $0.05$ & $0.30$ & $1.63$  \\
\hline
\end{tabular}
\tablefoot{
\tablefoottext{a}{Cepheid groups: \texttt{best\_class\_name = CEP, ACEP, T2CEP}}
\tablefoottext{b}{RR Lyrae star groups: \texttt{best\_class\_name = RRAB, RRC, RRD, ARRD}}
\tablefoottext{c}{Long-period var. is \texttt{best\_class\_name = MIRA\_SR}}
\tablefoottext{d}{$\delta$ Sct and SX Phe is: \texttt{best\_class\_name = DSCT\_SXPHE}}
\tablefoottext{e}{For the skewness and kurtosis, we included only stars with $>2$ and $>3$ FoV transits, respectively.}
}
\vfill
\end{minipage}
}\end{sidewaystable*}


\begin{sidewaystable*}
{\footnotesize\begin{minipage}[t][180mm]{\textwidth}
\caption{Statistical parameters for the SOS types published in DR2. The 1\%, 50\% (median), and 99\% percentile
    are provided for each type.\label{tbl:tsrStatSOS}}
\begin{tabular}{lr@{\,\,\,\,}r@{\,\,\,\,}rr@{\,\,\,\,}r@{\,\,\,\,}rr@{\,\,\,\,}r@{\,\,\,\,}rr@{\,\,\,\,}r@{\,\,\,\,}rr@{\,\,\,\,}r@{\,\,\,\,}r}
\hline\hline
\relax\\[-1.7ex]
$G$ band  & \multicolumn{3}{c@{}}{Cepheids} & \multicolumn{3}{c@{}}{RR Lyrae stars} & \multicolumn{3}{c@{}}{Long-period var.}
        & \multicolumn{3}{c@{}}{Rot.~Modulation} & \multicolumn{3}{c@{}}{Short timesc.}\\
\cmidrule(l){2-4}
\cmidrule(l){5-7}
\cmidrule(l){8-10}
\cmidrule(l){11-13}
\cmidrule(l){14-16}
Filtered number of FoV transits & $15$ & $28$ & $118$ & $13$ & $27$ & $81$ & $12$ & $27$ & $69$ & $24$ & $43$ & $88$ & $19$ & $32$ & $69$ \\ 
Time duration (d) & $514.00$ & $627.70$ & $667.48$ & $430.10$ & $593.78$ & $657.13$ & $427.22$ & $591.78$ & $638.78$ & $481.77$ & $580.21$ & $651.04$ & $475.32$ & $587.09$ & $642.99$ \\ 
Mean observation time (d) & $1741.28$ & $2018.29$ & $2114.19$ & $1761.88$ & $2022.56$ & $2163.10$ & $1887.55$ & $2036.56$ & $2160.34$ & $1835.54$ & $2027.26$ & $2183.81$ & $1844.59$ & $2026.86$ & $2163.03$ \\ 
Min. magnitude & $8.07$ & $15.82$ & $17.98$ & $13.08$ & $17.59$ & $20.18$ & $7.87$ & $13.08$ & $16.93$ & $12.38$ & $16.07$ & $18.86$ & $16.13$ & $16.90$ & $18.47$ \\ 
Max. magnitude & $8.57$ & $16.33$ & $18.56$ & $13.74$ & $18.24$ & $20.95$ & $8.97$ & $14.12$ & $17.81$ & $12.43$ & $16.13$ & $19.04$ & $16.64$ & $17.37$ & $19.76$ \\ 
Mean magnitude & $8.39$ & $16.11$ & $18.29$ & $13.49$ & $17.98$ & $20.61$ & $8.50$ & $13.60$ & $17.28$ & $12.40$ & $16.10$ & $18.95$ & $16.51$ & $17.10$ & $18.87$ \\ 
Median magnitude & $8.40$ & $16.13$ & $18.31$ & $13.51$ & $18.02$ & $20.64$ & $8.48$ & $13.59$ & $17.27$ & $12.40$ & $16.10$ & $18.96$ & $16.49$ & $17.08$ & $18.86$ \\ 
Magnitude range & $0.13$ & $0.49$ & $1.12$ & $0.26$ & $0.65$ & $1.33$ & $0.21$ & $0.53$ & $3.15$ & $0.01$ & $0.05$ & $0.30$ & $0.14$ & $0.44$ & $2.43$ \\ 
Standard deviation (mag) & $0.04$ & $0.16$ & $0.37$ & $0.08$ & $0.19$ & $0.37$ & $0.06$ & $0.16$ & $1.08$ & $0.00$ & $0.01$ & $0.08$ & $0.04$ & $0.12$ & $0.56$ \\ 
Skewness of magnitudes & $-1.56$ & $-0.29$ & $0.93$ & $-2.61$ & $-0.44$ & $0.57$ & $-1.26$ & $0.15$ & $1.69$ & $-2.21$ & $0.05$ & $2.24$ & $-0.91$ & $0.38$ & $2.40$ \\ 
Kurtosis of magnitudes & $-1.84$ & $-1.15$ & $2.19$ & $-1.83$ & $-0.91$ & $9.21$ & $-1.94$ & $-0.87$ & $3.40$ & $-1.48$ & $-0.28$ & $13.04$ & $-1.55$ & $-0.47$ & $6.49$ \\ 
MAD (mag) & $0.04$ & $0.17$ & $0.45$ & $0.06$ & $0.19$ & $0.42$ & $0.02$ & $0.15$ & $1.44$ & $0.00$ & $0.01$ & $0.08$ & $0.03$ & $0.11$ & $0.43$ \\ 
Abbe & $0.17$ & $0.67$ & $1.19$ & $0.50$ & $0.97$ & $1.41$ & $0.06$ & $0.28$ & $0.61$ & $0.14$ & $0.54$ & $1.20$ & $0.71$ & $0.99$ & $1.42$ \\ 
IQR (mag) & $0.07$ & $0.26$ & $0.69$ & $0.10$ & $0.30$ & $0.66$ & $0.04$ & $0.24$ & $2.12$ & $0.00$ & $0.02$ & $0.12$ & $0.05$ & $0.16$ & $0.65$  \\
\cmidrule(l){2-16}\\[-3mm]
$G_{\rm BP}$ band\tablefootmark{a}  \\
Filtered number of FoV transits & $12$ & $27$ & $118$ & $2$ & $24$ & $77$ & $9$ & $26$ & $67$ & $23$ & $42$ & $87$ & $18$ & $30$ & $68$ \\ 
Time duration (d) & $493.23$ & $627.60$ & $667.05$ & $182.20$ & $588.20$ & $655.81$ & $421.03$ & $590.19$ & $638.78$ & $461.80$ & $579.89$ & $650.80$ & $427.53$ & $585.24$ & $642.58$ \\ 
Mean observation time (d) & $1741.25$ & $2018.23$ & $2119.45$ & $1755.42$ & $2020.82$ & $2179.33$ & $1885.59$ & $2034.97$ & $2168.89$ & $1834.54$ & $2027.21$ & $2186.35$ & $1837.85$ & $2026.82$ & $2164.76$ \\ 
Min. magnitude & $8.64$ & $15.94$ & $17.88$ & $12.53$ & $17.11$ & $19.70$ & $9.99$ & $15.57$ & $18.88$ & $12.71$ & $16.27$ & $18.80$ & $14.84$ & $17.04$ & $18.47$ \\ 
Max. magnitude & $9.36$ & $16.71$ & $19.12$ & $14.02$ & $18.62$ & $22.37$ & $11.37$ & $17.35$ & $22.62$ & $12.89$ & $16.83$ & $21.59$ & $16.62$ & $17.89$ & $20.30$ \\ 
Mean magnitude & $9.06$ & $16.40$ & $18.54$ & $13.71$ & $18.17$ & $20.67$ & $10.85$ & $16.59$ & $20.46$ & $12.85$ & $16.72$ & $20.25$ & $16.40$ & $17.57$ & $19.25$ \\ 
Median magnitude & $9.04$ & $16.43$ & $18.56$ & $13.74$ & $18.21$ & $20.64$ & $10.80$ & $16.58$ & $20.45$ & $12.85$ & $16.73$ & $20.28$ & $16.41$ & $17.58$ & $19.26$ \\ 
Magnitude range & $0.21$ & $0.76$ & $3.03$ & $0.24$ & $1.20$ & $5.75$ & $0.36$ & $1.17$ & $6.25$ & $0.03$ & $0.35$ & $4.52$ & $0.20$ & $0.76$ & $4.07$ \\ 
Standard deviation (mag) & $0.06$ & $0.22$ & $0.58$ & $0.08$ & $0.33$ & $1.30$ & $0.10$ & $0.32$ & $1.94$ & $0.01$ & $0.06$ & $0.78$ & $0.05$ & $0.16$ & $0.82$ \\ 
Skewness of magnitudes & $-4.18$ & $-0.39$ & $1.15$ & $-4.19$ & $-0.58$ & $1.63$ & $-3.18$ & $0.01$ & $1.61$ & $-6.14$ & $-1.74$ & $2.55$ & $-5.00$ & $-0.60$ & $2.16$ \\ 
Kurtosis of magnitudes & $-1.77$ & $-0.78$ & $21.82$ & $-1.81$ & $0.16$ & $44.52$ & $-1.91$ & $-0.66$ & $14.30$ & $-1.04$ & $6.72$ & $42.92$ & $-1.33$ & $2.34$ & $28.09$ \\ 
MAD (mag) & $0.05$ & $0.21$ & $0.54$ & $0.06$ & $0.27$ & $0.91$ & $0.05$ & $0.29$ & $2.51$ & $0.01$ & $0.03$ & $0.44$ & $0.04$ & $0.11$ & $0.53$ \\ 
Abbe & $0.20$ & $0.72$ & $1.25$ & $0.47$ & $0.99$ & $1.47$ & $0.08$ & $0.35$ & $1.06$ & $0.31$ & $0.94$ & $1.28$ & $0.52$ & $1.02$ & $1.42$ \\ 
IQR (mag) & $0.08$ & $0.32$ & $0.83$ & $0.09$ & $0.41$ & $1.44$ & $0.08$ & $0.45$ & $3.70$ & $0.01$ & $0.04$ & $0.61$ & $0.05$ & $0.16$ & $0.77$  \\
\cmidrule(l){2-16}\\[-3mm]
$G_{\rm RP}$ band\tablefootmark{a}  \\
Filtered number of FoV transits & $12$ & $27$ & $117$ & $2$ & $25$ & $77$ & $10$ & $26$ & $67$ & $23$ & $42$ & $87$ & $18$ & $30$ & $68$ \\ 
Time duration (d) & $493.23$ & $627.60$ & $666.98$ & $184.18$ & $588.70$ & $655.97$ & $425.29$ & $590.45$ & $638.78$ & $474.04$ & $580.14$ & $650.80$ & $428.93$ & $585.42$ & $642.58$ \\ 
Mean observation time (d) & $1741.25$ & $2017.51$ & $2119.45$ & $1756.01$ & $2019.99$ & $2178.16$ & $1887.12$ & $2034.64$ & $2166.20$ & $1834.20$ & $2026.69$ & $2185.00$ & $1838.43$ & $2025.96$ & $2164.88$ \\ 
Min. magnitude & $7.37$ & $15.06$ & $17.21$ & $11.24$ & $16.05$ & $18.85$ & $6.51$ & $11.55$ & $15.17$ & $11.46$ & $14.93$ & $17.08$ & $11.54$ & $15.75$ & $17.30$ \\ 
Max. magnitude & $7.92$ & $15.76$ & $18.08$ & $13.27$ & $17.45$ & $20.97$ & $7.55$ & $12.58$ & $16.14$ & $11.84$ & $15.38$ & $17.89$ & $15.57$ & $16.43$ & $19.07$ \\ 
Mean magnitude & $7.61$ & $15.54$ & $17.70$ & $13.05$ & $17.14$ & $19.72$ & $7.10$ & $12.08$ & $15.64$ & $11.80$ & $15.32$ & $17.63$ & $15.42$ & $16.21$ & $18.11$ \\ 
Median magnitude & $7.60$ & $15.57$ & $17.75$ & $13.05$ & $17.18$ & $19.76$ & $7.06$ & $12.06$ & $15.64$ & $11.80$ & $15.33$ & $17.68$ & $15.43$ & $16.22$ & $18.13$ \\ 
Magnitude range & $0.15$ & $0.57$ & $5.26$ & $0.12$ & $0.91$ & $7.62$ & $0.17$ & $0.51$ & $3.08$ & $0.02$ & $0.21$ & $3.62$ & $0.12$ & $0.59$ & $4.83$ \\ 
Standard deviation (mag) & $0.05$ & $0.15$ & $0.94$ & $0.04$ & $0.24$ & $1.46$ & $0.05$ & $0.15$ & $1.01$ & $0.01$ & $0.04$ & $0.58$ & $0.03$ & $0.13$ & $0.90$ \\ 
Skewness of magnitudes & $-5.86$ & $-0.48$ & $1.02$ & $-5.41$ & $-0.72$ & $1.68$ & $-3.75$ & $0.13$ & $1.73$ & $-6.68$ & $-2.83$ & $2.76$ & $-5.48$ & $-0.74$ & $2.02$ \\ 
Kurtosis of magnitudes & $-1.77$ & $-0.41$ & $45.26$ & $-1.77$ & $1.15$ & $62.16$ & $-1.95$ & $-0.76$ & $17.23$ & $-1.11$ & $11.49$ & $49.34$ & $-1.57$ & $2.68$ & $33.68$ \\ 
MAD (mag) & $0.04$ & $0.14$ & $0.37$ & $0.03$ & $0.19$ & $0.55$ & $0.02$ & $0.14$ & $1.32$ & $0.00$ & $0.02$ & $0.11$ & $0.02$ & $0.09$ & $0.41$ \\ 
Abbe & $0.21$ & $0.78$ & $1.27$ & $0.48$ & $1.00$ & $1.46$ & $0.07$ & $0.32$ & $1.01$ & $0.31$ & $0.95$ & $1.24$ & $0.52$ & $1.04$ & $1.44$ \\ 
IQR (mag) & $0.06$ & $0.21$ & $0.57$ & $0.04$ & $0.27$ & $0.85$ & $0.04$ & $0.21$ & $1.96$ & $0.01$ & $0.02$ & $0.17$ & $0.03$ & $0.13$ & $0.61$  \\
\hline

\end{tabular}
\tablefoot{
\tablefoottext{a}{For RR Lyrae stars, the percentiles for $G_{\rm BP}$ and $G_{\rm RP}$ were computed for a number of FoV transits $>1$,
    although there are stars with $0$ or $1$ transit remaining in the band after our cleaning. For the skewness and kurtosis, we
    included only RR Lyrae stars with $>2$ and $>3$ transits, respectively.}
}
\vfill
\end{minipage}
}\end{sidewaystable*}

\begin{appendix}
\onecolumn
\section{Examples of \textit{Gaia} archive queries \label{sec:archiveQueries}}
This section describes Gaia archive queries in the ADQL format that return the various selections presented in this paper. These queries can be made online in the \textit{Gaia} archive at \href{http://gea.esac.esa.int/archive/}{\tt http://gea.esac.esa.int/archive/}:\\

\subsection{Total counts and table merging}

      Figure~\ref{fig:globalNumbers} total number of unique sources with variability results (bottom left unique \texttt{source\_id} count):
      \begin{footnotesize}\begin{verbatim}
      select count(*) from gaiadr2.gaia_source where phot_variable_flag='VARIABLE'
      \end{verbatim} \end{footnotesize}

      Retrieve various astrometric and general statistical variability fields for all published variable sources in \textit{Gaia} DR2:
      \begin{footnotesize}\begin{verbatim}
      select gs.source_id, ra, dec, parallax, parallax_error, pmra, pmdec, epoch_photometry_url, 
        num_selected_g_fov, mean_mag_g_fov, range_mag_g_fov, std_dev_mag_g_fov, time_duration_g_fov,
        num_selected_bp, mean_mag_bp, range_mag_bp, std_dev_mag_bp, time_duration_bp,
        num_selected_rp, mean_mag_rp, range_mag_rp, std_dev_mag_rp, time_duration_rp
        from gaiadr2.gaia_source as gs join gaiadr2.vari_time_series_statistics as varistats using(source_id) 
        where phot_variable_flag='VARIABLE'
      \end{verbatim} \end{footnotesize}

\subsection{Creating a variable summary table\label{summaryTable}}

It is very desirable to have an overview of the occurrence of a given source\_id in any of the vari\_* tables. The following query generates such a table in which each row contains a unique variable source \texttt{source\_id}, and each column represents one of the vari-tables. A column contains integer 1 when it contains the \texttt{source\_id} and \textit{NULL} when not. After logging into the Gaia archive  a user table can be created from the query result (suggested name: \texttt{gaiadr2\_vari\_summary}), after which it appears in the user tables and can be queried like any other DR2 table.
     \begin{footnotesize}
      \begin{verbatim}
 select
   gs.source_id
   ,vrrl.incl as vari_rrlyrae   -- 1 if RR Lyrae SOS table entry
   ,vcep.incl as vari_cepheid   -- 1 if Cepheid SOS table entry
   ,vrm.incl as vari_rotation_modulation   -- 1 if rotation modulation SOS table entry
   ,vsts.incl as vari_short_timescale   -- 1 if short timescale SOS table entry
   ,vlpv.incl as vari_long_period_variable   -- 1 if long period variable SOS table entry
   ,vcr_rrl.incl as vari_classifier_result_rrl   -- 1 if classif. entry of any RR Lyrae type
   ,vcr_cep.incl as vari_classifier_result_cep   -- 1 if classif. entry of any Cepheid type
   ,vcr_dscsxp.incl as vari_classifier_result_dsct_sxphe -- 1 if classif. entry of Delta Scuti
   ,vcr_mirasr.incl as vari_classifier_result_mira_sr   -- 1 if classif. entry of Mira or semi-regular type
   ,1 as vari_time_series_statistics   -- 1 if time series statistics entry (1 for all in table)
   ,1 as gaia_source_variable_flag   -- 1 if gaia_source phot_variable_flag='VARIABLE' (1 for all in table)
   from gaiadr2.gaia_source as gs 
   LEFT OUTER join (select source_id, 1 as incl from gaiadr2.vari_rrlyrae) as vrrl using(source_id)
   LEFT OUTER join (select source_id, 1 as incl from gaiadr2.vari_cepheid) as vcep using(source_id)
   LEFT OUTER join (select source_id, 1 as incl from gaiadr2.vari_rotation_modulation) as vrm using(source_id)
   LEFT OUTER join (select source_id, 1 as incl from gaiadr2.vari_short_timescale) as vsts using(source_id)
   LEFT OUTER join (select source_id, 1 as incl from gaiadr2.vari_long_period_variable) as vlpv using(source_id)
   LEFT OUTER join (select source_id, 1 as incl from gaiadr2.vari_classifier_result where 
                    best_class_name='RRAB' or best_class_name='RRC' or best_class_name='RRD' 
                    or best_class_name='ARRD') as vcr_rrl using(source_id)
   LEFT OUTER join (select source_id, 1 as incl from gaiadr2.vari_classifier_result where 
                    best_class_name='CEP' or best_class_name='ACEP' or best_class_name='T2CEP') 
                   as vcr_cep using(source_id)
   LEFT OUTER join (select source_id, 1 as incl from gaiadr2.vari_classifier_result where 
                    best_class_name='DSCT_SXPHE') as vcr_dscsxp using(source_id)
   LEFT OUTER join (select source_id, 1 as incl from gaiadr2.vari_classifier_result where 
                    best_class_name='MIRA_SR') as vcr_mirasr using(source_id)
   where gs.phot_variable_flag='VARIABLE'
   order by gs.source_id 
      \end{verbatim} 
      \end{footnotesize}

\subsection{HEALPix grouped data}
To reproduce sky-distribution data sets using HEALPix grouping as shown in Figures~\ref{fig:skyDensities} and \ref{fig:numTransitsHist}, the convenient  \texttt{gaia\_healpix\_index(norder, source\_id)} function in the archive that extracts the HEALPix pixel id that is encoded in the Gaia \texttt{source\_id} can be used. Using \texttt{norder=6} creates pixels of about 0.84~deg$^2$. A program like TOPCAT \citep{2005ASPC..347...29T} can read and visualise such data set with ease.
   
Figure~\ref{fig:skyDensities} top left panel `Classif.: RR Lyrae stars (RRAB, RRC, RRD, ARRD)' number of sources per square degree, and also the min, average, max, and standard deviation of the \texttt{vari\_classifier\_result.best\_class\_score} value per HEALPix pixel:
      \begin{footnotesize}\begin{verbatim}   
  select
    gaia_healpix_index(6, source_id) AS healpix_6,
    count(*) / 0.83929 as sources_per_sq_deg,
    min(best_class_score) AS min_best_class_score,
    avg(best_class_score) AS avg_best_class_score,
    max(best_class_score)  AS max_best_class_score,
    stddev(best_class_score) as stddec_int_average_g
    FROM gaiadr2.vari_classifier_result WHERE 
       best_class_name='RRAB' or best_class_name='RRC' or best_class_name='RRD' or best_class_name='ARRD'
    GROUP BY healpix_6  
    \end{verbatim} \end{footnotesize}

Figure~\ref{fig:skyDensities} top right panel `SOS: RR Lyrae stars': number of sources per square degree, and also the min, average, max, and standard deviation of the \texttt{vari\_rrlyrae.int\_average\_g} value per HEALPix pixel:
      \begin{footnotesize}\begin{verbatim}   
  select
    gaia_healpix_index(6, source_id) AS healpix_6,
    count(*) / 0.83929 as sources_per_sq_deg,
    min(int_average_g) AS min_int_average_g,
    avg(int_average_g) AS avg_int_average_g,
    max(int_average_g)  AS max_int_average_g,
    stddev(int_average_g) as stddec_int_average_g
    FROM gaiadr2.vari_rrlyrae
    GROUP BY healpix_6  
    \end{verbatim} \end{footnotesize}

Figure~\ref{fig:numTransitsHist} all three panels (`number of transits in time series of the $G$, $G_{\rm BP}$, and $G_{\rm RP}$ photometric bands') average, and also the min, max, and standard deviation of the values per HEALPix pixel:
      \begin{footnotesize}\begin{verbatim}   
   select
     gaia_healpix_index(6, source_id) AS healpix_6,
     count(*) / 0.83929 as sources_per_sq_deg,
     -- G-band FoV
     avg(num_selected_g_fov) AS avg_num_selected_g_fov,
     min(num_selected_g_fov) AS min_num_selected_g_fov,
     max(num_selected_g_fov)  AS max_num_selected_g_fov,
     stddev(num_selected_bp) as stddec_num_selected_g_fov,
     -- BP-band
     avg(num_selected_bp) AS avg_num_selected_bp,
     min(num_selected_bp) AS min_num_selected_bp,
     max(num_selected_bp)  AS max_num_selected_bp,
     stddev(num_selected_bp) as stddec_num_selected_bp,
     -- RP-band
     avg(num_selected_rp) AS avg_num_selected_rp,
     min(num_selected_rp) AS min_num_selected_rp,
     max(num_selected_rp)  AS max_num_selected_rp,
     stddev(num_selected_rp) as stddec_num_selected_rp
     FROM gaiadr2.vari_time_series_statistics
    GROUP BY healpix_6
    \end{verbatim} \end{footnotesize}

\subsection{Detailed counting\label{detailedCounting}}
This section shows how the counts presented in Figures~\ref{fig:globalNumbers} and \ref{fig:histMags} can be reproduced from the published DR2 tables. Alternatively, a summary table can first be created as described in appendix~\ref{summaryTable} and the relevant selections on the columns involved (not shown) can be  made.
      
      Figures~\ref{fig:globalNumbers} and \ref{fig:histMags} total number of sources with classification: all / $\delta$ Scuti and SX Phoenicis / RR Lyrae stars / Cepheids / LPV (Mira and semi-regular):
      \begin{footnotesize}\begin{verbatim}   
      select count(*) from gaiadr2.vari_classifier_result
      select count(*) from gaiadr2.vari_classifier_result where best_class_name='DSCT_SXPHE'
      select count(*) from gaiadr2.vari_classifier_result where best_class_name='RRAB'
          or best_class_name='RRC' or best_class_name='RRD' or best_class_name='ARRD'
      select count(*) from gaiadr2.vari_classifier_result where best_class_name='CEP'
          or best_class_name='ACEP' or best_class_name='T2CEP'
      select count(*) from gaiadr2.vari_classifier_result where best_class_name='MIRA_SR'
      \end{verbatim} \end{footnotesize}

      Figures~\ref{fig:globalNumbers} and \ref{fig:histMags} total number of sources with SOS: RR Lyrae / Cepheid / LPV / short timescale / rotation modulation results:
      \begin{footnotesize}\begin{verbatim}
      select count(*) from gaiadr2.vari_rrlyrae 
      select count(*) from gaiadr2.vari_cepheid
      select count(*) from gaiadr2.vari_long_period_variable
      select count(*) from gaiadr2.vari_short_timescale
      select count(*) from gaiadr2.vari_rotation_modulation
      \end{verbatim}\end{footnotesize}

      Figure~\ref{fig:globalNumbers} number of sources in common\textbf{\textbf{\textup{\textbf{}}}} between classification and SOS for: RR Lyrae  / Cepheid / LPV in \textit{Gaia} DR2,\\
      these are the numbers in the overlap\textbf{} blocks of the Venn-diagrams:
      \begin{footnotesize}\begin{verbatim}
      select count(*) from gaiadr2.vari_classifier_result as clasrrl 
        INNER join gaiadr2.vari_rrlyrae as sosrrl on clasrrl.source_id=sosrrl.source_id 
        where best_class_name='RRAB' or best_class_name='RRC' or best_class_name='RRD' 
        or best_class_name='ARRD'
      select count(*) from gaiadr2.vari_classifier_result as clascep 
        INNER join gaiadr2.vari_cepheid as soscep on clascep.source_id=soscep.source_id 
        where best_class_name='CEP' or best_class_name='ACEP' or best_class_name='T2CEP'
      select count(*) from gaiadr2.vari_classifier_result as claslpv
        INNER join gaiadr2.vari_long_period_variable as soslpv 
        on claslpv.source_id=soslpv.source_id 
        where best_class_name='MIRA_SR'
      \end{verbatim} \end{footnotesize}
      
      Figure~\ref{fig:globalNumbers} number of sources that are
only found in classification and not\textbf{} in SOS for RR Lyrae  / Cepheid / LPV in \textit{Gaia} DR2,\\
      these are the numbers in the left\textbf{} blocks of the Venn-diagrams:
      \begin{footnotesize}\begin{verbatim}
      select count(*) from gaiadr2.vari_classifier_result as clasrrl 
        LEFT OUTER join gaiadr2.vari_rrlyrae as sosrrl on clasrrl.source_id=sosrrl.source_id 
        where (best_class_name='RRAB' or best_class_name='RRC' or best_class_name='RRD' 
        or best_class_name='ARRD') and sosrrl.source_id is NULL
      select count(*) from gaiadr2.vari_classifier_result as clascep 
        LEFT OUTER join gaiadr2.vari_cepheid as soscep on clascep.source_id=soscep.source_id 
        where (best_class_name='CEP' or best_class_name='ACEP' or best_class_name='T2CEP')
        and soscep.source_id is NULL
      select count(*) from gaiadr2.vari_classifier_result as claslpv 
        LEFT OUTER join gaiadr2.vari_long_period_variable as soslpv 
        on claslpv.source_id=soslpv.source_id 
        where best_class_name='MIRA_SR' and soslpv.source_id is NULL
      \end{verbatim} \end{footnotesize}
      
     Figure~\ref{fig:globalNumbers}  number of sources that are
only found in SOS and not\textbf{} in classification for: RR Lyrae  / Cepheid / LPV in \textit{Gaia} DR2,\\
      these are the numbers in the right\textbf{} blocks of the Venn-diagrams:
      \begin{footnotesize}\begin{verbatim}
      select count(*) from gaiadr2.vari_classifier_result as clasrrl 
        RIGHT OUTER join gaiadr2.vari_rrlyrae as sosrrl on clasrrl.source_id=sosrrl.source_id 
        where (best_class_name!='RRAB' and best_class_name!='RRC' and best_class_name!='RRD' 
        and best_class_name!='ARRD') or clasrrl.source_id is NULL
      select count(*) from gaiadr2.vari_classifier_result as clascep 
        RIGHT OUTER join gaiadr2.vari_cepheid as soscep on clascep.source_id=soscep.source_id 
        where (best_class_name!='CEP' and best_class_name!='ACEP' and best_class_name!='T2CEP')
        or clascep.source_id is NULL
      select count(*) from gaiadr2.vari_classifier_result as claslpv 
        RIGHT OUTER join gaiadr2.vari_long_period_variable as soslpv 
        on claslpv.source_id=soslpv.source_id 
        where best_class_name!='MIRA_SR' or claslpv.source_id is NULL
      \end{verbatim} \end{footnotesize}

\end{appendix}

\end{document}